\documentclass[a4paper,11pt,notitlepage]{article}

\usepackage{a4wide,epsfig,graphics,amsthm,amsfonts,amssymb,amsopn,amsmath,longtable,dcolumn,calc,verbatim,threeparttable,eurosym,lscape,rotating,multirow, color, ulem, threeparttablex, environ, booktabs, hyphenat}
\usepackage[paper=a4paper,left=25mm,right=25mm,top=28mm,bottom=28mm]{geometry}
\usepackage[utf8]{inputenc}
\usepackage{bm} 
\usepackage{lscape}
\usepackage{natbib}
\usepackage{caption}
\usepackage{pdfpages}
\usepackage{ragged2e,array,longtable}
\usepackage{arydshln}
\usepackage{booktabs}
\usepackage[onehalfspacing]{setspace}
\usepackage{bbm}
\newcolumntype{.}{D{.}{.}{-1}}
\newcolumntype{d}[1]{D{.}{.}{#1}}

\renewcommand{\baselinestretch}{1.5} 
\theoremstyle{definition}

\theoremstyle{plain}
\usepackage{diagbox}
\usepackage{graphicx}
\usepackage[colorlinks=true,linkcolor=blue,citecolor=blue,backref=page]{hyperref}
\usepackage{paralist}
\usepackage[bottom]{footmisc}
\usepackage{subcaption}

\newcommand{\bi}{\begin{itemize}}
\newcommand{\ei}{\end{itemize}}

\newcommand{\bc}{\begin{center}}
\newcommand{\ec}{\end{center}}

\newcommand{\bs}{\begin{scriptsize}}
\newcommand{\es}{\end{scriptsize}}
\newcommand{\beq}{\begin{equation}}
\newcommand{\eeq}{\end{equation}}
\newcommand{\ben}{\begin{enumerate}}
\newcommand{\een}{\end{enumerate}}
\newcommand{\bean}{\begin{eqnarray}}
\newcommand{\eean}{\end{eqnarray}}

\DeclareMathSymbol{\R}{\mathalpha}{AMSb}{"52}
\DeclareMathSymbol{\E}{\mathalpha}{AMSb}{"45}
\newcommand{\sym}[1]{$^{#1}$} 

\usepackage{amssymb}
\usepackage{pifont}
\usepackage{float}

\def\sym#1{\ifmmode^{#1}\else\(^{#1}\)\fi}

\usepackage[section]{placeins}
\newlength{\asdf}
\usepackage{xargs}   
\usepackage[pdftex,dvipsnames]{xcolor}  
\usepackage[colorinlistoftodos,prependcaption,textsize=tiny]{todonotes}

\begin{document}

\renewcommand{\baselinestretch}{1.12}
\title{\vspace{-10mm}\textbf{Economic Shocks, Opportunity Costs, and the Supply of Politicians}\thanks{We are grateful to Claudio Ferraz, Patrick Francois, Krisztina Kis-Katos, Bouke Klein Teeselink, Tommy Krieger, Quentin Lippmann, Pierre-Guillaume M\'eon, Konstantinos Matakos, Nathan Nunn, Mounu Prem, Simon Quinn, Johanna Rickne, Manuel Santos Silva, and participants at the Göttingen-Braunschweig Internal Workshop, BeNA Winter Workshop, Iberoamerican Brown-Bag Seminar, Harvard Political Economy Graduate Workshop, Göttingen Development Economics Seminar, Applied Young Economist Webinar, German Development Economics Conference, RIDGE/LACEA Workshop on Political Economy, Verein für Socialpolitik Conference, EJPE-CEPR Political Economy Conference, Research Seminar in Economics at the Free University of Berlin, Workshop on Economics \& Politics, Political Economy of Democracy and Dictatorship Conference, Brown Bag Dev/Pol Vancouver School of Economics, Early Career Workshop in Quantitative Political Economy, and LACEA Annual Meeting for helpful comments and suggestions. We thank Gustavo de Britto Rocha, Lucas Resende de Carvalho, Frederico Gonzaga Jayme Junior, and the Center for Development and Regional Planning (Cedeplar) for supporting us with data access. Aiko Schmeißer also gratefully acknowledges funding from the Joachim Herz Foundation.} }

\author{
	\textbf{Laura Barros}  
	\thanks{Georg-August University Göttingen; address: Platz der Göttinger Sieben 5, 37073 Göttingen, Germany; phone: +49 551 39 27354; e-mail: \texttt{laura.barros@uni-goettingen.de}.} \hspace{2em}
	\textbf{Aiko Schmeißer} 
	\thanks{Columbia University; address: Center for Political Economy, 601 W 125th St, New York, NY 10027, USA; phone: +1 604 409 7683 ; e-mail: \texttt{as8091@columbia.edu}.} \\
	\bigskip
}

\date{
    December 9, 2025
	\\ 
	\vspace{-5mm}
}

\maketitle

\begin{abstract}

\noindent Adverse economic shocks are known to reshape voter behavior -- the demand side of politics. Much less is known about their consequences for the supply side: how such shocks affect who becomes a politician. This paper examines how job losses influence individuals' decisions to enter politics and the implications for political selection. Using administrative data linking political participation records to matched employer-employee data covering all formal workers in Brazil, and exploiting mass layoffs for causal identification, we find that job loss significantly increases the likelihood of joining a political party and running for local office. Layoff-induced candidates are positively selected on various competence measures, indicating that economic shocks can improve the quality of political entrants. The increase in candidacies is strongest among laid-off individuals with greater financial incentives from holding office and higher predicted income losses. A regression discontinuity design further shows that eligibility for unemployment benefits increases political entry. These results are consistent with a reduction in individuals' opportunity costs -- both in terms of reduced private-sector income and increased time resources -- facilitating greater political engagement.

	\vspace{0.5cm}
	
 
\end{abstract}

\thispagestyle{empty}
\clearpage
\setcounter{page}{1}

\section{Introduction}\label{sec:intro}
A large empirical literature documents that adverse economic conditions can undermine democratic stability: rising unemployment and declining incomes reduce trust in institutions, exacerbate political polarization, and increase support for populist and extremist parties \citep[e.g.,][]{Funke2016, Algan2017a, Autor, Guiso2023}. This research has primarily focused on the \textit{demand} side of democratic erosion -- that is, how economic hardship reshapes the political preferences and electoral behaviors of voters. Yet an equally important, but less explored, dimension concerns the \textit{supply} side of politics: How do economic shocks shape individuals’ incentives and opportunities to become a politician?

Politicians are perhaps the most important agents affecting the quality of democratic institutions. They aggregate citizens' preferences in legislatures and influence which policies are enacted and how well they are implemented \citep[e.g.,][]{Pande2003, Jones2005, Besley2011, ClotsFigueras2012}. Understanding how economic conditions influence who enters politics is therefore crucial. Individuals considering a political career face opportunity costs that depend on conditions in the private labor market. When private sector opportunities are abundant, talented individuals may be drawn away from political service. Conversely, when these opportunities contract, political careers may become more attractive, potentially increasing both the number and the quality of those who choose to seek office. Thus, in contrast to their role for the demand side of politics, adverse economic shocks may have positive consequences for the political system by improving the supply of politicians. 

In this paper, we shed new light on the political consequences of economic shocks by studying how experiencing a job loss affects individuals' decision to enter politics. Using rich administrative data from Brazil, we focus on party memberships and candidacies for local councils, the lowest-level positions in Brazilian politics. We rely on individual-level information on the universe of party members and local candidates between 2000 and 2020, which we match to employer-employee data on all formally employed workers between 2004 and 2018. Various individual competence measures allow us to document the implications of job loss for the selection of those who enter politics. Moreover, we investigate effect heterogeneity along individual, local, and party characteristics and explore the consequences of eligibility for unemployment benefits after the job loss, providing novel insights into the mechanisms behind individual decisions to engage with politics following the displacement. 

To estimate the effects of job loss on the probability of entering politics, we exploit firm-level mass layoffs as plausibly exogenous variation for individual-level job loss. We compare individuals laid off in a mass layoff firm to a matched control group of similar workers from other firms who were not dismissed in the same year. The richness of our data allows us to exactly match treated and control workers on several characteristics, such as gender, education, age, wage, tenure, and firm size, and to control for fine-grained local- and industry-specific shocks that may jointly affect mass layoffs and political behavior. Importantly, accounting for changes in the local political environment ensures that our results capture individual supply responses and are not influenced by changes in the local demand for politicians. We use a difference-in-differences (DiD) design to estimate dynamic treatment effects for up to three election cycles (12 years) after job loss. The underlying parallel-trends assumption is supported by the absence of differential trends in political outcomes between treated and control individuals before the layoff, and we also provide evidence that selection into (mass-)layoffs does not drive our results.\footnote{Our empirical strategy closely follows recent papers studying the effects of mass layoffs in Brazil on crime \citep{Britto2022}, domestic violence \citep{Bhalotra2021}, health \citep{Amorim2023}, and children's education \citep{Britto2022a}.} 

Our main results show that experiencing a job loss increases the probability of joining a party by more than 12\% compared to the control mean. This increase in party membership is long-lasting, starting in the election cycle after job loss and remaining statistically significant in the two subsequent election cycles. For the probability of running for local office, the effect increases over time, ranging from 6\% in the first election cycle after layoff up to 28\% in the two subsequent election cycles. The delayed increase in candidacies likely reflects the time needed to plan a campaign and the absence of term limits for local councilors. The persistence of effects is also consistent with the long-term scarring effects of job displacement, as we show that earnings in the formal labor market remain durably lower after the layoff.

The positive effects of layoffs on political engagement are remarkably widespread. First, in robustness tests, we show that the effects do not differ strongly by the size of a mass layoff, mitigating concerns regarding external validity to regular layoffs. Second, we examine heterogeneity across a range of individual covariates (education, age, gender) and local characteristics (average income, unemployment rates, inequality, homicide rates, left vs. center-right presidential vote shares). We find significant positive effects on party memberships and candidacies in all the considered individual and local subsamples. Third, we show that the effects of job loss extend beyond the realms of party memberships and councilor candidacies, encompassing various facets of engagement in politics. Specifically, we also find positive effects on the likelihood of running for mayor, donating to a political candidate, and working for a political campaign.

What explains the positive effects of job loss on political entry? Economic theories of political selection model the decision to run for office as a trade-off between the expected benefits from holding office and the costs of running \citep{Black1972, Caselli2004, Messner2004}. A layoff decreases the opportunity costs of entering politics by lowering the private-sector outside options and increasing the time available for political engagement. The size and granularity of our data enable us to examine heterogeneity in treatment effects and to test for the role of the opportunity cost channel in driving the increase in political entry after job loss. We start by exploiting individual variation in workers' predicted income losses from the layoff and local variation in councilor salaries which depend on the size of Brazilian municipalities \citep{Ferraz2011}. The former captures the severity of the drop in private-sector opportunities predicted from individual observables and labor market characteristics, while the latter captures the financial attractiveness of a political career. We find larger layoff effects on candidacies among individuals who are predicted to face the greatest income loss due to a layoff. Consistently, the increase in candidacies is highest in municipalities with higher wage caps, with no corresponding effects for memberships. Together, these results show the importance of monetary incentives for individuals' decision to enter politics following a job loss. 

In order to explore the role of time resources, we analyze the effects of eligibility for benefits from unemployment insurance (UI). During benefit receipt, individuals have lower incentives to find new employment and thus more time available for engaging with politics. We conduct a regression discontinuity analysis exploiting the fact that benefit eligibility varies discontinuously with the time since the last layoff (a minimum of 16 months is required to be able to file a new UI claim). We find that benefit eligibility reduces employment \textit{and} increases the likelihood of becoming a party member or running for local office. The effects are larger for individuals with higher education, amounting to 6.2\% increase in memberships and a 13.0\% increase in candidacies (relative to the control mean) in the two election cycles after the layoff. These results are robust to varying bandwidth choices, polynomial specifications, and permutation tests, and point towards the importance of time resources in enabling more educated individuals to enter politics after layoffs.

An alternative explanation for the increase in political engagement following layoffs is that laid-off individuals may be mobilized through a change in party support. Individuals can be activated by parties promising to better represent their new economic policy preferences. Evidence suggests that more extreme parties, both on the left and the right, are often more successful in mobilizing voters who have experienced economic hardship \citep[e.g.,][]{Algan2017a, Autor}. Moreover, the layoff may spark individuals into action as they seek to overturn the incumbent who is blamed for the change in their own economic circumstances \citep[e.g.,][]{Tilley2018, Ahlquist2020}. Another mobilizing force may arise from labor unions which often take part in mass layoff negotiations and have been shown to exert influence over the political behavior of employees \citep[e.g.,][]{kim2017, matzat2023}. 

To test for these mobilization mechanisms, we disaggregate memberships and candidacies in different parties. We start by examining ideological differences, classifying parties as left, center, and right. We find that job loss increases memberships and candidacies in parties across the ideological spectrum.\footnote{The only notable heterogeneity is a larger positive effect on right-wing party memberships in the 2020 cycle, the term of far-right president Jair Bolsonaro.} We also disentangle the party of the local incumbent mayor and other non-incumbent parties. The results reveal positive participation effects for memberships and candidacies in both incumbent and non-incumbent parties. Finally, we look at the role of union affiliation by identifying parties historically aligned with labor unions. We document that following layoffs, there is an increase in individuals' likelihood of becoming members and running for office both at union-affiliated and non-union-affiliated parties. Overall, the heterogeneity analyses indicate that incumbent, center and non-union-affiliated parties also experience increased memberships and candidacies following layoffs, suggesting that mobilization and politicization cannot fully explain the rise in political participation. 

A crucial question that follows is how layoffs affect the selection of politicians. Our results suggest that individuals' decisions to engage with politics following layoffs are at least partly driven by economic rather than purely ideological motives. If economic incentives vary across individuals' skill levels, economic shocks may also influence the average competence of politicians and, in turn, the functioning of the political system. To study the implications for the selection of political participants, we provide two pieces of evidence. First, we compare the characteristics of individuals who become party members and candidates due to the layoff (compliers) with those who would engage with politics regardless (always-takers) and those who would not engage with politics under any circumstances (never-takers). We find that compliers are positively selected in terms of education, previous wages, and AKM person wage effects, compared to always-takers and never-takers. For candidates, we also document that compliers have higher levels of education relative to the universe of candidates. Importantly, this does not come at the expense of political representation, as our results indicate that individuals who decide to run for office due to the layoff are younger and more likely to be female in comparison to always-takers and the population of candidates. Second, to study aggregate implications, we estimate the effects of municipal employment rates on the average education, gender, and age of local councilors. Exploiting local labor demand shocks from a \cite{bartik1991} instrument, we find that lower employment rates increase the average years of education not only among those individuals who run for office but also among those who are ultimately elected as councilors. Additionally, we observe that lower employment rates lead to a higher share of female and young councilors. Overall, these results suggest that economic downturns can lead to more positive political selection without compromising political representation. 

Our paper contributes to two main strands of research. First, we speak to the literature on the political consequences of economic shocks. A large body of work shows that voting outcomes are influenced by macro shocks, such as import competition \citep{colantone2018, Autor, dippel2022}, economic crisis \citep{deBromhead2013, Funke2016, Barros2019}, automation \citep{Frey2018, anelli2019, Im2019}, and austerity policies \citep{Fetzer2019, GalofreVila2021,  DuqueGabriel2025}.\footnote{For a comprehensive review of the literature linking economic shocks to the rise of populism, see \cite{Guriev2022}.} Similarly, at the individual level, many studies document that changes in economic conditions affect citizens' policy preferences, including support for redistribution and welfare provision \citep{margalit2013, marten2019, Ahlquist2020, ballardrosa2021}. In terms of political participation, the literature has focused mostly on voter turnout and has found mixed results, ranging from turnout increasing with local unemployment rates \citep{charles2013, burden2014}, decreasing after individual job loss \citep{emmenegger2017, oestermann2023}, or not being affected at all by income changes \citep{jungkunz2022, geys2024}. 

We add to this strand of literature by revealing that economic shocks affect not only the \textit{demands} of voters, but also impact the \textit{supply} side of politics. Specifically, we document that layoffs increase individuals' propensity to enter the political system, both as members of a political party and as candidates in local elections. The effects increase with several measures of individual competence, suggesting that negative labor market shocks can improve the quality of political entrants. To the best of our knowledge, we are the first to study the causal impact of changes in individual economic conditions on political participation using administrative data and an identification strategy that exploits individuals' exposure to exogenous layoff shocks. 

Second, we contribute to the literature on the determinants of political selection.\footnote{See \cite{besley2005}, \cite{DalBo2018} and \cite{Gulzar2021} for reviews of this literature.} In the tradition of the citizen-candidate framework \citep{osborne1996, besley1997}, this literature highlights the endogenous nature of entry into politics. Descriptive studies document how politicians differ from the general population: in Sweden, office holders are positively selected along various competence measures \citep{DalBo2017}, while in Brazil, politicians are more likely to have a criminal background \citep{britto2024}. Causal analyses of political selection focus mostly on the role of institutional determinants, such as electoral and fiscal rules \citep{arora2022,beath2016,gamalerio2025}, gender quotas \citep{besley2017}, campaign spending limits \citep{avis2022}, and political training programs \citep{dal2024}. Most closely related to our paper, \cite{Ferraz2011} and \cite{gagliarducci2013} find that higher politician salaries attract more educated candidates for local political offices in Brazil and Italy, respectively. Both studies demonstrate how financial incentives influence who enters politics by altering the returns to holding office. 

We extend these studies by examining the reverse mechanism: how reductions in private-sector opportunities -- and thus reductions in the opportunity cost of holding office -- affect political entry. 
In doing so, we provide novel evidence on the economic trade-offs that motivate individuals to engage in politics. Leveraging rich administrative data, we shed light on the channels through which layoffs increase political entry: they lower individuals' opportunity costs in terms of foregone income and time constraints. We further show that layoffs improve the selection of political candidates along several competence measures, and that access to unemployment benefits amplifies the entry of better-educated individuals into politics. Finally, the large-scale individual-level data allow us to tackle the rare-event problem that typically constrains empirical work on political selection.

This paper is structured as follows. Section \ref{sec:background} provides background information on party members and local council candidates in Brazil, and discusses theoretical predictions about the role of opportunity costs for political entry. Section \ref{sec:data} introduces the datasets used in the analyses. Section \ref{sec:results} presents the main results and heterogeneity analyses for the effect of job loss. In Section \ref{sec:ui}, we investigate the role of UI benefit eligibility. Finally, Section \ref{sec:implications} examines potential implications for political selection, and Section \ref{sec:conclusion} concludes.

\section{Institutional and Theoretical Background}\label{sec:background}
In Brazil, a significant portion of the population is engaged in politics. As mandated by the 1988 Brazilian constitution, voting is compulsory for all literate citizens aged 18 to 69.\footnote{The average turnout rate in Federal Elections in the last 20 years was about 80\% \citep{IDEA2024}.} Apart from voting, more than 16 million Brazilians were registered as members of a political party in 2020. This corresponds to about 11\% of the electorate, among the highest shares observed across democratic countries. In addition, between 2000 and 2020, about 1.7 million individuals ran for office in local elections in Brazil. In this section, we describe the institutional environment in which individuals decide about their involvement as party members or local candidates and present a theoretical framework that illustrates the role of opportunity costs in this decision. 

\subsection{Party members}\label{sec:background_members}
Brazil has a multi-party political system, encompassing 33 registered parties as of 2024. The effective functioning of political parties relies heavily on an engaged membership base. According to the Brazilian electoral law, a party must have a minimum number of members equal to 0.5\% of the votes cast in the previous national election. Challenging the notion that political affiliation in Brazil is solely driven by clientelistic motives, \cite{ribeiro2019} show that most political parties are well-organized and have a strong base of activists. Members carry out the day-to-day organizational work, participate in electoral campaigns, raise funding, attend party meetings, help maintain ties with the electorate outside of election years, and serve as the pool from which parties recruit their candidates and officials.

In principle, all eligible voters can join a party. In practice, however, registration and membership rules vary substantially across parties. On top of the opportunity costs related to registration and effective participation, in some parties, members have to pay registration fees and monthly dues.\footnote{For example, NOVO requires members to contribute a minimum of 460 BRL (93 USD in 2024 values) yearly, with a fee waiver for low-income individuals. PT has sliding-scale fees, with a minimum yearly fee of 30 BRL (6 USD) for low-income individuals and a maximum of 12\% of members' net salary for those individuals earning more than 6 times the minimum wage. PSDB, on the other hand, does not charge any fee for their members.} Some parties use interviews and formal tests for the selection of members, while others, especially more recently, require a simple online registration. 

Membership recruitment in Brazil is predominantly a local phenomenon, fueled by internal competition at parties' municipal conventions. Local elites, aiming to secure the party's nomination, attract supporters by enlisting them in the party to build a loyal voting base at the convention and signal their potential as candidates \citep{sells2020, frey2022}. Supporters who become members of a winning political party are often rewarded with employment opportunities within the government bureaucracy \citep{brollo2017,barbosa2023}. As suggestive evidence of such benefits in our setting, Appendix Figure \ref{fig:es_employment_memberpre} shows that laid-off individuals are more likely to start a public-sector job if they were members of a political party before the layoff, with no comparable increase observed for private-sector employment. The strategic importance of party members in local politics is further reflected in the cyclical pattern of new affiliations: as shown in Appendix Figure \ref{fig:hist_affil}, parties significantly expand their ranks in the year before local elections.

\subsection{Local councilors}\label{sec:background_elections}
Brazil has a highly decentralized political system, with municipal governments managing substantial budgets to provide public services, such as local infrastructure, transportation, education, and health care. Each of Brazil's 5,568 municipalities is governed by a mayor (\textit{prefeito}) and a council of local legislators (\textit{vereadores}). Every four years, during the municipal elections, citizens cast one vote for a mayoral candidate and one vote for a council candidate.\footnote{Local elections are held at the same time for all municipalities across the country, usually in October. Elections for the President, governors, and federal and state legislators also take place every four years but are staggered by two years relative to local elections.} Local councilors are elected based on a system of open-list proportional representation and serve four-year terms, with no term limits. The size of local councils varies depending on population, with a minimum of 9 councilors elected in municipalities with up to 15,000 inhabitants and a maximum of 55 councilors elected in municipalities with more than 8 million inhabitants. Local legislators are responsible for drafting laws, voting bills, proposing public works, and monitoring the actions of the executive. In addition, they must approve the municipal budgets and thereby have great influence over the delivery of public goods.

As in many other contexts, serving as a local councilor in Brazil can be financially rewarding.\footnote{According to PoliticalSalaries.com, a not-for-profit open data project, being a legislator is financially attractive in many countries, as measured by legislator salaries as a percentage of GDP per capita. In Brazil, legislators earn salaries equivalent to 444\% of GDP per capita, a ratio similar to other upper-middle-income countries, such as Colombia (490\%) and South Africa (443\%).} The wages of councilors are substantial on average -- amounting to about 2.5 times the average wage in their municipalities. Councilor wages vary strongly by municipality size -- being capped at 20\% of state deputies' wages in municipalities with less than 10,000 inhabitants and at 75\% in municipalities with more than 500,000 inhabitants. Importantly, serving as a councilor is not a full-time activity and allows legislators to retain their private-sector jobs after being elected. On average, legislators need to be present in the council only four days per month \citep{Ferraz2011}. Besides the salaries received by elected councilors, patronage channels can also generate returns for candidates who are not elected. \cite{colonnelli2020} find that non-elected councilor candidates are about 15 percentage points more likely to be employed in the public sector and receive 40\% higher earnings in the formal economy if their supported party comes into power.

To become a candidate for local council, individuals must be affiliated with a political party one year prior to the election.\footnote{Additionally, candidates must be Brazilian citizens, aged at least 18 years old, have residency in the municipality where they are running for office for at least one year, possess full political rights, and possess a certificate of completion of military service (for men).} Each political party (coalition) nominates up to 1.5 (2) times the total number of available seats in the local council. Political parties are required to meet a gender quota, with at most 70\% of candidates being of the same gender. The nomination of political candidates occurs at the parties' municipal conventions which take place between July and August of each electoral year. Due to the large number of available seats and parties' incentives to fill as many candidacies as possible, running for local council is relatively common, even for individuals without prior political experience.\footnote{The number of seats the party wins depends on the total number of votes it receives, creating an incentive for parties to fill as many candidacies as possible.} 

How large are the financial costs associated with candidacies for local council? To finance their campaigns, candidates rely on party funds, financial contributions from individuals and firms, and their own resources.\footnote{Campaign spending limits were introduced in 2015 \citep{avis2022}. In addition to banning financial contributions from private firms, the electoral law also set spending limits for political campaigns based on the population of the municipality. Since 2019, candidates for local council are only allowed to fund up to 10\% of their municipality's spending limit with their own resources, ensuring more equitable running conditions.} We collected data from the Superior Electoral Court (\textit{Tribunal Superior Eleitoral}, TSE) on all campaign contributions received by councilor candidates in the electoral cycles 2004 to 2020. Appendix Table \ref{tab:sumstat_donations} presents detailed summary statistics for each year. Across all years, the median candidate raised a total of 1,587 BRL (392 USD), with only 168 BRL (34 USD) coming from own resources. Only about half of all candidates use any private funds for the campaign. These figures suggest that personal financial contributions play a limited role in local campaigns, suggesting that the financial barriers to running for councilor are relatively low.

\subsection{Theoretical considerations}\label{sec:background_theory}
We now introduce a simple economic framework to illustrate how a layoff may affect an individual's likelihood of becoming a party member or councilor candidate.\footnote{For economic models of candidate selection, see, for example,
\cite{Black1972}, \cite{Caselli2004}, and \cite{Messner2004}.} Consider an individual who decides how to divide her non-leisure time between pursuing a career in the private sector or in politics. Let $h^P$ denote the share of time devoted to politics. If the individual does not engage in politics ($h^P=0$), she earns a market hourly wage of $w^M$. When entering politics ($0 < h^P \leq 1$), she foregoes a share of her earnings that is proportional to $h^P$.\footnote{Since neither party membership nor local councilor is a full-time political commitment, we consider the possibility of simultaneously working in the private sector, albeit with a reduced number of hours.} At the same time, political engagement provides her with an array of expected benefits, $b^P$, which increase with the time devoted to political activities. These benefits include monetary rewards, such as councilor salaries when elected to office \citep{Ferraz2011}, public sector employment opportunities resulting from political connections to the winning party or candidate \citep{brollo2017, colonnelli2020, barbosa2023}, and improved employment prospects in the private labor market under co-partisan employers \citep{Colonnelli2022}. Non-monetary rewards, such as ego rents or prosocial benefits, may also play a role \citep{gulzar2024good}. Finally, to engage in politics, the individual incurs entry costs $c^P$ (e.g., campaign costs or membership registration fees). 

An individual who is currently employed in the private sector decides to enter politics if the benefits of engaging in politics are greater than the benefits of focusing exclusively on market work, i.e., if $w^M (1-h^P) + b^P h^P - c^P \geq w^M$, or equivalently, if the net benefits are positive:
\begin{equation*}
    \pi_{\text{pre}} = h^P (b^P - w^M) - c^P \geq 0.
\end{equation*}

If the individual is laid off, she experiences an earnings loss that may result from both lower employment prospects and a wage penalty upon reemployment. Let $\lambda$ denote the reemployment probability and $\tilde{w}^M$ the new potential market wage, such that the expected labor income is given by $\lambda \tilde{w}^M$. If the individual finds reemployment, her expected benefits from engaging in politics become $\tilde{w}^M (1-h^P) + b^P h^P - c^P$. If not reemployed, she earns no labor income and can devote her entire time to politics, yielding benefits of $b^P - c^P$.\footnote{In practice, unemployed individuals may not allocate their entire non-leisure time to politics, but also engage in other activities such as job search. The model's qualitative predictions remain unchanged as long as the time devoted to politics while unemployed exceeds the level of political time investment under reemployment -- that is, it exceeds $h^P$.} The individual will enter politics if its returns are greater than the expected value of pure market work, i.e., if $\lambda\left(\tilde{w}^M (1-h^P) + b^P h^P - c^P \right) + (1-\lambda)\left(b^P - c^P\right) \geq \lambda \tilde{w}^M$, which leads to the following condition on net benefits:
\begin{equation*}
    \pi_{\text{post}} = \lambda \left(h^P (b^P - \tilde{w}^M) - c^P\right) + (1-\lambda)\left(b^P - c^P \right) \geq 0.
\end{equation*}

To assess the effect of a layoff on the likelihood of entering politics, we can compare the net benefits before and after the layoff: 
\begin{equation*}
    \pi_{\text{post}} - \pi_{\text{pre}} = h^P (w^M - \lambda \tilde{w}^M) + (1-h^P) (1-\lambda) b^P.
\end{equation*}

With $w^M \geq \tilde{w}^M \geq 0$, $b^P \geq 0$, $0 \leq h^P \leq 1$, and $0 \leq \lambda \leq 1$, this difference is weakly positive, implying that layoffs may induce more individuals to enter politics. The change in net benefits resulting from the layoff has two components. First, it is proportional to the expected income loss, $w^M - \lambda \tilde{w}^M$, weighted by the share of time not allocated to private-sector work, $h^P$. This term captures the decline in monetary opportunity costs arising from the layoff: as earnings potential in the private sector falls, individuals have to give up less income when engaging in political activities, particularly those that are time-intensive. 
To empirically test the relevance of this mechanism, we will analyze heterogeneity in treatment effects by individuals' expected income losses due to the layoff, which are predicted based on a rich set of individual and local labor market characteristics. 

Second, the effects increase with the expected benefits of political engagement during unemployment, $(1-\lambda) b^P$, weighted by the time not already allocated to politics, $1-h^P$. Intuitively, this term reflects the notion that, when remaining unemployed, individuals can reallocate time that would otherwise be spent in market work towards political activities, thereby gaining additional political benefits. To shed light on one of the monetary returns from engaging with politics, we will investigate whether the treatment effects are larger in localities where councilors earn higher salaries.\footnote{In addition to monetary returns, political involvement may also provide policy payoffs. In citizen-candidate models \citep{osborne1996, besley1997}, political candidates represent particular policies and care about winning \textit{per se}. As argued by \cite{hall2019} and \cite{thomsen2017}, the policy returns may be especially relevant for ideologically extreme candidates. Thus, we will examine potential heterogeneity in the effects on memberships and candidacies in parties of different ideological positions.} Moreover, we will test the role of time resources by exploiting discontinuities in eligibility for UI benefits that increase laid-off workers' incentives to remain unemployed.

Note that the relative importance of both channels -- reduced income opportunities and increased time availability -- is determined by the time intensity of political engagement, $h^P$. For forms of engagement that require substantial time investments, such as running for office, the income-loss channel is likely more important. By contrast, for lower-commitment forms of participation, such as party membership, the time-availability channel may play a larger role. 

The framework also provides predictions about the potential implications of layoffs for the quality of individuals who select into politics. Given that the participation effects of layoffs are predicted to be larger for individuals with higher income losses and higher expected benefits from political engagement, the consequences for selection depend on how these factors are related to individual competence. High-ability individuals -- such as those with higher education -- tend to earn higher wages in the private sector. Thus, a layoff may represent a larger shock to their opportunity costs, leading to a more pronounced increase in their incentives to enter politics. In addition, the expected benefits of political engagement may vary with individuals' competence. For instance, the returns to running for office rise with the probability of winning, which may be higher for more capable individuals. Moreover, the expected likelihood of gaining access to a public-sector job through political connections may depend on ability, either negatively if political support acts as a substitute for individual competence \citep{colonnelli2020} or positively if politicians use their discretionary power to appoint more competent individuals \citep{brollo2017, bazzi2023}.

\section{Data}\label{sec:data}
In our analysis, we combine detailed information on the universe of party members and local political candidates with administrative data on the population of formal employees in Brazil. This section describes the data sources as well as the steps we follow to construct and merge our estimation sample. By having identified records on individuals' employment histories as well as political outcomes, we can uniquely identify whether individuals' political behavior changes after they lose their jobs. Additionally, detailed information on individual, municipality, and party characteristics allows us to disentangle potential mechanisms. 

\subsection{Political outcomes}\label{sec:data_tse}
To measure individuals' decision to enter politics, we use two datasets provided by the TSE. Information on the universe of political members comes from the \textit{Sistema de Filiação Partid\'{a}ria}, a national registry containing records of all affiliations to political parties, including individuals' names, affiliated party, as well as date and municipality of registration. The data cover the period 2000 to 2020. Second, we gather data on the universe of individuals running for local council in the six election cycles of 2000, 2004, 2008, 2012, 2016, and 2020. The data contain information on candidates' taxpayer registry numbers (\textit{Cadastro de Pessoas F\'{i}sicas}, CPF), socio-demographic characteristics, party of candidacy, whether they have been elected, and municipality of candidacy.

\subsection{Labor market data}\label{sec:data_rais}
We obtain matched employer-employee data covering the universe of formal employees between 2004 and 2018 from the \textit{Rela\c{c}\~{a}o Anual de Informa\c{c}\~{o}es Sociais} (RAIS), an administrative database collected annually by the Ministry of Labor and Employment.\footnote{In 2015, formally employed individuals accounted for approximately 55\% of the Brazilian labor force \citep{derenoncourt2021}.} The database was created in 1975 for statistical and administrative purposes and contains detailed information such as hiring and separation date, average monthly earnings, contracted hours, occupation, firms' location and industry, and workers' socio-demographic characteristics. For our analysis, we make use of the identified dataset, containing individuals' names, taxpayer registry numbers (CPF), as well as firms' tax identification numbers (\textit{Cadastro Nacional da Pessoa Jur\'{i}dica}, CNPJ). This enables us to identify individuals' employment histories over time and across firms.

\subsection{Merging political and labor market data}\label{sec:data_merge}
We merge the information on political candidates with the employment data using an individual's taxpayer registry number (CPF). As for party membership, we merge this information to the RAIS data based on each individual's full name. To minimize measurement error when exactly matching on names, we follow \cite{Britto2022a, Britto2022} and \cite{Bhalotra2021} and restrict our sample to individuals who have a unique name in Brazil. Individuals with unique names cover roughly half of the Brazilian population (50.2\% in our sample), given that individuals commonly have several surnames, including at least one from both their father and mother.\footnote{To identify unique names, we build a dataset of names in Brazil. We collect the names of all workers who appear in the RAIS data in any year between 2004 and 2018, and supplement this list with information on the universe of beneficiaries in the social programs \textit{Bolsa Família} and \textit{Benefício de Presta\c{c}\~{a}o Continuada} between 2013 and 2021 and on all party members between 2000 and 2020. The resulting list covers about 90\% of the Brazilian adult population. Not observing the names of a small part of the population will introduce some measurement error in our merged outcome variables, which is, however, not expected to be related to the likelihood of being laid off.} 

Appendix Table \ref{tab:sum_stat_nameuniqueness} compares the characteristics of laid-off individuals in our estimation sample who have a unique and non-unique name, respectively. Those with a unique name are slightly positively selected (9\% higher earnings, 8\% more years of education, 0.9 percentage points more likely to be a manager), and are more likely to be female (8.1 percentage points) and white (5.3 percentage points). Given that we merge our second political outcome -- local candidacies -- using the CPF, we have this information for individuals with and without a unique name. Before the layoff, we do not observe differences in the likelihood of running for councilor across both groups. While we focus on the sample of individuals with unique names (in which we have merged political members and candidates) in our baseline specification, we also analyze the external validity of the results by estimating layoff effects on candidacies in the sample of individuals with non-unique names.

\section{Effects of Job Loss on Political Entry}\label{sec:results}

\subsection{Empirical strategy}\label{sec:empirics}
We aim to estimate the impact of adverse economic shocks on individuals' likelihood of entering politics. To do so, we study individuals who are laid off from formal employment in the Brazilian labor market. Brazil's private labor market exhibits very high job turnover \citep{Gonzaga2003}, with layoffs without just cause accounting for about three-quarters of all separations (and voluntary quits by workers accounting for most of the remaining quarter). Among all workers who appear in RAIS between 2004 and 2018, 66.6\% experienced at least one layoff during this period.\footnote{The yearly layoff rate (among those with a formal job in a given year) is 19.1\%.} Political entrants are not shielded from these shocks: among all individuals who newly became party members (ran for local councilor at least once) between 2000 and 2020, 36.8\% (41.5\%) experienced at least one layoff.\footnote{Note that only 55.6\% (70.3\%) of all new party members (councilor candidates) ever appeared in RAIS with a formal contract over the 2004-2018 period.} Thus, layoffs are common labor market shocks faced by individuals, including those who engage in politics.

The main difficulty in estimating the effect of layoffs on individuals' entry into politics is the fact that layoffs are not random. On the one hand, individual shocks could simultaneously influence both political participation and the likelihood of losing a job, leading to a spurious relationship. On the other hand, individuals' political behavior could also affect their probability of being dismissed. \cite{Colonnelli2022} show, for instance, that political discrimination affects labor market outcomes in Brazil. To address these identification challenges, we exploit the timing of firm-level mass layoffs as a source of plausibly exogenous variation in individual-level job loss using a difference-in-difference design in a matched sample of workers. The timing of mass layoffs is arguably orthogonal to individual workers' political behavior and other confounding worker-level shocks, and has been widely used to estimate the effects of job loss on various outcomes. In particular, our empirical strategy closely follows recent papers that study the effect of dismissals in Brazil \citep{Bhalotra2021, Britto2022, Britto2022a, Amorim2023}. This subsection describes how we construct the estimation sample and then presents the estimation model.

\vspace{1em} \noindent \textbf{Sample selection.} We restrict our sample to workers aged 25 to 50 with a full-time (i.e., at least 30 hours per week), open-ended, private-sector job of at least six months tenure. Among those, the treatment group comprises all workers displaced without just cause from a mass layoff firm between 2009 and 2011.\footnote{We exclude individuals who are recalled by their initial employer within five years after the layoff. In cases where individuals experienced multiple mass layoffs between 2009 and 2011, we keep only the first one.} This treatment period allows us to study long-run effects over three election cycles (12 years) after the layoff, and also to check for pre-trends over two election cycles (8 years) before the layoff. In our baseline specification, we define a mass layoff firm as a firm with at least 30 workers dismissing at least 30\% of its workforce in a given year, and later we show that our results are robust to various thresholds of firm size and layoff share.\footnote{We always drop firms from the mass layoff sample if they reallocate under a new identifier, specifically if at least 30\% of their dismissed workers transition to the same new firm identifier in the following year.} 

We then match each treated individual with a control worker who is not dismissed in the same year and who works in a non-mass layoff firm.\footnote{Note that even though control workers are not laid off in the matching year, they may be laid off in later years. Given that we estimate effects for up to 12 years after the layoff, restricting control workers to those who were continuously employed throughout the whole observation period would result in a small and highly selected pool of control workers. Defining control workers without conditioning on future employment follows a large share of papers in the job loss literature \citep[e.g.,][]{Britto2022, bertheau2023, schmieder2023}. In addition, we do not impose any restrictions on potential layoffs prior to the matching year, other than the minimum six-month tenure requirement. Given the high turnover rates in Brazil, stricter tenure requirements would yield a substantially selected sample. In Appendix Figure \ref{fig:es_separations}, we show that treated and matched control individuals were not differently exposed to mass layoffs before the treatment period.} To obtain a comparable control group, we perform an exact matching on gender, education (5 categories), age (deciles), monthly earnings (deciles), tenure (deciles), firm size (deciles), and state (27 categories).\footnote{We allow control individuals to be matched to more than one treated individual. In cases where more than one control individual was matched to one treated individual, we randomly selected one control individual.} For 79.5\% of the workers in the initial treatment pool, we are able to find a comparable control worker. We end up with an estimation sample that consists of nearly one million treated and matched control workers. 

\begin{table}[tb]\centering \caption{Summary statistics} \label{tab:sum_stat}
\begin{threeparttable}
\begin{tabular}{l*{3}{c}} 
\toprule[1.5pt]
					& (1) & (2) & (3)  \\
                    &     Treated&     Control&    Std Diff\\
\midrule
\multicolumn{3}{l}{\textbf{Socio-demographic characteristics}} \\
Years of education  &       10.72&       10.77&        -0.02\\
Age                 &       33.84&       33.85&        -0.00\\
Male (\%)           &       65.78&       65.78&        -0.00\\
[1em]
\multicolumn{3}{l}{\textbf{Job and firm characteristics}} \\ 
Earnings (per month, BRL, CPI 2018)&     2158.15&     2140.37&       0.01\\
Tenure (months)     &       32.62&       32.85&        -0.01\\
Manager (\%)        &        2.70&        2.20&       0.03\\
Firm size           &      610.80&      625.65&        -0.01\\
[1em]
\multicolumn{3}{l}{\textbf{Political outcomes before layoff}} \\ 
Party membership (2008, \%)  &        7.47&        7.30&       0.01\\ 
New party membership (2005-2008, \%) &        1.99&        1.89&       0.01\\
Candidate (2005-2008, \%) &        0.11&        0.10&       0.00\\
[1em]
Observations		& 944,214 & 944,214 & \\
\hline\hline
\end{tabular}
\begin{tablenotes} \scriptsize 
    \item \textit{Note:} 
    The table reports the average characteristics of treated workers who are displaced from a mass layoff firm (column (1)) and matched control workers who are not displaced in the same calendar year (column (2)), and the standardized difference between the two groups (column (3)). The sample only includes workers with a unique name within the country. Appendix Table \ref{tab:sum_stat_party} reports results for additional pre-layoff political outcomes split by party characteristics.
\end{tablenotes}
\end{threeparttable}
\end{table}

Table \ref{tab:sum_stat} presents summary statistics for treated and control individuals and shows that both groups are similar in socio-demographic, job, and firm characteristics. For all considered variables, the standardized mean difference between the two groups is well below the threshold of 0.20 recommended by \cite{imbens2015}. Importantly, treated and control units are also balanced in terms of pre-treatment political outcome variables which were not part of the matching process. In both groups, approximately 7\% of individuals were members of a political party in 2008. Nearly 2\% newly registered as a member in the election cycle between 2005 and 2008, and 0.1\% ran for local councilor in the 2008 election.\footnote{Appendix Table \ref{tab:sum_stat_party} also shows that there are no differences in the ideology, local incumbency, and union affiliation of the parties that treated and control observations affiliated with or ran for before the layoff.} The validity of our DiD design does not require that treated and control individuals are similar in levels, but covariate balance makes it more plausible that both groups exhibit similar trends in counterfactual outcomes. 

\vspace{1em} \noindent \textbf{Estimation model.} As our main specification for the effect of job loss on political participation, we estimate the following dynamic DiD (event-study) model:
\begin{equation}\label{model:es}
Y_{ic} = \alpha + \beta Treat_i + \sum_{c=2000, c\neq2008}^{2020} \delta_c (Treat_i \times Cycle_c) + \mu_{m(i)k(i)c} + \epsilon_{ic},
\end{equation} 
where $i$ denotes workers and $c$ denotes four-year municipal election cycles (2000, 2004, 2008, 2012, 2016, 2020). Our main outcomes, $Y_{ic}$, are binary variables capturing new party memberships and candidacies for local councils in a given election cycle. The effects of interest are captured by $\delta_c$, the coefficients for interactions between a treatment group dummy, $Treat_i$, and dummies for each election cycle, $Cycle_c$. For $c= \{2012, 2016, 2020\}$, the parameters identify the dynamic treatment effects (relative to the baseline period $c=2008$), and for $c=\{2000, 2004\}$ they check whether outcomes evolved in parallel in the pre-treatment period. We also estimate a static version of the DiD model:
\begin{equation}\label{model:did}
Y_{ic} = \alpha + \beta Treat_i + \delta_{DiD} (Treat_i \times \mathbbm{1}[c\geq2012]) + \mu_{m(i)k(i)c} + \epsilon_{ic},
\end{equation}
which summarizes the average treatment effects over all periods in the coefficient $\delta_{DiD}$. In both models, we include municipality $\times$ two-digit industry $\times$ cycle fixed effects, $\mu_{m(i)k(i)c}$, which flexibly control for local and industry shocks that may simultaneously cause mass layoffs and changes in political behavior. Importantly, the municipality $\times$ cycle fixed effects also hold constant the local political environment and allow us to separate individual-level supply-side effects from potential effects of mass layoffs on the demand for politicians among the local electorate.\footnote{Note that the estimation models (\ref{model:es}) and (\ref{model:did}) rely on a single treatment period, as all treated workers were laid off in the 2012 cycle. Thus, our main specification is not susceptible to weighting issues that arise in staggered difference-in-differences models with heterogeneous treatment effects \citep{dechaisemartin2020, goodmanbacon2021, sun2021}.} 

While our main specification analyzes party memberships and candidacies in each election cycle, for party memberships and labor market outcomes we can also exploit yearly variation. This allows us to test for pre-trends closer to the layoff event and assess responses in the shorter run.\footnote{We have information on the exact date individuals affiliate with a political party. Still, we focus on cycle-level variation in our main specification, since new affiliations are highly concentrated around election years (see Appendix Figure \ref{fig:hist_affil}), and use the year-level model as a supplementary analysis of effect dynamics. For labor market outcomes, we always show results with the year-level model, because we do not observe outcomes for all election cycles between 2000 and 2020 (we only have access to RAIS between 2004 and 2018).} We estimate the following event-study model for outcomes measured in each year relative to the layoff:
\begin{equation}\label{model:es_yearly}
Y_{it} = \alpha + \beta Treat_i + \sum_{k\neq-1} \delta_k (Treat_i \times \mathbbm{1}[t=t^* + k]) + \mu_{m(i)k(i)t} +\lambda_{tt^*} + \epsilon_{it},
\end{equation}
where $i$, $t$, and $t^*$ refer to workers, calendar years, and layoff years, respectively.\footnote{For control workers, we assign the layoff year of their matched treated worker.} $\delta_k$ capture the dynamic treatment effects for each event year after the layoff $k$. The model includes calendar year $\times$ layoff year fixed effects, $\lambda_{tt^*}$, to ensure that the identifying variation only comes from comparing treated to control individuals who have been matched in the same year.\footnote{This stacking approach avoids the ``forbidden comparisons'' between late and early-treated individuals that can cause negative weights under heterogeneous treatment effects \citep{cengiz2019, dube2023}.}

\subsection{Main results}\label{sec:results_main}
\begin{figure}[h!]
	\caption{Effect of job loss on labor earnings and employment}  \label{fig:es_laboroutcomes}
	\centerline{
		\begin{threeparttable}
			\begin{footnotesize}
				\begin{tabular}{cc}
 					\textbf{A. Labor earnings} & \textbf{B. Employment}\\
					\includegraphics[width=0.5\linewidth]{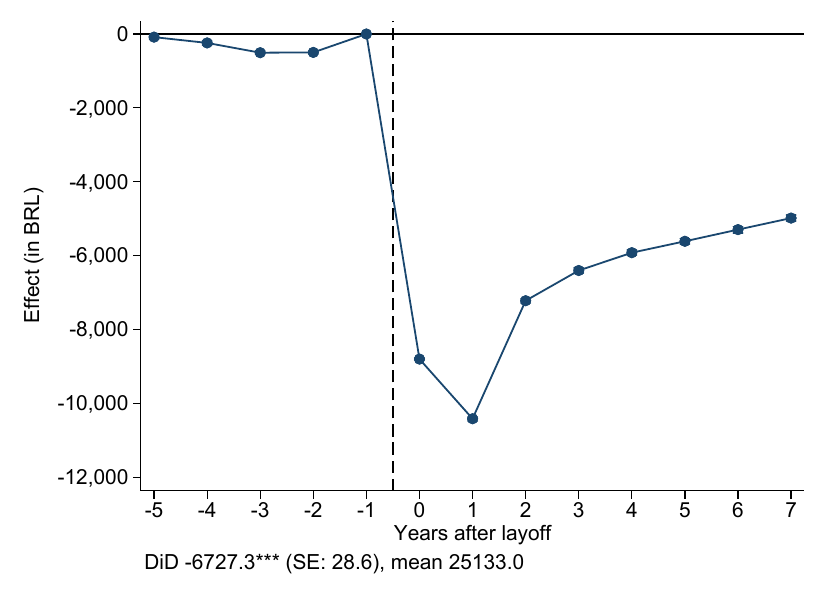} &\includegraphics[width=0.5\linewidth]{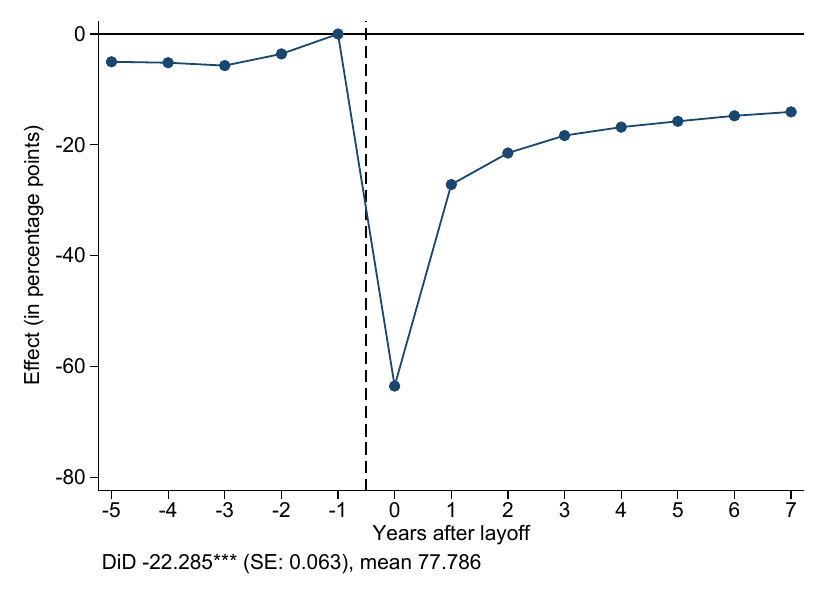} \\     
				\end{tabular}
			\end{footnotesize}
			\begin{tablenotes} \scriptsize
			\item \textit{Note:}
                The figure reports event-study coefficients $\delta_k$, estimated in model (\ref{model:es_yearly}), for the effect of job loss on yearly formal labor earnings (Panel A) and the likelihood of being formally employed at the end of the year (Panel B). $N = 24,549,564$ individual-year observations. The graphs include vertical bars that depict 95\% confidence intervals based on standard errors clustered at the individual level (too small to be visible). Below each graph, the DiD coefficient from a static version of model (\ref{model:es_yearly}), its standard error, and the mean of the control group across all post-treatment periods are reported.
                \end{tablenotes}
	\end{threeparttable}}
\end{figure}

We begin by discussing the effects of job loss on labor market trajectories. Figure \ref{fig:es_laboroutcomes} displays yearly event-study estimates based on model (\ref{model:es_yearly}). In the year of the layoff, workers experience a sharp decline in labor earnings of around 10,000 BRL (2,028 USD), which correspondents to a 45\% reduction relative to baseline earnings in the year prior to the layoff. Moreover, the probability to be formally employed at the end of the year falls by about 60 percentage points. Starting two years after the layoff, both earnings and employment start to continuously recover; yet, the recovery is only partial and does not return to pre-layoff levels. Seven years after the layoff -- the longest horizon we can observe in the RAIS data -- dismissed workers still earn about 5,000 BRL (1,014 USD) less per year and are 15 percentage points less likely to be formally employed compared to the control group. Appendix Figure \ref{fig:es_separations} further shows that job loss reduces subsequent job stability, as reflected in a higher frequency of job separations. These findings align with the job loss literature that has widely documented the long-term scarring effects of job displacements \citep[e.g.,][]{jacobson1993, eliason2006, davis2011, lachowska2020, schmieder2023}.\footnote{\cite{Britto2022} also show, using survey data on Brazilian workers, that accounting for income from the informal sector reduces the estimated earnings losses following layoffs by only about 10\%.} 

\smallskip

\begin{figure}[h!]
	\caption{Effect of job loss on party membership and running for local councilor}  \label{fig:es_main}
	\centerline{
		\begin{threeparttable}
			\begin{footnotesize}
				\begin{tabular}{cc}
					\textbf{A. New party membership} & \textbf{B. Candidacy}\\
					\includegraphics[width=0.55\linewidth]{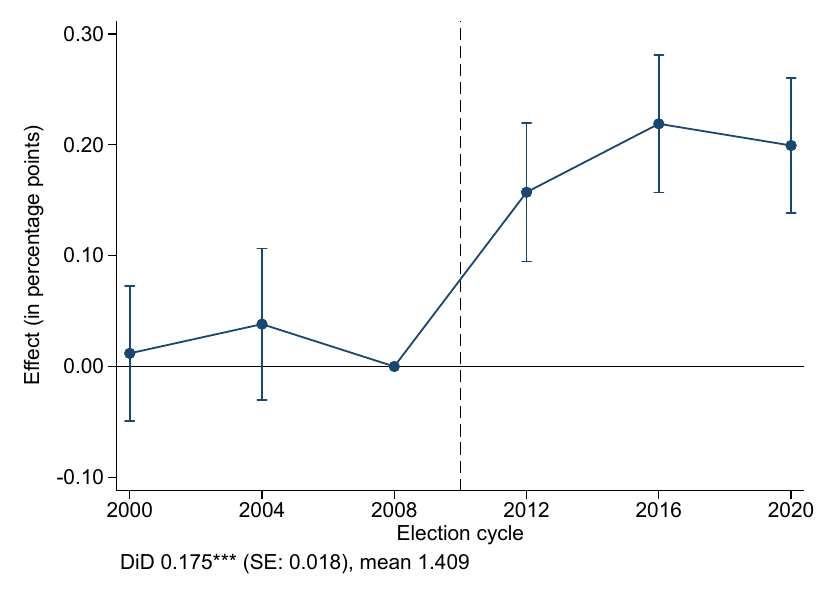} &\includegraphics[width=0.55\linewidth]{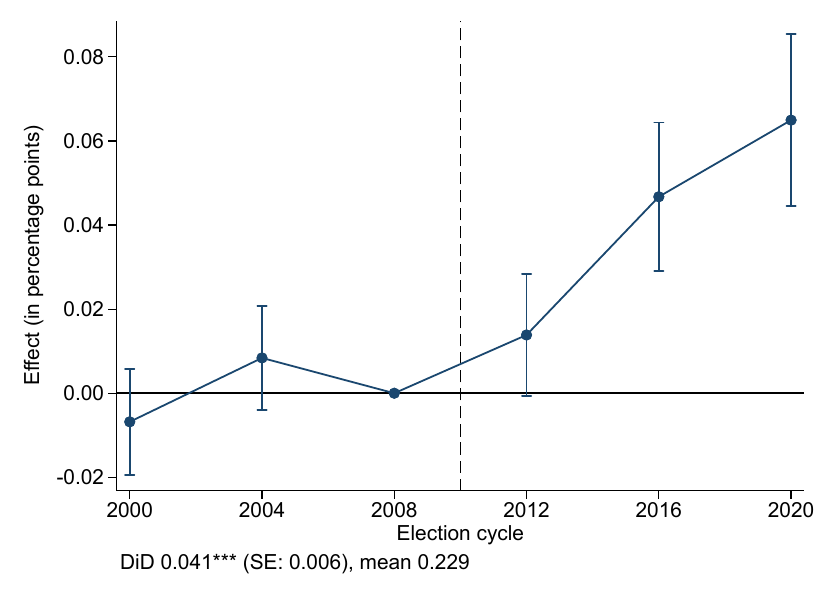} \\ 
				\end{tabular}
			\end{footnotesize}
			\begin{tablenotes} \scriptsize
			\item \textit{Note:}
				The figure reports event-study coefficients $\delta_c$, estimated in model (\ref{model:es}), for the effect of job loss on the likelihood of newly registering as a party member (Panel A) and running for local councilor (Panel B). The treated group contains workers who were displaced from a mass layoff firm between 2009 and 2011, and the matched control group comprises workers who were not displaced in the same calendar year and who worked in a non-mass layoff firm. $N = 11,317,632$ individual-cycle observations. The vertical bars depict 95\% confidence intervals based on standard errors clustered at the individual level. Below each graph, the DiD coefficient from model (\ref{model:did}), its standard error, and the mean of the control group in the post-treatment period (average across 2012, 2016, and 2020) are reported. 
			\end{tablenotes}
	\end{threeparttable}}
\end{figure}

Figure \ref{fig:es_main} presents our main results on the effects of job loss on newly registering as a party member and running for local council, based on estimates from models (\ref{model:es}) and (\ref{model:did}). We observe that both outcomes do not develop significantly differently across treated and control individuals before the layoff, lending support to the parallel-trends assumption. After the layoff, outcomes sharply diverge. We estimate that job loss increases the likelihood of registering as a party member by about 0.20 percentage points, which is a 12\% increase relative to the control group mean.\footnote{We compare effect sizes to the mean of the control group in the post-treatment period (average across 2012, 2016, and 2020).} The effect emerges in the election cycle immediately after the job loss and remains stable in the subsequent two cycles. For the probability of running for local council, we also find a significantly positive effect that increases from 0.01 percentage points (6\% relative to the control mean) in the cycle of the layoff to about 0.06 percentage points (28\%) two cycles later. The gradual increase in effects on candidacies may reflect the preparation time required for election campaigns, coupled with the fact that local councilors have no term limit and often seek re-election after their initial candidacy. Moreover, the persistence of effects over three election cycles (12 years) is consistent with the long-term decline in private-sector employment prospects documented in Figure \ref{fig:es_laboroutcomes}, which may increase the relative attractiveness of political careers. In sum, the results indicate that job loss has a strong and long-lasting positive effect on individuals' probability of engaging with politics, either as members of political parties or as local candidates.

\vspace{1em} \noindent \textbf{Robustness and external validity.}\quad Under our baseline definition of mass layoffs, there may still be considerable scope for firms to select which 30\% of workers to dismiss. Appendix Table \ref{tab:robust_mldefinition} evaluates the robustness of our main findings when applying more stringent mass layoff definitions. Irrespective of whether we increase the minimum share of dismissed workers, use plant closures, increase the minimum firm size, or focus on firms with stable employment growth before the mass layoff, we still find similarly large positive effects of job loss on party memberships and candidacies. These results also mitigate concerns regarding the external validity of our analyses. The effects of mass layoffs could differ from those of regular layoffs, for example, due to spillover effects across laid-off workers or increased attention to mass layoffs by politicians or the media. The absence of differential effects by the share of displaced workers suggests that the size of a mass layoff is not an important factor driving our results. In Appendix Table \ref{tab:robust_spec}, we also explore the sensitivity when including different sets of municipality and industry fixed effects. We show that our results are largely unaffected by accounting for local and temporal variation, suggesting that our matching strategy already compares treated and control workers in similar labor markets. Finally, for the candidacy outcome, we check the external validity of our results for individuals without a unique name. As shown in Appendix Figure \ref{fig:es_cand_unique}, the increase in the probability of running for office following job loss is very similar across individuals with and without unique names. Altogether, the robustness checks reduce concerns that our estimated effects are driven by a particular mass-layoff definition, model specification, or sample restriction. 

\vspace{1em} \noindent \textbf{Effect timing.}\quad 
Next, we take a closer look at the dynamics in the estimated treatment effects. Given that deciding to become a candidate and preparing for an election campaign requires time and planning, we start by examining whether our results differ by proximity to the next local election. Appendix Figure \ref{fig:es_layoffyear} shows results when splitting the sample by the layoff year (2009, 2010, or 2011). We find that the positive effects of job loss on party memberships and candidacies are not driven by one specific treatment year, suggesting that temporal proximity to the local election does not matter. This dynamic also supports the external validity of our results. For instance, there may be the concern that political responses following layoffs in 2009 could differ from other layoffs, given the proximity to the 2008 financial crisis. However, our results show similar effects for layoffs occurring in 2010 and 2011, once the crisis had already faded out in Brazil \citep{barbosa2010}. 

\begin{figure}[tb]
	\caption{Effect of job loss on new party membership - yearly-level analysis}  \label{fig:es_member_yearly}
	\centerline{
		\begin{threeparttable}
            \centering
			\includegraphics[width=0.65\linewidth]{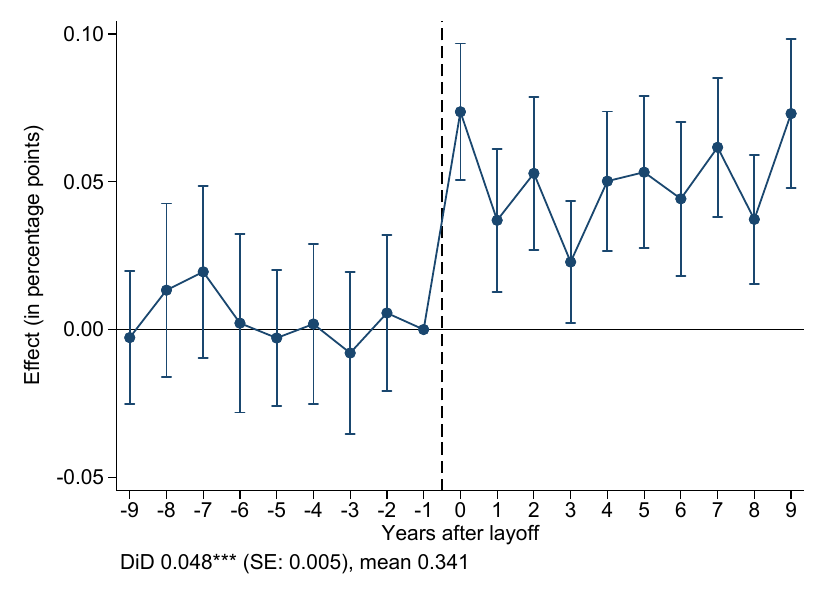}
			\begin{tablenotes} \scriptsize
			\item \textit{Note:}
				The figure reports event-study coefficients $\delta_k$, estimated in model (\ref{model:es_yearly}), for the effect of job loss on the likelihood of newly registering as a party member. The vertical bars depict 95\% confidence intervals based on standard errors clustered at the individual level. $N = 35,880,132$ individual-year observations. Below the graph, the DiD coefficient from a static version of model (\ref{model:es_yearly}), its standard error, and the mean of the control group in the post-treatment period are reported. 
			\end{tablenotes}
	\end{threeparttable}}
\end{figure}

In addition, we conduct a year-level analysis exploiting more detailed information on the exact timing when individuals join a political party. Figure \ref{fig:es_member_yearly} presents results from the yearly event-study estimated using model (\ref{model:es_yearly}). We find positive and statistically significant effects on individuals' probability of newly registering as a party member that start in the year directly following the layoff and persist up to 9 years after the shock. Reassuringly, we detect no evidence of pre-trends prior to the layoff -- neither in the year immediately preceding the layoff nor in earlier years. Taken together, the short-term dynamics are consistent with the patterns observed at the cycle level.

\begin{figure}[h!]
	\caption{Effect of job loss on additional participation outcomes}  \label{fig:es_mayor}
	\centerline{
		\begin{threeparttable}
  			\begin{footnotesize}
				\begin{tabular}{cc}
					\textbf{A. Running for mayor} & \textbf{B. Campaign donation}\\
					\includegraphics[width=0.55\linewidth]{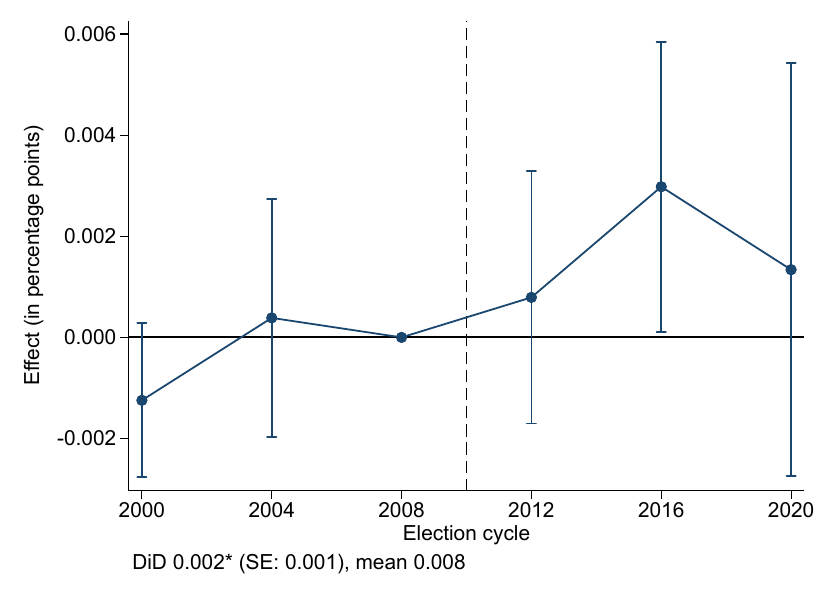} &\includegraphics[width=0.55\linewidth]{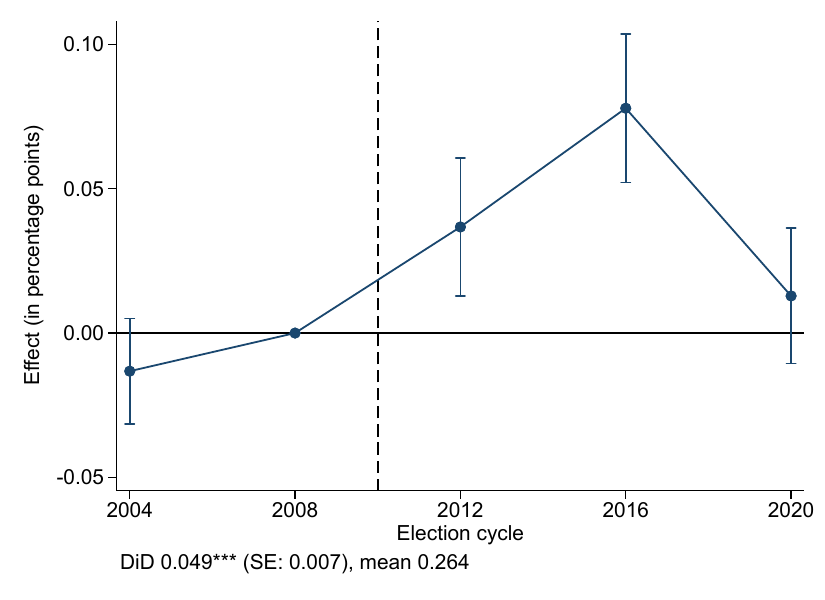} \\ 
				\end{tabular}
                    \begin{center}
                    \textbf{C. Campaign work} \\
    			\includegraphics[width=0.55\linewidth]{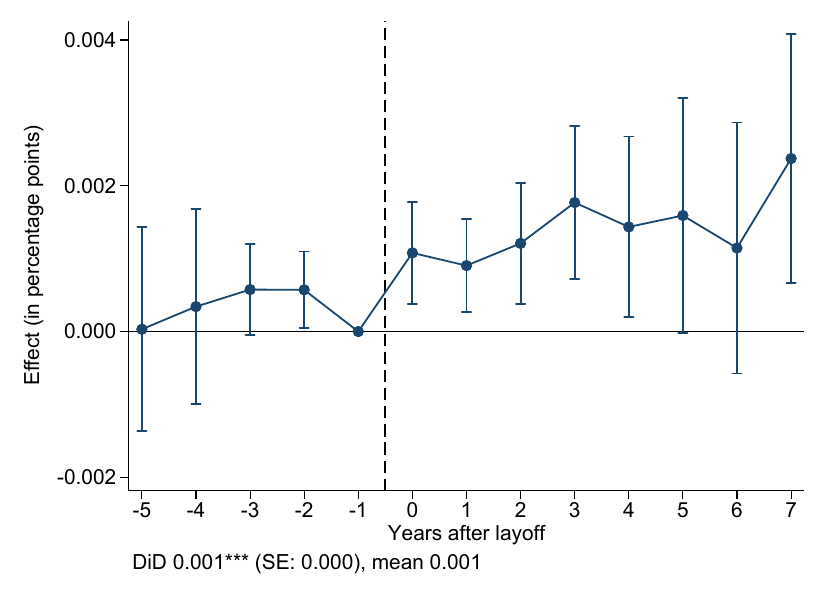}
                    \end{center}
			\end{footnotesize}
			\begin{tablenotes} \scriptsize
			\item \textit{Note:}
			The figure shows results for the effect of job loss on the likelihood of running for municipal mayor (Panel A), donating to a political campaign (Panel B), and working for a political campaign (Panel C). Information on mayoral candidates and campaign donations are obtained from the \textit{Tribunal Superior Eleitoral}. The data on campaign donations is only available starting from the 2004 election cycle. Following \cite{bazzi2023}, campaign workers are identified in RAIS as individuals who have a full-time contract at a firm in industry \textit{Atividades de Organizações Políticas} (CNAE 1.0: 91928, CNAE 2.0: 94928) or with legal nature \textit{Partido Político} (3123) or \textit{Candidato a Cargo Político Eletivo} (4090). Panels A and B report cycle-specific event-study coefficients $\delta_c$ from model (\ref{model:es}), while Panel C reports year-specific event-study coefficients from model (\ref{model:es_yearly}). The vertical bars depict 95\% confidence intervals based on standard errors clustered at the individual level. Below the graph, the DiD coefficient from static versions of models (\ref{model:es}) and (\ref{model:es_yearly}), its standard error, and the mean of the control group in the post-treatment period are reported. 
			\end{tablenotes}
	\end{threeparttable}}
\end{figure}

\vspace{1em} \noindent \textbf{Additional participation outcomes.}\quad 
In addition to estimating the effects on new party memberships and candidacies for local council, in Figure \ref{fig:es_mayor} we also investigate the consequences of job loss for three additional forms of political participation. First, we consider candidacies for municipal mayor. Compared to councilor candidacies, running for mayor requires a more elaborate election campaign to win the support of party members during the party's municipal convention. Moreover, mayoral candidacies typically involve greater resources and party investments, increasing the constraints associated with individual candidacies. Nevertheless, we also find positive effects of job loss on the likelihood of running for mayor. Given the rarity of mayoral candidacies, the results are not estimated very precisely. The positive effect is highest and significant only in the medium term (in the 2016 cycle).

Second, we examine the effect of job loss on donations to political candidates. In Brazil, individuals are allowed to donate up to 10\% of their annual income to a political campaign. For candidates, individual contributions serve as an important source for their campaign spending. Consequently, political candidates frequently reward their supporters by offering jobs in the public sector \citep{colonnelli2020}. Our findings reveal a positive effect on campaign donations in the two cycles after job loss.\footnote{Data on campaign contributions is also obtained from the \textit{Tribunal Superior Eleitoral}. We exclude contributions to individuals' own campaigns to ensure that the results are not only driven by own candidacies.} The estimates are both highly significant and substantial in magnitude. The DiD coefficient indicates a 0.049 percentage point increase in individuals' probability of donating to a political campaign, corresponding to an 18\% increase relative to the control mean. The results suggest that despite experiencing financial setbacks from job layoffs, individuals are not deterred from participating in political activities that require financial contributions. This also aligns with the observed positive impact on party memberships and candidacies, both of which may entail some  financial investments in the form of membership fees or campaign funding.

Third, we study the possibility of working for a political campaign. \cite{bazzi2023} show that working for a winning mayoral campaign in Brazil yields substantial private income gains from improved access to jobs in the local bureaucracy. Following their approach, we identify dedicated salaried campaign staff in the RAIS data using information on the industry and legal nature of the employers.\footnote{Specifically, we define campaign workers as individuals who have a full-time contract at a firm in industry \textit{Atividades de Organizações Políticas} (CNAE 1.0: 91928, CNAE 2.0: 94928) or with legal nature \textit{Partido Político} (3123) or \textit{Candidato a Cargo Político Eletivo} (4090).} Our results show an increase in individuals' probability of working for a political campaign after job loss. Despite also being a rare event in our sample, the estimates are highly significant and suggest roughly a doubling in the likelihood of becoming a campaign worker. The effects are also highly persistent up to seven years after the layoff which is in line with our findings on party memberships and councilor candidacies.

\subsection{Individual and local heterogeneity}\label{sec:results_individual}
As a next step, we analyze heterogeneity in the effects of job loss along key socio-demographic characteristics. Specifically, we test for differential effects by pre-layoff education (categorized into three groups: middle school or less, high school, university education), age (in terciles), and gender. For each subgroup, we separately estimate the DiD model from equation (\ref{model:did}) and scale the treatment effect by the group-specific mean of the control group. 

\begin{figure}[tb]
	\caption{Heterogeneity by individual characteristics}  \label{fig:did_hetero_individual}
	\centerline{
		\begin{threeparttable}
			\begin{footnotesize}
				\begin{tabular}{cc}
					\textbf{A. New party membership} & \textbf{B. Candidacy}\\
					\includegraphics[width=0.55\linewidth]{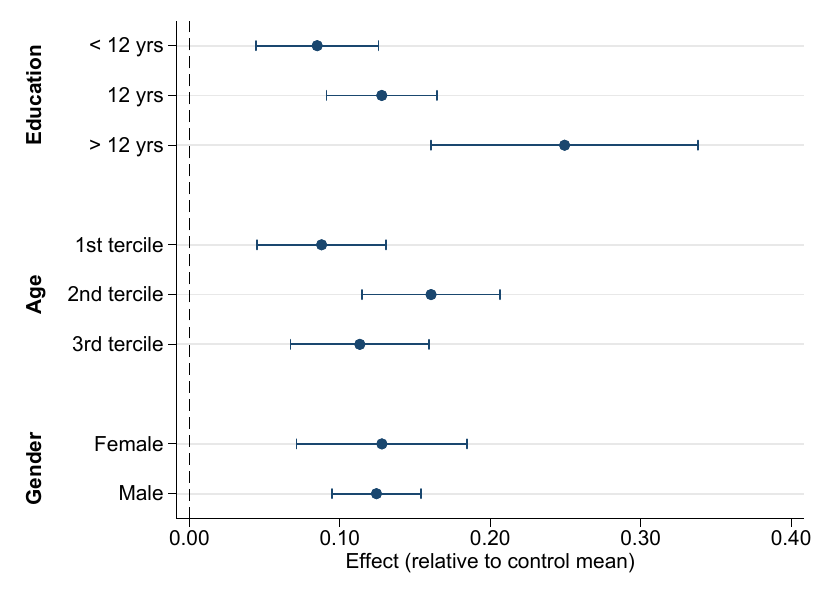} &\includegraphics[width=0.55\linewidth]{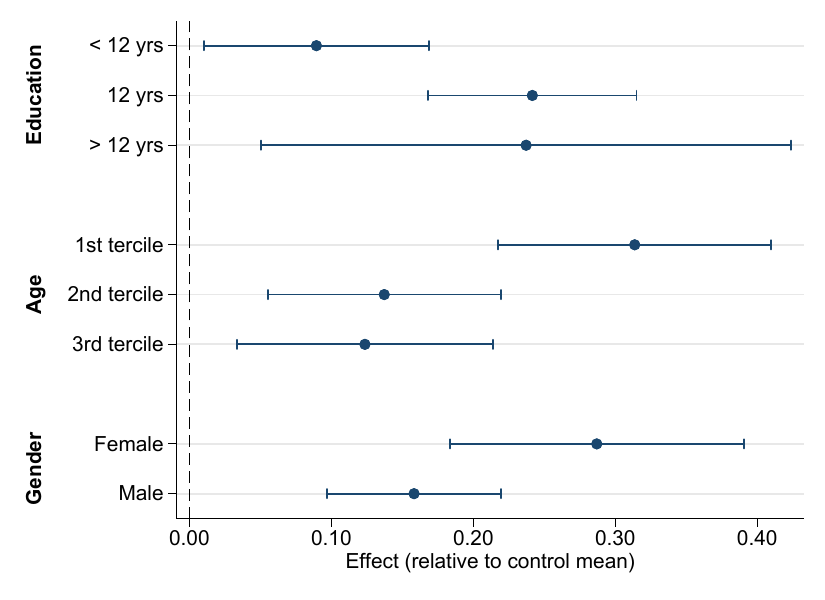} \\ 
				\end{tabular}
			\end{footnotesize}
			\begin{tablenotes} \scriptsize
			\item \textit{Note:}
                The figure reports DiD coefficients, estimated in model (\ref{model:did}), for the effect of job loss on the likelihood of newly registering as a party member (Panel A) and running for local councilor (Panel B) for different subgroups of individuals. For education, we differentiate no degree, elementary school, or middle school ($<$ 12 years), high school (12 years), and university education ($>$ 12 years). For age, the sample is split by terciles. The horizontal bars depict 95\% confidence intervals based on standard errors clustered at the individual level. All coefficients and standard errors are scaled by the group-specific mean of the control group in the post-treatment period (average across 2012, 2016, and 2020). 			\end{tablenotes}
	\end{threeparttable}}
\end{figure}

Results are shown in Figure \ref{fig:did_hetero_individual}. One striking pattern that emerges is that the effects of job loss on party memberships and candidacies are remarkably pervasive, being significantly positive in all individual subsamples considered. Appendix Figure \ref{fig:did_hetero_municipal} shows that the effects are also pervasive across several municipality characteristics. These include the local average household income, unemployment rate, Gini index, homicide rate, as well as the electorate's ideological leaning, which is measured as the vote shares for the left- vs. center-right candidate (Luiz In\'acio Lula da Silva vs. Geraldo Alckmin) in the last pre-treatment Presidential election of 2006. Despite the large heterogeneity across regions in Brazil, we find significant positive effects of job loss on political entry along the distribution of all these local characteristics.

Beyond the overall positive effects, Figure \ref{fig:did_hetero_individual} reveals substantial differences across education levels, with larger effects on party membership and candidacy among more educated individuals. In quantitative terms, we estimate that individuals with a university degree ($>$ 12 years of education) experience an increase of about 25\% in the probabilities of both joining a political party and running for office, while for individuals without a high school degree ($<$ 12 years of education), the estimated effects on both outcomes are below 10\%. These patterns are consistent with the opportunity-cost framework presented earlier. Highly educated individuals tend to have better earnings prospects in the private sector, so being displaced in a mass layoff may represent a larger shock to their labor income. Furthermore, the benefits from political engagement -- whether in terms of prestige, perceived chances of winning, or future job opportunities -- may be higher for more educated individuals, making a political career more attractive. To shed light on these mechanisms, Appendix Figure \ref{fig:es_laboroutcomes_education} examines post-layoff earnings and employment trajectories by education level. The results show that highly-educated individuals indeed suffer substantially larger total earnings losses after the layoff. While their overall employment rate does not improve more rapidly, they are more likely to transition into public-sector jobs, suggesting that political connections may offer them greater potential returns through improved access to such positions.

In terms of the other two individual characteristics considered -- age and gender -- we do not observe a clear gradient in the effects on new party memberships. In contrast, the impact on candidacies is more pronounced among younger individuals and women. These patterns may reflect differences in labor market attachment and career flexibility. Younger workers often face less entrenched career paths, which can lower the barriers to start a political position after a layoff.\footnote{The larger effects for young workers are also consistent with the ``impressionable years'' hypothesis according to which political attitudes are more malleable at younger ages \citep[e.g.,][]{krosnick1989}.} Similarly, given persistent gender gaps in labor market opportunities, women may also face fewer opportunity costs when shifting toward a political career, contributing to their higher likelihood of running for office following displacement.

\subsection{Monetary incentives}\label{sec:results_monetary}
Holding a political position often entails substantial monetary rewards, making it an appealing strategy for mitigating the income loss resulting from being laid off in the private sector. To examine the role of financial incentives in shaping the decision to enter politics after a job loss, we conduct two heterogeneity analyses. First, to capture changes in private-sector opportunity costs, we construct a measure of the predicted income loss from displacement and test whether individuals facing larger losses are more likely to enter politics. Second, we study the attractiveness of local councilor jobs by exploiting variation in councilor wages across Brazilian municipalities.  

\vspace{1em} \noindent \textbf{Predicted income loss.}\quad We start by examining whether predicted income losses due to the layoff are related to the effects of job loss on political entry.\footnote{We focus on predicted, rather than actual earnings loss, since the latter may be directly affected by individuals' behavior after the layoff, such as their decision to run for office. \cite{hilger2016} and \cite{Britto2022a} conduct similar exercises for the effect of parental job loss on children's outcomes.} The income effects of job loss are predicted as follows. Similar to \cite{schmieder2023}, we first construct for each laid-off worker in our sample an individual-specific measure of the relative earnings losses after layoff by exploiting the fact that we have matched each treated worker to a similar control worker:
\begin{equation*}
    \Delta_{dd} y_{it} = \frac{\bar{y}_{i, post} - \bar{y}_{i, pre}}{\bar{y}_{i, pre}} - \frac{\bar{y}_{i', post} - \bar{y}_{i', pre}}{\bar{y}_{i', pre}} 
\end{equation*}
where $\bar{y}_{i, post}$ is the average earnings in the 5 years after the layoff of treated worker $i$, or her matched control worker $i'$. $\bar{y}_{i, pre}$ is the average earnings in the 5 years before the layoff. This double difference can be thought of as the share of earnings that each worker loses in the medium run because of the layoff. We then regress $\Delta_{dd} y_{it}$ on a rich set of pre-layoff characteristics and compute the predicted relative earnings loss. The regression includes the following individual covariates: gender, education, age decile, earnings decile, tenure decile, firm size decile, and layoff year. We also add municipality $\times$ industry dummies to capture variation in labor market conditions. Finally, we re-estimate the effects of job loss on new party memberships and candidacies in model (\ref{model:did}), by splitting the sample into quintiles of the predicted earnings loss distribution. 

\begin{figure}[tb]
	\caption{Heterogeneity by predicted earnings loss}  \label{fig:did_hetero_earningsloss}
	\centerline{
		\begin{threeparttable}
			\begin{footnotesize}
				\begin{tabular}{cc}
					\textbf{A. New party membership} & \textbf{B. Candidacy}\\
					\includegraphics[width=0.55\linewidth]{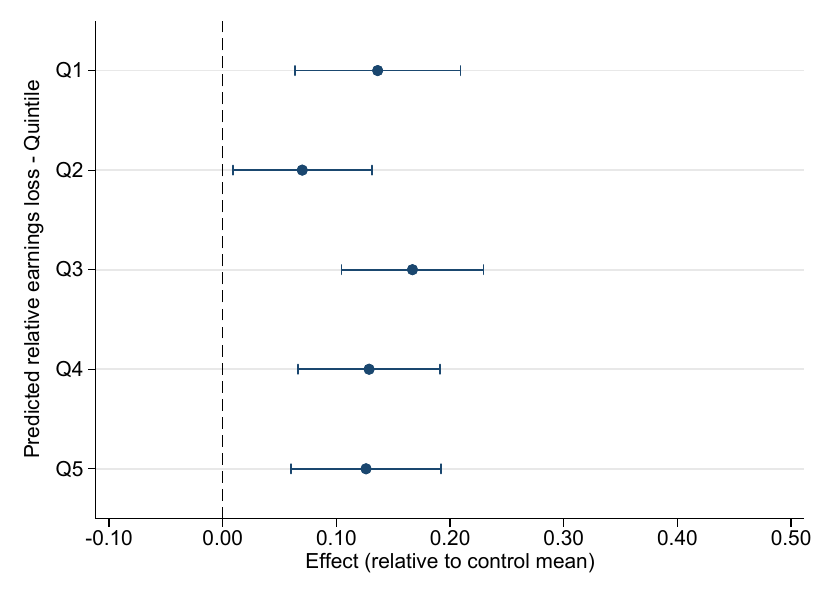} &\includegraphics[width=0.55\linewidth]{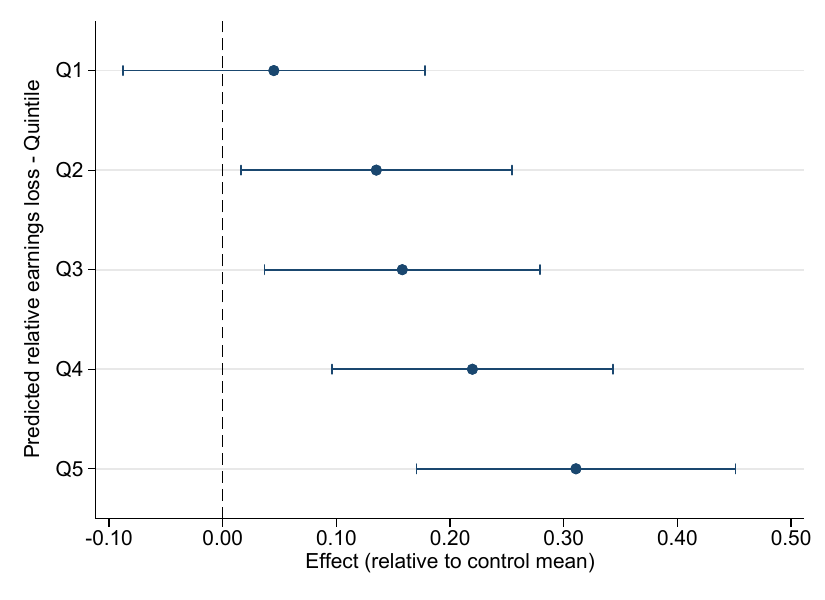} \\ 
				\end{tabular}
			\end{footnotesize}
			\begin{tablenotes} \scriptsize
			\item \textit{Note:}
				The figure reports DiD coefficients, estimated in model (\ref{model:did}), for the effect of job loss on the likelihood of newly registering as a party member (Panel A) and running for local councilor (Panel B). We split the sample by quintiles of the treated individuals' predicted relative earnings loss due to the layoff. See the text for details on how earnings losses are predicted. Control workers are assigned to their matched treated workers in each subsample. The horizontal bars depict 95\% confidence intervals based on standard errors clustered at the individual level. All coefficients and standard errors are scaled by the group-specific mean of the control group in the post-treatment period (average across 2012, 2016, and 2020). 
			\end{tablenotes}
	\end{threeparttable}}
\end{figure}

Results are shown in Figure \ref{fig:did_hetero_earningsloss}. The predicted income losses are clearly related to the estimated effects of job loss on councilor candidacies. In the lowest quintile of earnings losses, we find null effects, whereas in the highest quintile job loss is estimated to raise the probability of running for councilor by more than 30\%. In contrast, we do not find heterogeneity in the effects on new party memberships. These results support the idea that the reduction in private-sector opportunity costs is a key mechanism behind the rise in political entry after layoffs -- particularly for time-intensive forms of political engagement such as running for office, where the opportunity cost of foregone earnings plays a larger role than for lower-commitment activities like party membership.  

\vspace{1em} \noindent \textbf{Councilor wages.}\quad Since 2004, a constitutional amendment has capped councilor salaries based on the population size of the municipality. In municipalities with less than 10,000 inhabitants, councilors can receive a maximum of 20\% of state deputy salaries. This share increases for larger municipalities, up to 75\% in municipalities with more than 500,000 inhabitants. State deputy salaries, in turn, can get at most 75\% of the federal deputy salary. Appendix Table \ref{tab:counc_wage_caps} reports the maximum allowed councilor salary for each population size group, as estimated by \cite{Ferraz2011}. In absolute terms, maximum councilor wages vary between 1,927 BRL (391 USD) and 7,227 BRL (1,466 USD) per month, demonstrating that there is substantial local variation in the financial incentives to run for local councilor.

\begin{figure}[tb]
	\caption{Heterogeneity by councilor wages}  \label{fig:did_hetero_wages}
	\centerline{
		\begin{threeparttable}
			\begin{footnotesize}
				\begin{tabular}{cc}
					\textbf{A. New party membership} & \textbf{B. Candidacy}\\
					\includegraphics[width=0.55\linewidth]{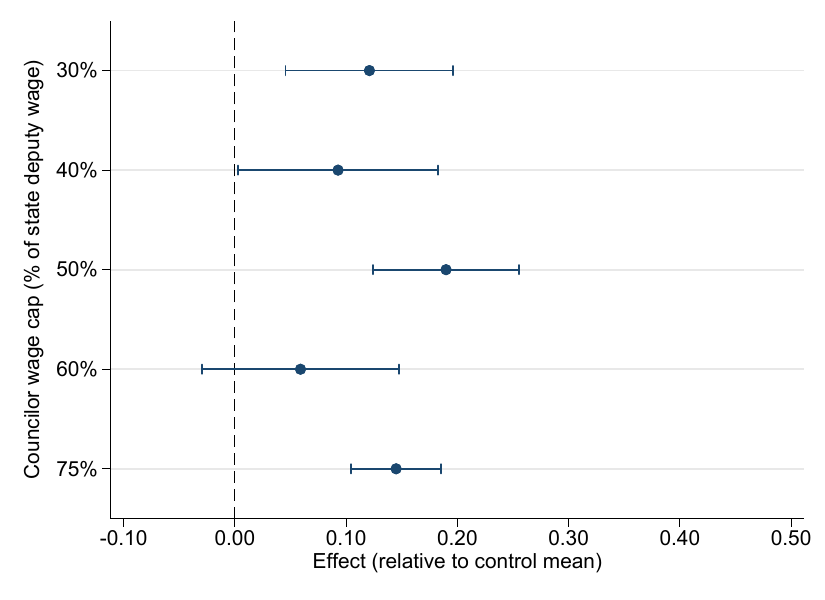} &\includegraphics[width=0.55\linewidth]{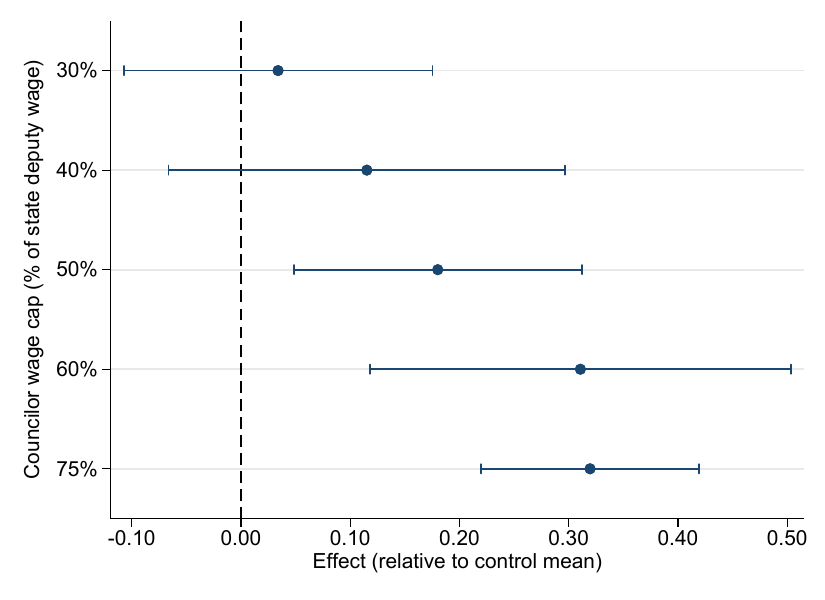} \\ 
				\end{tabular}
			\end{footnotesize}
			\begin{tablenotes} \scriptsize
			\item \textit{Note:}
				The figure reports DiD coefficients, estimated in model (\ref{model:did}), for the effect of job loss on the likelihood of newly registering as a party member (Panel A) and running for local councilor (Panel B). We split the sample by the maximum monthly wage that councilors can earn (in \% of state deputy salaries) in the municipality that the individual has worked in before the layoff. The wage cap groups are determined by municipality population counts of 2008 (see cutoffs in Appendix Table \ref{tab:counc_wage_caps}). We pool the two lowest population groups, referring to a maximum councilor wage of 20\% and 30\%, as the lowest population group is too small to yield meaningful results. The horizontal bars depict 95\% confidence intervals based on standard errors clustered at the individual level. All coefficients and standard errors are scaled by the group-specific mean of the control group in the post-treatment period (average across 2012, 2016, and 2020). 
			\end{tablenotes}
	\end{threeparttable}}
\end{figure}

In Figure \ref{fig:did_hetero_wages}, we examine heterogeneity in the effects of job loss by the maximum wage that councilors can receive in a municipality. We find a strong gradient in the effects on running for councilor. In municipalities with the lowest wage caps, job loss does not significantly affect local candidacies. In contrast, in municipalities in which councilors can earn up to 75\% of state deputy salaries, job loss is estimated to increase the likelihood of running for councilor by more than 30\% relative to the control mean. Again, we do not find similar patterns for the effects on party memberships. As party members do not receive wages, this result is expected and gives us confidence that we are capturing the role of financial incentives, rather than other differences between small and large municipalities, in influencing the decision to enter politics after being laid off.   

\subsection{Party disaggregations}\label{sec:results_party}
Job loss may not only affect overall political participation but also lead to changes in individuals' support of different political parties. We, therefore, study heterogeneities in the effects on memberships and candidacies by party characteristics, focusing on parties' ideology, local incumbency, and alignment with labor unions. Estimates from the static DiD model (\ref{model:did}) are presented in Table \ref{tab:hetero_party}, while dynamic event-study results from model (\ref{model:es}) are shown in Appendix Figures \ref{fig:es_ideo} - \ref{fig:es_union}. We also report results from model (\ref{model:did}) separately for each of Brazil's major parties in Appendix Figure \ref{fig:did_party}.

\begin{table}[tb]
	\centerline{
		\begin{threeparttable}
			\caption{Heterogeneity by party characteristics}
			\label{tab:hetero_party}
			\begin{footnotesize}
				\begin{tabular}{lcccccccc}
					\toprule[1.5pt] 
					& (1) & (2) & (3)& (4)& (5)& (6)& (7) & (8) \\
					\midrule \\[-2.0ex]
					& All & \multicolumn{3}{c}{Ideology} & \multicolumn{2}{c}{Local incumbency} & \multicolumn{2}{c}{Union affiliation} \\ \cmidrule(lr){3-5} \cmidrule(lr){6-7} \cmidrule(lr){8-9}
                        &    & Left & Center & Right & Yes & No & Yes & No \\
					\midrule
					\multicolumn{9}{l}{\textbf{[A] Outcome: new party membership (\%)}} \\ 
					[0.5em]
$\delta_{DiD}$ 	&       0.175\sym{***}&       0.045\sym{***}&       0.039\sym{***}&       0.091\sym{***}&       0.024\sym{***}&       0.151\sym{***}&       0.033\sym{***}&       0.051\sym{***}\\
               &     (0.018)         &     (0.011)         &     (0.010)         &     (0.011)         &     (0.006)         &     (0.018)         &     (0.011)         &     (0.010)         \\
[0.5em]
Control mean   &       1.407         &       0.537         &       0.353         &       0.517         &       0.099         &       1.308         &       0.442         &       0.448         \\
Relative effect&        12.4\%         &         8.4\%         &        11.0\%         &        17.6\%         &        24.0\%         &        11.6\%         &         7.5\%         &        11.3\%         \\
Observations   &  11,317,632         &  11,317,632         &  11,317,632         &  11,317,632         &  11,317,632         &  11,317,632         &  11,317,632         &  11,317,632         \\
					[0.5em]
					\midrule
					\multicolumn{9}{l}{\textbf{[B] Outcome: candidacy (\%)}} \\ 
					[0.5em]
$\delta_{DiD}$  &       0.041\sym{***}&       0.012\sym{***}&       0.010\sym{***}&       0.019\sym{***}&       0.005\sym{***}&       0.036\sym{***}&       0.010\sym{***}&       0.012\sym{***}\\
               &     (0.006)         &     (0.003)         &     (0.003)         &     (0.004)         &     (0.001)         &     (0.006)         &     (0.003)         &     (0.003)         \\
[0.5em]
Control mean   &       0.229         &       0.078         &       0.060         &       0.090         &       0.013         &       0.216         &       0.066         &       0.072         \\
Relative effect&        18.1\%         &        15.3\%         &        17.2\%         &        21.1\%         &        43.1\%         &        16.6\%         &        15.5\%         &        16.7\%         \\
Observations   &  11,317,632         &  11,317,632         &  11,317,632         &  11,317,632         &  11,317,632         &  11,317,632         &  11,317,632         &  11,317,632         \\
[0.5em]
\hline\hline
				\end{tabular}
			\end{footnotesize}
			\begin{tablenotes} \scriptsize 
				\item \textit{Note:} 
			The table reports DiD coefficients, estimated in model (\ref{model:did}), for the effect of job loss on the likelihood of newly registering as a party member (Panel A) and running for local councilor (Panel B). Column (1) presents the baseline results for all parties. Columns (2) to (4) differentiate left, center, and right parties, following the ideological classification of \cite{Colonnelli2022} which is shown in Appendix Table \ref{tab:class_parties_ideo}. Columns (5) and (6) distinguish memberships and candidacies in the party of the elected mayor in the individual's municipality in 2008 and all other non-incumbent parties. Columns (7) and (8) show results separately for parties historically affiliated with labor unions (PT, PDT, PSB, PCB, PSD, and MDB) and all other left or center parties. All coefficients, standard errors, and control means have been scaled by 100, such that effects are interpreted in terms of percentage points. \sym{*} \(p<0.10\), \sym{**} \(p<0.05\), \sym{***} \(p<0.01\)
			\end{tablenotes}
		\end{threeparttable}
	}
\end{table}

\vspace{1em} \noindent \textbf{Ideology.}\quad Layoffs may lead individuals to align with parties whose political programs better address workers' revised economic interests, for example, through redistributive policies. Moreover, even in the absence of economic platforms, extreme parties may be more successful in exploiting individuals' resentment of the establishment following economic downturns \citep[e.g.,][]{Algan2017a, Autor}. To analyze whether layoffs lead to ideological shifts, we distinguish left, center, and right parties, following the classification of \cite{Colonnelli2022}.\footnote{The ideological classification of parties is shown in Appendix Table \ref{tab:class_parties_ideo}.} Our results, shown in columns (2) to (4) of Table \ref{tab:hetero_party}, indicate that job loss leads to an increase in party memberships and candidacies across the ideological spectrum. Relative to the control mean, we find an increase of 8.4\%, 11.0\% and 17.6\% in the likelihood of individuals newly affiliating with left, center and right parties, respectively. The effects on candidacy reveal a similar pattern, with likelihood increases of 15.1\%, 17.3\%, and 21.3\% of running for office in a left, center, or right-wing party. Although we observe larger effects for right-wing parties, our results show that center and left-wing parties also receive more members and candidates following layoffs. Interestingly, the event-study results presented in Appendix Figure \ref{fig:es_ideo} show that the higher increase in affiliations with right-wing parties is driven only by the 2020 cycle, the first local election after Bolsonaro took power in 2019. The rise in right-wing support following economic hardships, which is well documented in the literature \citep[e.g.,][]{colantone2018, dippel2022, Barros2019}, appears to be contingent on the current political climate.  

\vspace{1em} \noindent \textbf{Local incumbency.}\quad Next, we examine whether individuals punish the party of the local incumbent government after being laid off. Retrospective voting has been widely documented to rationalize the behavior of voters following economic shocks. In particular, individuals may blame incumbent politicians for local economic policies that result in mass layoffs, or for insufficient support policies that mitigate economic hardship after a layoff \citep[e.g.,][]{Tilley2018, Ahlquist2020}. In columns (5) and (6) of Table \ref{tab:hetero_party}, we compare the effects for the party of the elected mayor in each individual's municipality before the layoff in 2008 and all other non-incumbent parties. The results show that job loss leads to increased participation in both incumbent and non-incumbent parties. Relative to the control mean, the effects are considerably stronger for incumbent parties, with an increase of 24.0\% in the probability of being affiliated with the party of the incumbent and an increase of 43.7\% in the likelihood of running for office in the party of the incumbent (as opposed to 11.6\% and 16.7\% in non-incumbent parties). The results do not support incumbent punishment as a main driver of our results. Instead, running under the incumbent party may raise the likelihood of electoral success, thereby increasing the expected benefits of participation (consistent with our theoretical framework in Section \ref{sec:background_theory}).

\vspace{1em} \noindent \textbf{Union affiliation.}\quad Lastly, we investigate whether the effects depend on parties' alignment with labor unions. Unions can influence employees' political behavior through various activities, such as educating, informing, and mobilizing their members, endorsing selected candidates and policies, and organizing get-out-the-vote campaigns \cite[e.g.,][]{kim2017, matzat2023}. In Brazil, labor unions have historically been linked to left and centrist parties and have played an important role in advocating for worker interests in domestic politics. Until 2017, unions were required by law to participate in negotiations to formalize dismissals in the context of mass layoffs.\footnote{In 2017, the Brazilian labor reform (Law 13,467/17) established that it was no longer mandatory for unions to take part in collective bargaining agreements during mass layoffs.} We define union-affiliated parties following the classification of \cite{ogeda2021} and compare them to all other left-wing or center parties that have no historical ties with labor unions in Brazil.\footnote{Union-affiliated parties are PT, PDT, PSB, PCB, PSD, and MDB, while non-union left and center parties are CIDADANIA, PCDOB, PV, PMN, PSOL, PSTU, PCO, SD, PROS, PPL, PMB, REDE, UP, PSDB, PTB, and AVANTE.} Our results show an increase in the probability of individuals becoming members or local candidates in both union-affiliated and non-union-affiliated parties (columns (7) and (8) of Table \ref{tab:hetero_party}). In quantitative terms, we estimate a 7.5\% (11.3\%) increase in the probability of newly affiliating with union-aligned (non-union-aligned) parties and a 15.4\% (16.7\%) increase in candidacy probabilities in union-affiliated (non-union-affiliated) parties following mass layoffs. Thus, we find no evidence that unions play a major role in politically mobilizing laid-off individuals.

\section{Effects of Unemployment Insurance on Political Entry}\label{sec:ui}
The main results presented in the previous section show that job loss significantly increases individuals' likelihood of joining a party and running for local office. As discussed in Section~\ref{sec:background_theory}, these effects may be explained by two mechanisms: income losses that reduce the opportunity costs of political entry, and increased time resources that allow individuals to engage more intensively in politics. Both channels may be influenced by individuals' eligibility for unemployment insurance benefits.
In Brazil, workers who are dismissed without just cause can receive UI benefits for 3 to 5 months, with an average replacement rate of about 80\%.\footnote{The length of UI payments depends on the length of employment in the 36 months before the layoff, and the replacement rate depends on the previous earnings (varying between 100\% for individuals earning the minimum wage and 67\% at the benefit cap, which is at 2.65 times the minimum wage).} Benefit receipt mitigates the income losses from layoffs and may thereby reduce the financial incentives to pursue a political position. At the same time, during benefit receipt dismissed workers have lower incentives to find reemployment.\footnote{The negative effects of UI eligibility on formal employment have been widely documented, including for Brazil \citep{gerard2020, gerard2021, Britto2022}. \cite{britto2022_restat} also shows that an extension of the UI eligibility period in Brazil decreased total employment when taking into account the positive effect on informal employment.} The negative employment effects of UI eligibility decrease labor earnings, which may offset the benefit payments, and increase individuals' time availability. The increased time resources, in turn, may facilitate individuals' political engagement. In this section, we investigate whether eligibility for UI benefits affects laid-off individuals' decision to enter politics, which can help shed light on the mechanisms driving the effect of job loss on political entry.

\subsection{Empirical strategy}\label{sec:ui_strategy}

We conduct a regression discontinuity (RD) analysis exploiting the fact that eligibility for UI benefits varies discontinuously with the time since the last layoff. Specifically, workers who are displaced without just cause from a formal job with at least six months tenure are eligible for UI benefits if at least 16 months have elapsed between the worker's current layoff date and the most recent layoff date used to claim UI benefits in the past. We compare workers displaced shortly before and shortly after the 16-month cutoff by estimating local linear regressions of the following form:
\begin{equation}\label{model:rdd}
Y_i=\alpha+\beta D_i+\gamma_1 X_i+\gamma_2 D_i X_i +\epsilon_i,
\end{equation}
where $Y_i$ is an outcome of worker $i$. $X_i$ denotes the time since the last layoff (re-centered around the 16-month cutoff), and $D_i$ is a dummy for being eligible for UI benefits, i.e., $D_i = 1(X_i\geq0)$. $\beta$ is the coefficient of interest that identifies the effect of UI eligibility.\footnote{As we lack information on UI take-up, our focus is on benefit eligibility. Consequently, our estimates should be interpreted as an intent-to-treat effect. According to \cite{Britto2022}, the probability of UI take-up jumps by about 60 percentage points at the 16-month eligibility cutoff.} To ensure that eligible and non-eligible workers are comparable, our main results are based on a local linear model with a narrow bandwidth of 60 days around the cutoff.\footnote{This choice follows the main specifications in \cite{Britto2022}, \cite{Britto2022a}, and \cite{Amorim2023}.} In Appendix Tables \ref{tab:ui_robust_member} and \ref{tab:ui_robust_candidate} we show that our main results are robust to alternative polynomial orders and bandwidth choices, including the optimal bandwidth proposed by \cite{calonico2014}. In Appendix Figure \ref{fig:ui_ptests} we also perform permutation tests that compare our main estimates to the distribution of RD estimates obtained from placebo cutoff points. 

As in the previous section, the estimation sample consists of all workers aged 25 to 50 displaced from a full-time, open-ended, private-sector job of at least six months tenure. To increase statistical power, we extend our layoff period to cover the years 2006 to 2014.\footnote{We start in 2006 because this allows us to observe past layoffs up to 24 months prior to the current layoff (our employment data begin in 2004) and to measure placebo political outcomes in the two cycles prior to the current layoff (our political data begin in the 2000 cycle). We stop in 2014 because of the numerous changes introduced to the UI system after that year.} This allows us to measure political outcomes over two election cycles after the layoff and labor market outcomes over up to 4 years after the layoff.\footnote{While we must restrict the sample to individuals with unique names for the party membership outcome, we do not impose this restriction when analyzing candidacies, which are less common and thus require larger sample sizes.} In addition, we exclude from the sample all workers who were displaced on the first or the last day of each month in their last layoff. Appendix Figure \ref{fig:ui_cycles} shows that the number of dismissals is higher at the very beginning and the very end of the month. Dropping these observations ensures that our RD cutoff does not coincide with spikes in the density of dismissals that occur every 30 days, i.e., that are not specific to the 16-month cutoff. 

Appendix Figure \ref{fig:ui_density} shows that, when using the restricted sample, there is no evidence of discontinuity in the density of the running variable at the 16-month cutoff. We also confirm the continuity using the bias-robust test by \cite{cattaneo2018, cattaneo2020}. In addition, in Appendix Figure \ref{fig:ui_covariates} we show that individuals around the cutoff are similar in a rich set of pre-determined covariates, including gender, education, age, earnings, tenure, and sector shares. Overall, these results provide strong support for the continuity assumption required by the RD approach.

\subsection{Results}\label{sec:ui_results}
\begin{figure}[h!]
	\caption{Effect of UI eligibility}  \label{fig:ui_main}
	\centerline{
		\begin{threeparttable}
  			\begin{footnotesize}
				\begin{tabular}{ccc}
                        \multicolumn{3}{c}{\textbf{A. Full sample}}\\ [0.5em]
					Months employed & New party membership & Candidacy \\
					\includegraphics[width=0.35\linewidth]{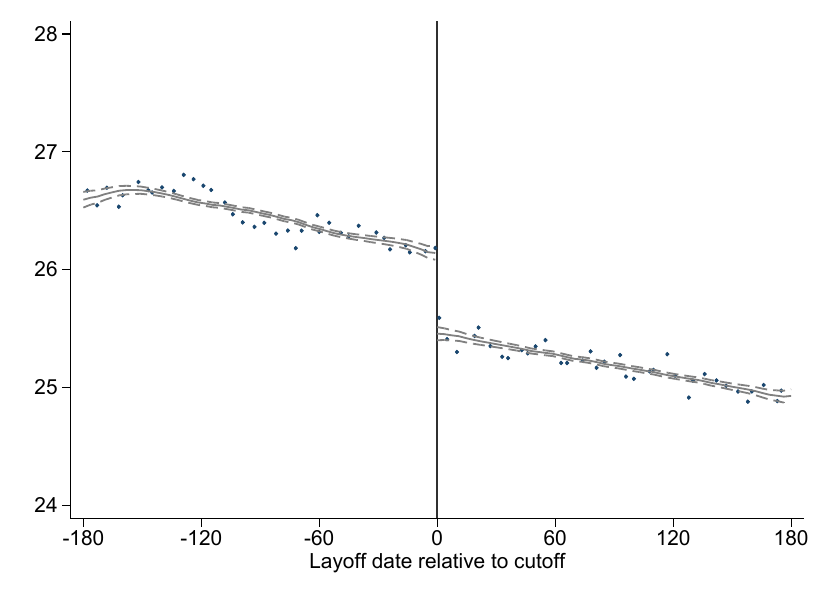}
                        & \includegraphics[width=0.35\linewidth]{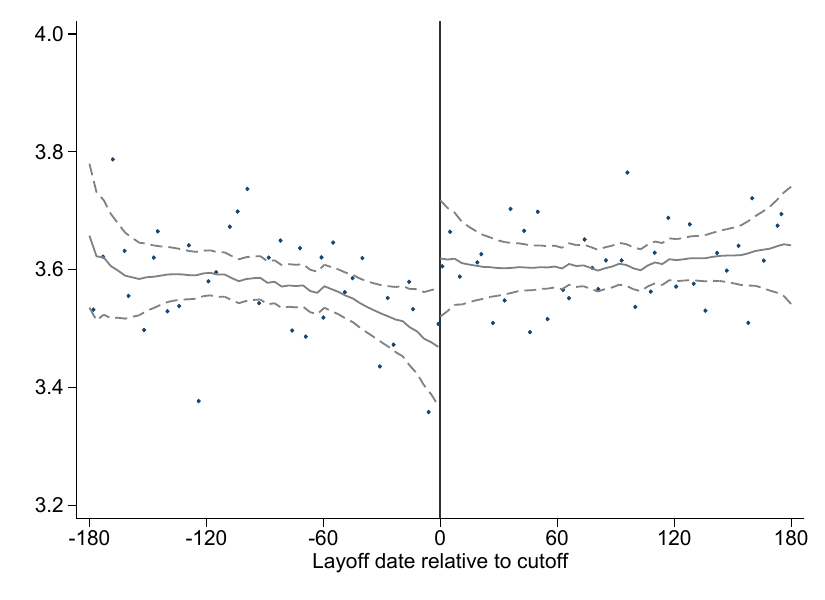}  
                        & \includegraphics[width=0.35\linewidth]{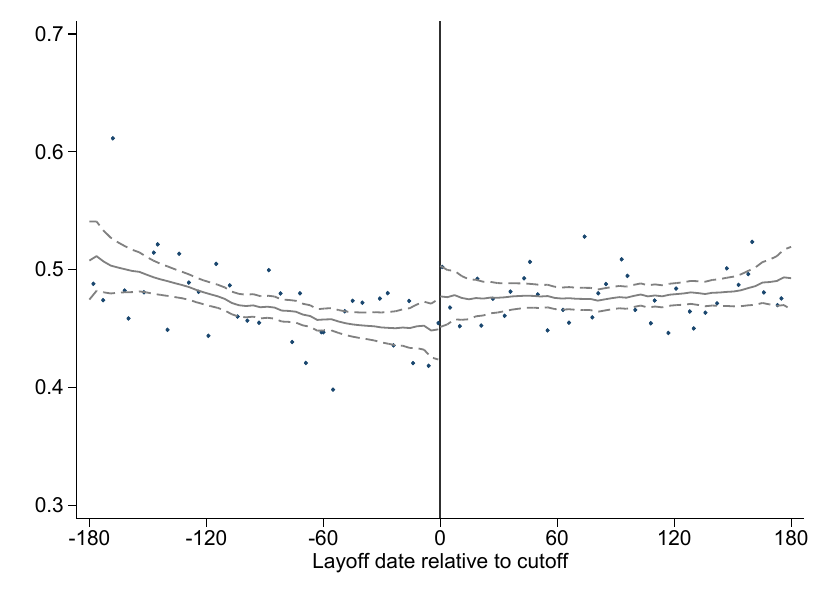} \\ [1em]
                        \multicolumn{3}{c}{\textbf{B. Low educated ($<$ 12 years of education)}}\\ [0.5em]
					Months employed & New party membership & Candidacy \\
					\includegraphics[width=0.35\linewidth]{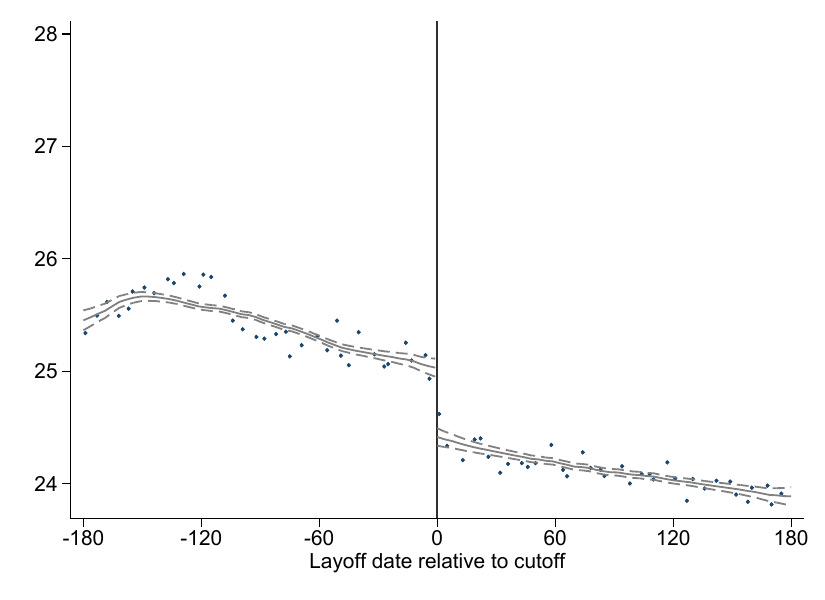}
                        & \includegraphics[width=0.35\linewidth]{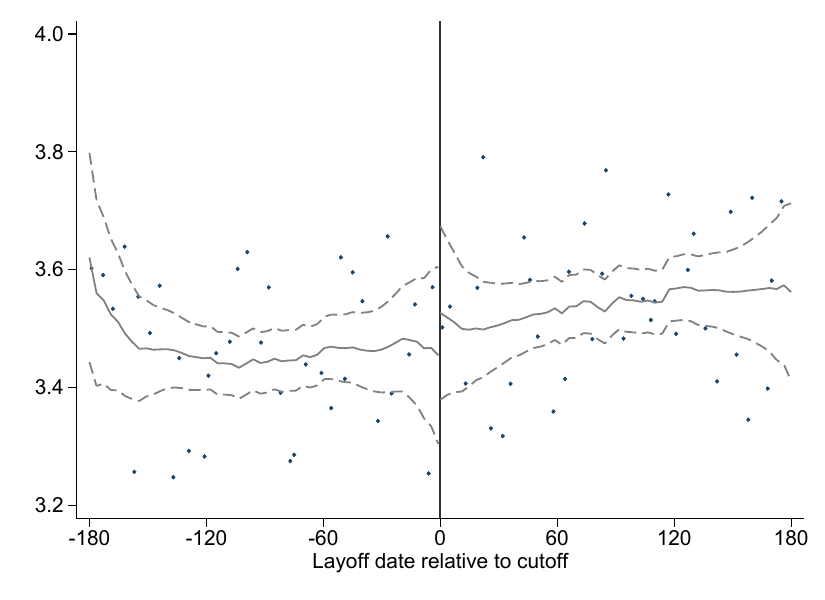}  
                        & \includegraphics[width=0.35\linewidth]{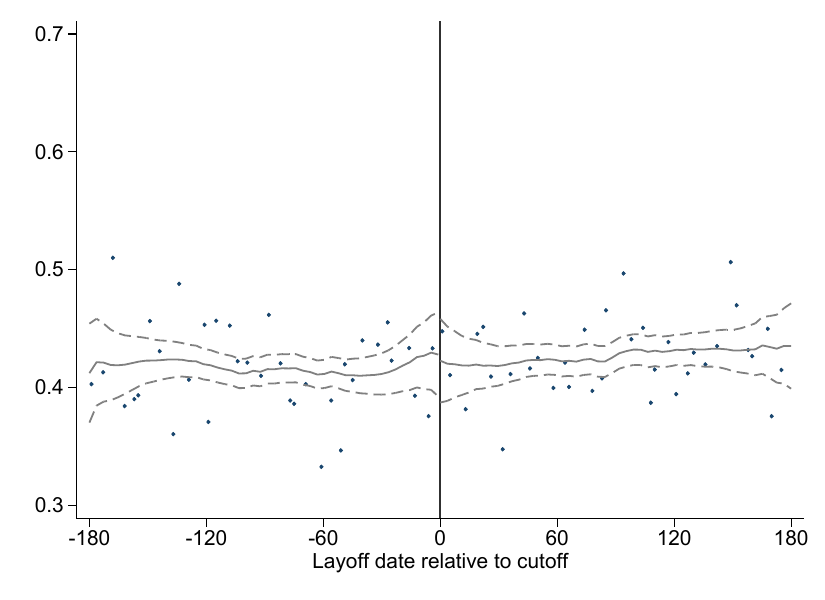} \\  [1em]         
                        \multicolumn{3}{c}{\textbf{C. High educated ($\geq$ 12 years of education)}}\\ [0.5em]
					Months employed & New party membership & Candidacy \\
					\includegraphics[width=0.35\linewidth]{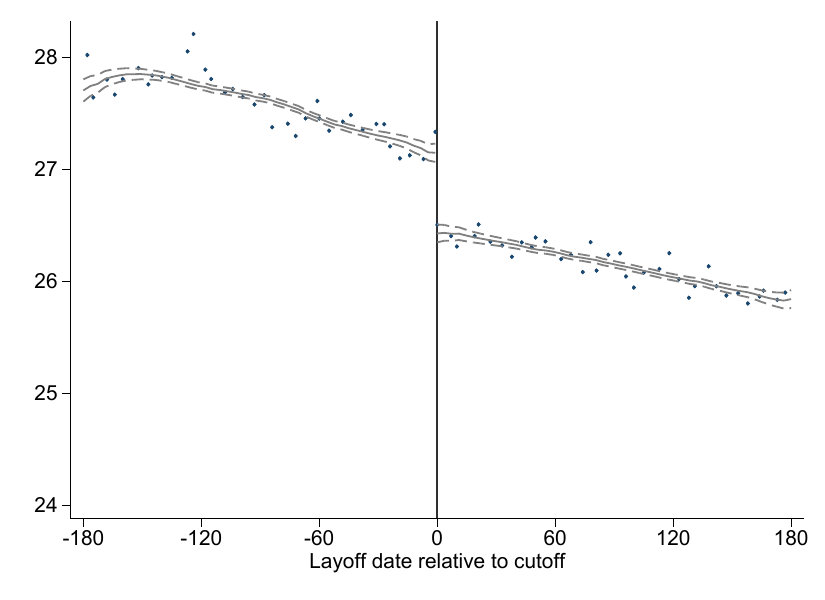}
                        & \includegraphics[width=0.35\linewidth]{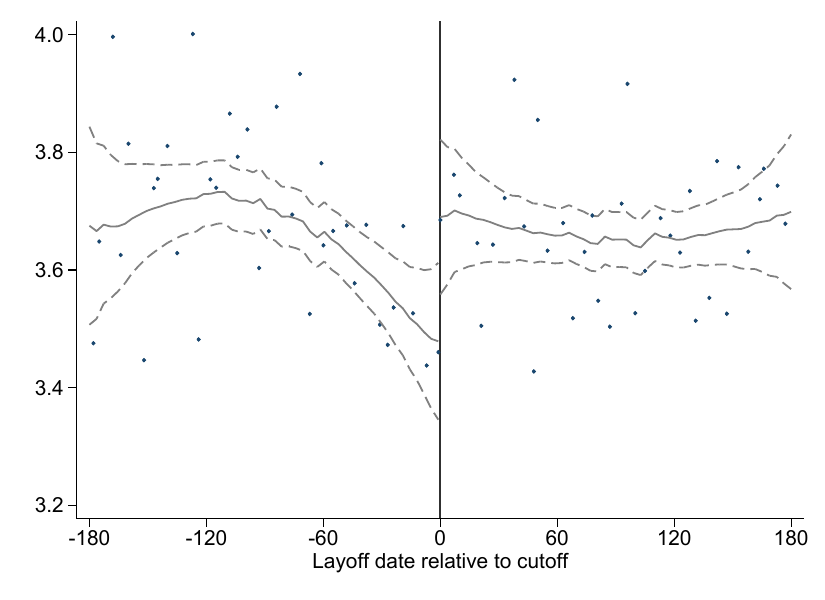}  
                        & \includegraphics[width=0.35\linewidth]{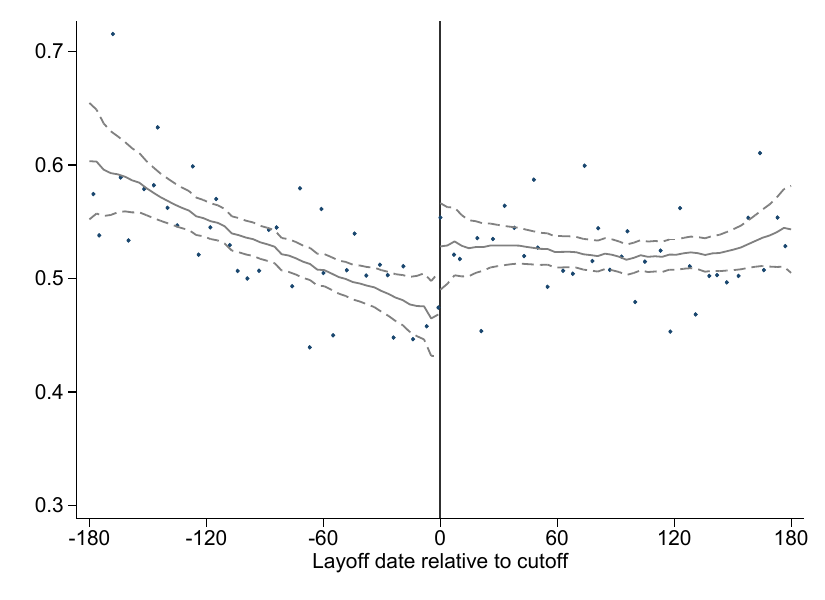} \\          
				\end{tabular}
			\end{footnotesize}
			\begin{tablenotes} \scriptsize
			\item \textit{Note:}
			The figure plots the average outcomes of workers around the 16-month cutoff date for eligibility for unemployment benefits, for different groups of workers. Outcomes are the number of months employed in the four years after the layoff, the probability of newly registering as a party member in the two election cycles after the layoff (in \%), and the probability of running for local councilor in the two election cycles after the layoff (in \%). Panel A shows results among all workers in the sample, and Panels B and C distinguish workers with less than vs. at least 12 years of education. Dots show averages for 5-day bins. The lines represent a local linear polynomial smoothing using a 60-day bandwidth, together with 95\% confidence intervals.
			\end{tablenotes}
	\end{threeparttable}}
\end{figure}

Results from model (\ref{model:rdd}) are presented in Table \ref{tab:ui_main}, and the corresponding RD plots are shown in Figure \ref{fig:ui_main}. As a first step, we estimate the effects of UI eligibility on labor market outcomes (Panel A, columns (1) and (2) of Table \ref{tab:ui_main}). We find a significant negative effect on employment that amounts to 0.68 fewer months employed over the four years after the layoff. Moreover, labor earnings decrease by 2,192 BRL. The drop in labor earnings is likely compensated by the UI payments that eligible individuals receive.\footnote{\cite{Britto2022} find that individuals barely meeting the 16-months cutoff receive additional UI benefits of about 2,100 BRL.} Thus, the results suggest that UI eligibility leads to a rise in time availability following the job loss without strongly affecting the total available income.


\begin{table}[h!]
	\centerline{
		\begin{threeparttable}
			\caption{Effects of UI eligibility}
			\label{tab:ui_main}
				\begin{tabular}{lcccccc}
					\toprule[1.5pt] 
					& (1) & (2) & (3) & (4) & (5) & (6) \\
					\midrule \\[-2.0ex]
				Outcome:	& Months   & Labor     & \multicolumn{2}{c}{New party membership (\%)} & \multicolumn{2}{c}{Candidacy (\%)} \\ \cmidrule{4-5} \cmidrule{6-7}
					& employed & earnings  & before layoff & after layoff & before layoff & after layoff \\	
					\midrule
					\multicolumn{7}{l}{\textbf{[A] Full sample}} \\ 
					[0.5em]
$\beta$ 	   &     -0.679\sym{***}& -2,192.5\sym{***}&      -0.002         &       0.150\sym{**} &      -0.011         &       0.028         \\
               &     (0.041)         & (149.9)         &     (0.080)         &     (0.072)         &     (0.013)         &     (0.019)         \\
[0.5em]
Control mean   &      26.138         &  56,281.8         &       4.330         &       3.467         &       0.236         &       0.449         \\
Relative effect&        -2.6\%         &        -3.9\%         &        -0.1\%         &         4.3\%         &        -4.6\%         &         6.2\%         \\
Observations   &   2,048,674         &   2,048,674         &   1,047,683         &   1,047,683         &   2,048,674         &   2,048,674         \\
					[0.5em]
					\midrule
					\multicolumn{7}{l}{\textbf{[B] Low educated ($<$ 12 years of education)}} \\ 
					[0.5em]
$\beta$        &     -0.617\sym{***}& -1,822.6\sym{***}&       0.071         &       0.066         &      -0.015         &      -0.006         \\
               &     (0.057)         & (166.1)         &     (0.121)         &     (0.109)         &     (0.019)         &     (0.026)         \\
[0.5em]
Control mean   &      25.029         &  47,526.6         &       4.264         &       3.454         &       0.221         &       0.428         \\
Relative effect&        -2.5\%         &        -3.8\%         &         1.7\%         &         1.9\%         &        -6.9\%         &        -1.4\%         \\
Observations   &     986,755         &     986,755         &     454,665         &     454,665         &     986,755         &     986,755         \\
					[0.5em]
					\midrule
					\multicolumn{7}{l}{\textbf{[C] High educated ($\geq$ 12 years of education)}} \\ 
					[0.5em]
$\beta$        &     -0.708\sym{***}& -2,311.2\sym{***}&      -0.059         &       0.216\sym{**} &      -0.007         &       0.061\sym{**} \\
               &     (0.059)         & (240.6)         &     (0.106)         &     (0.097)         &     (0.019)         &     (0.027)         \\
[0.5em]
Control mean   &      27.143         &  64,213.1         &       4.381         &       3.476         &       0.250         &       0.468         \\
Relative effect&        -2.6\%         &        -3.6\%         &        -1.3\%         &         6.2\%         &        -2.7\%         &        13.0\%         \\
Observations   &   1,061,919         &   1,061,919         &     593,018         &     593,018         &   1,061,919         &   1,061,919         \\
[0.5em]
\hline\hline
				\end{tabular}
			\begin{tablenotes} \scriptsize 
				\item \textit{Note:} 
				The table presents results from model (\ref{model:rdd}) for the effect of unemployment benefit eligibility on the number of months employed and the total formal labor earnings in the four years after the layoff (columns (1) and (2)), on the likelihood of newly registering as a party member in the two election cycles before and after the layoff (columns (3) and (4)), and on the likelihood of running for local councilor in the two election cycles before and after the layoff (columns (5) and (6)). The sample includes all workers displaced within a bandwidth of 60 days around the cutoff date that determines benefit eligibility, which is 16 months since the last layoff. Standard errors clustered at the individual level are in parentheses. The table also reports the control mean of the outcome at the cutoff and the effect sizes scaled by the control mean. All coefficients, standard errors, and control means in columns (3) to (6) have been scaled by 100, such that effects are interpreted in terms of percentage points. \sym{*} \(p<0.10\), \sym{**} \(p<0.05\), \sym{***} \(p<0.01\)
			\end{tablenotes}
		\end{threeparttable}
	}
\end{table}

In the second step, we examine the effects of UI benefits on individuals' entry into politics. In columns (4) and (6), we show the effects on the likelihood of newly registering to a political party and running for local council in the two cycles after the layoff. Columns (3) and (5) show that both outcomes were balanced in the two cycles before the layoff, providing additional support for the RD design. After the layoff, we find that UI eligibility increases the likelihood of becoming a party member by 0.15 percentage points (4.3\% relative to the control mean). For candidacies, we find an increase of 0.03 percentage points (6.2\% relative to the control mean). However, the estimate is not statistically significant due to the limited power that we have for the candidacy outcome.\footnote{As shown in Appendix Table \ref{tab:ui_robust_candidate}, the effect is significant for some alternative bandwidth and polynomial order choices.}

Given that earlier we have documented differential effects of job loss by education, Panels B and C of Table \ref{tab:ui_main} also show the effects of UI benefits separately for low-educated workers ($<$ 12 years of schooling) and high-educated workers ($\geq$ 12 years of schooling). We find positive effects on party memberships and candidacies among the highly educated but no significant impact among the low educated. The effects on workers with high levels of education are statistically significant and meaningful in size, indicating a 6.2\% increase in the likelihood of becoming a party member and a 13.0\% increase in the likelihood of running for office (both relative to the control mean). These findings are consistent with the time-resources mechanism: by reducing reemployment incentives, eligibility for UI benefits frees up time that can be redirected toward political engagement. As suggested by our theoretical framework in Section~\ref{sec:background_theory}, the importance of this channel increases with the expected benefits of political participation. These benefits may vary with education --  for example, if more educated individuals have greater chances of electoral success or are more likely to leverage political connections for gaining access to public-sector jobs. Such mechanisms can explain why both job loss and UI eligibility have stronger effects on political entry among the highly educated.

\section{Implications for Political Selection}\label{sec:implications}
Our main results show that job loss increases individuals' likelihood of becoming a party member or a local political candidate. An important question that follows is how these new entrants affect the political system. The increased number of party members and candidates means that parties have a larger pool from which they can select competent politicians for leadership roles. In addition, layoffs may drive individuals with different motivations and different skill levels into politics, and thereby change the average quality of politicians. As discussed in Section~\ref{sec:background_theory}, layoffs can have positive consequences for political selection if income losses or the expected returns from political entry increase with individuals' ability level. In particular, if more competent individuals experience greater reductions in private-sector opportunity costs or anticipate higher benefits from political engagement during unemployment, layoffs may raise the average competence of the candidate pool.

\subsection{Individual-level evidence} \label{sec:implications_individual}
To explore the impact of layoffs on political selection, we examine the individual characteristics of candidates and political members who enter politics as a result of being laid off. We use a combination of sample moments in an instrumental variable framework to estimate the average characteristics of three groups: (1) compliers -- individuals who enter politics because of the layoff, (2) always-takers -- those who enter politics regardless of whether they are laid off or not, and (3) never-takers -- those who do not enter politics regardless of their layoff status. To identify compliers, we build on the kappa weighting approach introduced by \cite{abadie2003}. We estimate the expected value of covariate $X_i$ for compliers as 
\begin{equation*}
           E[X_i | compliers] = \frac{E[\kappa_i X_i]}{E[\kappa_i]},
\end{equation*}
where $\kappa_i$ is defined as
\begin{equation*}
            \kappa_i = Y_i \frac{Treat_i - P(X_i)}{P(X_i)(1 - P(X_i))}.
\end{equation*}

Here, $Y_i$ indicates whether an individual enters politics in a post-layoff cycle (either as a party member or a councilor candidate), and $Treat_i$ represents their layoff status. We estimate the propensity score, $P(X_i) = P(Treat_i = 1 | X_i)$, using the set of all covariates considered in the complier analysis.\footnote{\cite{sloczynski2025} provide a comprehensive treatment of various kappa weighting estimators that are implemented in the Stata command \texttt{kappalate}. We use the estimator based on their $\kappa_{i1}$ definition. The authors recommend these kappa weights for applications without always-takers (one-sided noncompliance) to avoid near-zero denominators. Similarly, in our case, the number of always-takers is very small relative to the number of never-takers.}

The complier analysis relies on two main assumptions. First, we assume instrument independence. This is supported by the use of the mass-layoff sample described in Section \ref{sec:results}, in which treated and control workers are matched on various pre-layoff characteristics (see, again, Table \ref{tab:sum_stat} for balance checks). Moreover, we account for any remaining differences in covariates in the propensity score estimation. Second, we assume treatment monotonicity. In our context, this means that individuals are not discouraged from entering politics due to a layoff. While we cannot entirely rule out this possibility, the heterogeneity analyses across individual, local, and party characteristics reveal significant positive effects across all considered subgroups, suggesting that the presence of defiers -- if any -- is likely limited (see, again, Figure \ref{fig:did_hetero_individual}, Appendix Figure \ref{fig:did_hetero_municipal}, and Table \ref{tab:hetero_party}).

When analyzing the characteristics of political entrants, we focus on measures that capture individual competence, in addition to socio-demographic characteristics. First, we consider years of education as a standard measure of competence.\footnote{For Brazil, \cite{melo2024} provide evidence that more educated mayors are more effective managers of public resources, as they commit fewer municipal administration infringements. However, they are not found to engage in fewer corrupt practices or improve the provision of local public goods.} Second, we examine individuals' log hourly wage before the layoff, which, following \cite{dalbo2013}, can be treated as a measure of skill, as the private sector typically rewards highly skilled workers with higher wages. Third, we consider person wage effects from an AKM model that regresses log wages on person fixed effects and firm fixed effects \citep{abowd1999}. The person fixed effects measure the wage premia that workers receive independent of the firm that they work at, likely reflecting workers' ability that is rewarded across all firms.\footnote{To our knowledge, we are the first to use AKM effects to study the competence of politicians. The AKM model is fitted on all RAIS contracts in the pre-layoff years 2004 to 2008. Appendix Table \ref{tab:akm_summary} reports a summary of the AKM estimation results.} Moreover, we consider residualized AKM person effects, controlling for gender, race, age deciles, education categories, as well as the full set of interactions between these variables. The resulting residuals capture wage premia relative to observably similar workers.\footnote{\cite{besley2017} and \cite{DalBo2017} also consider residuals from Mincer wage regressions. Our AKM measures additionally purge wages from firm variation in wage-setting, which may not reflect individual ability.} To validate our competence measures, we correlate them with local councilors' misconduct detected during the electoral process and with their electoral performance. Results are reported in Appendix Table \ref{tab:validation} and show that all our competence measures are significantly negatively correlated with the probability of misbehaving in the electoral process and significantly positively related the the probability of being elected.\footnote{The misbehavior variable captures a broad range of irregularities detected by the Brazilian Electoral Court during the elections in 2016 and 2020 that resulted in candidacy annulment. These include abuse of power, failure to meet registration requirements, vote buying, prohibited conduct, ineligibility under the ``Clean Record Act'' (\textit{Lei da Ficha Limpa}), illicit use of campaign resources, objections to the candidacy, rejection by the party or coalition, and invalidation of the party.} Finally, in addition to the competence measures, we also study socio-demographic representation in terms of gender and age. 

\clearpage
\begin{figure}[h!]
	\caption{Characteristics of political entrants}  \label{fig:compliers}
	\centerline{
		\begin{threeparttable}
  			\begin{footnotesize}
                    \centering \textbf{A. New party membership}\\ 
				\includegraphics[width=0.85\linewidth]{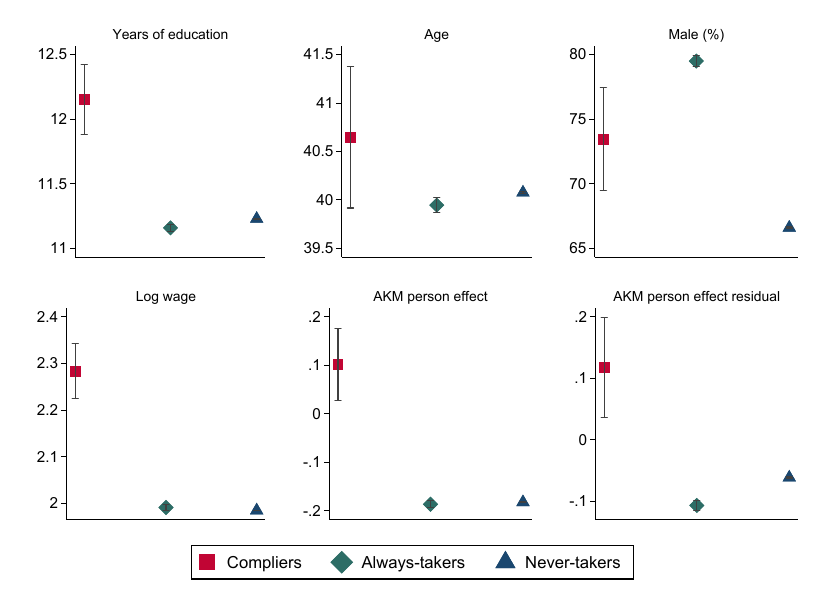}\\ [1em] 
                    \centering \textbf{B. Candidacy}\\ 
				\includegraphics[width=0.85\linewidth]{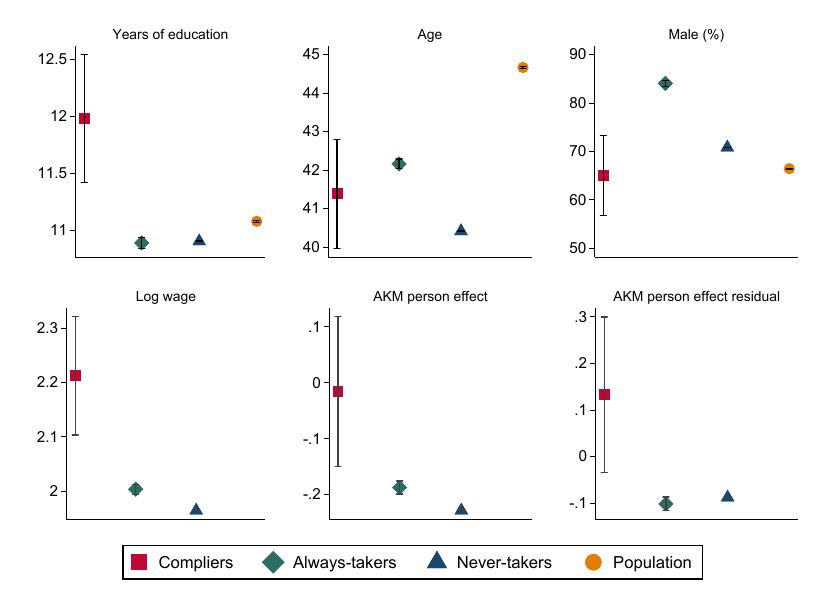} \\          
			\end{footnotesize}
			\begin{tablenotes} \scriptsize
			\item \textit{Note:}
The figure shows the average pre-layoff characteristics of individuals in our estimation sample who are compliers (those induced to enter politics due to the layoff), always-takers (those who enter politics regardless of being laid off or not), and never-takers (those who do not enter politics regardless of being laid off or not). Panel A and B report results separately for our two outcomes: registering as a party member and running for local councilor in a post-layoff cycle (2012, 2016, and 2020). While the estimation sample is restricted to individuals with unique names for the party membership outcome, we do not impose this restriction when analyzing candidacies. For the characteristics provided by the TSE, we also show the average characteristics of the total population of candidates (also in 2012, 2016, and 2020). In Panel A, $N$ = 11,389 compliers, 70,088 always-takers, and 4,868,859 never-takers. In Panel B, $N$ = 3,350 compliers, 24,227 always-takers, 9,855,945 never-takers, and 1,428,485 candidates in the total population. Log wages and AKM person effects are measured based on individuals' contracts in RAIS between 2004 and 2008, and we also show residualized AKM person effects that control for gender, race, age deciles, education categories, as well as the full set of interactions between these variables. AKM person effects and residuals are standardized across all workers in RAIS between 2004 and 2008. 
			\end{tablenotes}
	\end{threeparttable}}
\end{figure}
\clearpage

Figure \ref{fig:compliers} presents the results. Overall, we find that individuals who enter politics because of a layoff (compliers) tend to be positively selected in terms of competence relative to always-takers and never-takers. For example, layoff-induced councilor candidates have an average of 12.0 years of education, compared to about 10.9 years among individuals who always or never run for local office, regardless of whether they are laid off or not.\footnote{Interestingly, we observe relatively small differences between always-takers and never-takers on all measures of competence, suggesting that political selection may be weak, on average.} For layoff-induced party members, we find similarly higher education, compared to always- and never-takers. Moreover, there are clear differences in log wages and AKM person effects (and their residuals) that also indicate higher competence among compliers. For instance, layoff-induced candidates have AKM person effects about 0.2 standard deviations higher than those of always- and never-takers, and for layoff-induced party members the difference is even larger. 

These results suggest that layoffs improve the competence of political entrants relative to always-takers in our estimation sample. To better assess how layoff-induced entrants can change the overall pool of politicians -- beyond our estimation sample -- we can compare our complier estimates to information on all councilor candidates reported by the TSE. The average education of all councilor candidates between 2012 and 2020 is 11.1 years, showing that layoff-induced candidates have about one more year of education than the average candidate.\footnote{Unfortunately, for the full candidate population, we observe only education and other socio-demographic characteristics. Wage data are unavailable in the TSE data, and similar information is not provided for the population of party members.}

Importantly, the higher levels of competence do not come at the expense of reduced representation. Our results indicate that individuals who run for office because of the layoff are more likely to be female and younger than always-takers and the full candidate population.\footnote{For party members, compliers are also more likely to be female, but slightly older than always-takers.} In sum, these findings suggest that job loss increases political entry more strongly among individuals with higher competence, and is also accompanied by improved political representation.

\subsection{Local-level evidence} \label{sec:implications_local}
Finally, we provide evidence on how labor demand shocks affect the selection of politicians at the local level. It is not clear \textit{a priori} whether the individual-level results translate into aggregate changes. For instance, if new entrants after a layoff simply crowd-out politicians with similar levels of education, the share of highly educated individuals would remain unchanged. To gain a better understanding of the aggregate implications, we compile municipality-level data on formal employment rates and the average years of education of all individuals who ran for and were elected to local councils in the 2004 to 2016 elections. We estimate the effects of employment rates on councilor education in the following first-difference model:
\begin{equation}\label{model:local}
\Delta Y_{mc} = \alpha + \beta \Delta E_{mc} + \mu_c + \epsilon_{mc},
\end{equation}
where $m$ and $c$ denote municipalities and election cycles, respectively, and $\mu_c$ are year fixed effects (or, alternatively, state $\times$ year fixed effects). First-differencing implicitly controls for any time-invariant heterogeneity across municipalities, while the year fixed effects capture changes in the outcome that are common to all municipalities.

In order to exploit plausibly exogenous changes in local labor demand, we follow \cite{bartik1991} and instrument $\Delta E_{mc}$ using variation in national industry employment growth rates interacted with baseline cross-sectional differences in municipalities' industry composition. If industry employment shocks at the national level are exogenous, this strategy can be used to identify the effects of demand-driven variation in local employment \citep{borusyak2022}. We construct the Bartik instruments as follows:
\begin{equation}\label{bartik_iv}
\hat{\Delta} E_{mc} = \sum_k s_{m,k,2004} \left(\ \frac{e_{-m, k, c}-e_{-m, k, c-4}}{e_{-m, k, c-4}} \right),
\end{equation}
where $s_{m,k,2004}$ is the initial employment share of industry $k$ in municipality $m$ and $e_{-m, k, c}$ is the national employment of industry $k$ leaving out municipality $m$.\footnote{Industry- and municipality-level employment is constructed by counting all formal contracts in RAIS. We use 581 four-digit CNAE 1.0 industries.}   

\begin{table}[tb]
	\centerline{
		\begin{threeparttable}
			\caption{Labor demand and political selection at the municipality level}
			\label{tab:bartik}
				\begin{tabular}{lcccc}
					\toprule[1.5pt] 
					& (1) & (2) & (3)& (4) \\
					\midrule \\[-2.0ex]
                        Outcome: & \multicolumn{4}{c}{$\Delta$ Average years of education of ...} \\\cmidrule(lr){2-5}
					& \multicolumn{2}{c}{Councilor candidates} & \multicolumn{2}{c}{Elected councilors}\\	
					\midrule
					\multicolumn{5}{l}{\textbf{[A] OLS}} \\ 
					[0.5em]
$\Delta$ Employment rate (ppt)  &      -0.035\sym{***}&      -0.029\sym{***}&      -0.030\sym{***}&      -0.027\sym{***}\\
                                &     (0.003)         &     (0.004)         &     (0.006)         &     (0.007)         \\
[0.5em]
Observations                    &  16,650             &  16,650             &  16,650             & 16,650         \\
					[0.5em]
					\midrule
					\multicolumn{5}{l}{\textbf{[B] 2SLS (Bartik IV)}} \\ 
					[0.5em]
$\Delta$ Employment rate (ppt)	&      -0.160\sym{***}&      -0.150\sym{***}&      -0.140\sym{***}&      -0.122\sym{***}\\
                                &     (0.014)         &     (0.016)         &     (0.022)         &     (0.025)         \\
[0.5em]
KP F-stat                       &       301.7         &       250.3         &       301.7         &     250.3         \\
Observations                    &  16,650             &  16,650             &  16,650             &  16,650        \\
[0.5em]
\midrule
Fixed effects  & Year & State $\times$ year & Year & State $\times$ year \\
[0.5em]
\hline\hline
				\end{tabular}
			\begin{tablenotes} \scriptsize 
				\item \textit{Note:} 
				The table presents OLS and 2SLS results from model (\ref{model:local}) for the effect of formal employment rates on the average education of local councilors. The sample contains all municipal elections in 2004, 2008, 2012, and 2016. The dependent variable measures the change across subsequent elections in the average years of education of all individuals running (columns (1) and (2)) and being elected (columns (3) and (4)) for local council. The explanatory variable is the percentage point change in the municipal formal employment-to-population rate. For the 2SLS results, changes in employment rates are instrumented with Bartik instruments that are constructed by interacting baseline differences in municipalities' industry composition with national industry employment growth rates (see equation (\ref{bartik_iv})). All regressions are weighted by municipal population. Standard errors clustered at the municipality level are in parentheses. \sym{*} \(p<0.10\), \sym{**} \(p<0.05\), \sym{***} \(p<0.01\)
			\end{tablenotes}
		\end{threeparttable}
	}
\end{table}

Table \ref{tab:bartik} presents the results of OLS and 2SLS regressions for model (\ref{model:local}). In all specifications, we see a significantly negative relation between local employment rates and the average education of those who run for local council. The Bartik IV regression implies that a 1 percentage point lower employment rate increases candidates' education by 0.16 years. This result is consistent with our findings at the individual level, which indicate that job loss has larger positive effects on candidacies among better-educated individuals. Importantly, Table \ref{tab:bartik} also shows that local labor demand shocks affect not only the selection of those who run for office but also those who are ultimately elected to office. A 1 percentage point lower employment rate is estimated to increase the education of elected councilors by 0.14 years. 

Appendix Tables \ref{tab:bartik_gender} and \ref{tab:bartik_age} further show that negative local labor demand shocks can also improve political representation. Consistent with our individual-level analysis of compliers, we find that lower employment rates lead to an increase in the share of female and younger individuals who run for and are elected to councilor office. Together, these results highlight that negative economic shocks can enhance the competence of local politicians without compromising their socio-demographic representation.

\section{Conclusion}
\label{sec:conclusion}
The quality of democratic institutions largely depends on individuals' willingness to engage in politics. While political careers plausibly compete with careers in the private sector, we lack causal evidence on how changes in private labor market opportunities affect the trade-offs faced by individuals when deciding to enter politics. Combining rich information on the universe of party members and local candidates with matched employer-employee data covering all formal workers in Brazil, we study the effects of mass layoffs on individuals' decisions to start a political career and the consequences of economic shocks for the selection of politicians.

Our results show that layoffs significantly increase individuals' likelihood of joining a political party or running for office in local elections. These positive effects persist over time, are prevalent across a wide range of individual, regional, and party characteristics, and extend to alternative forms of political participation. Several of our results suggest that the increase in political entry can be rationalized by an opportunity-cost mechanism, whereby job loss reduces individuals' outside options in the private sector and increases their available time resources. In particular, we observe larger increases in candidacies among individuals with higher financial incentives -- measured either by the predicted income loss from the layoff or the wage received in public office -- and among those with more time resources which result from UI benefit incentives.

In terms of the implications of layoffs for the selection of political entrants, we document that treated individuals who become party members or candidates because of the layoff are positively selected across a number of competence measures. This positive selection is also evident at the aggregate level, where we find a positive effect of municipal labor demand shocks on the education of individuals running for and being elected to local councils. Taken together, our paper provides novel evidence that layoffs increase the likelihood that highly capable individuals enter politics. This suggests that political selection can be counter-cyclical: negative economic shocks may lead to an increase in politician quality.

While our results emphasize how economic trade-offs can influence who joins politics, they also raise important questions about how private-sector opportunities shape the behavior of politicians once they attain power. In particular, do politicians' economic circumstances influence their performance in office, including legislative productivity, propensity for personal rent extraction, and ability to effectively provide public goods? Furthermore, how does experiencing a job loss impact the policy priorities set by politicians and the representation of voters' preferences in policy-making? These remain unresolved questions that are crucial for understanding the broader consequences of economic shocks for the supply side of the political system.
 
\clearpage
\newpage\clearpage
\renewcommand{\baselinestretch}{1.05}
\setlength{\baselineskip}{12pt}
\addcontentsline{toc}{section}{References}
\bibliography{extended_abstract.bib}

\clearpage 
\renewcommand{\baselinestretch}{1.5}
\setlength{\baselineskip}{20pt}


\clearpage
\newpage

\appendix
\begin{center}
	\section*{{\LARGE Online Appendix}}
\end{center}

\renewcommand{\thesection}{A}
\renewcommand{\thetable}{A.\arabic{table}}
\setcounter{table}{0}
\renewcommand{\thefigure}{A.\arabic{figure}}
\setcounter{figure}{0}

\section{Appendix to Sections \ref{sec:background} to \ref{sec:data}}

\begin{figure}[h!]
	\caption{Employment effects of job loss by party membership before the layoff}  \label{fig:es_employment_memberpre}
	\centerline{
		\begin{threeparttable}
			\begin{footnotesize}
				\begin{tabular}{cc}
 					\textbf{A. Public-sector employment} & \textbf{B. Private-sector employment}\\
					\includegraphics[width=0.5\linewidth]{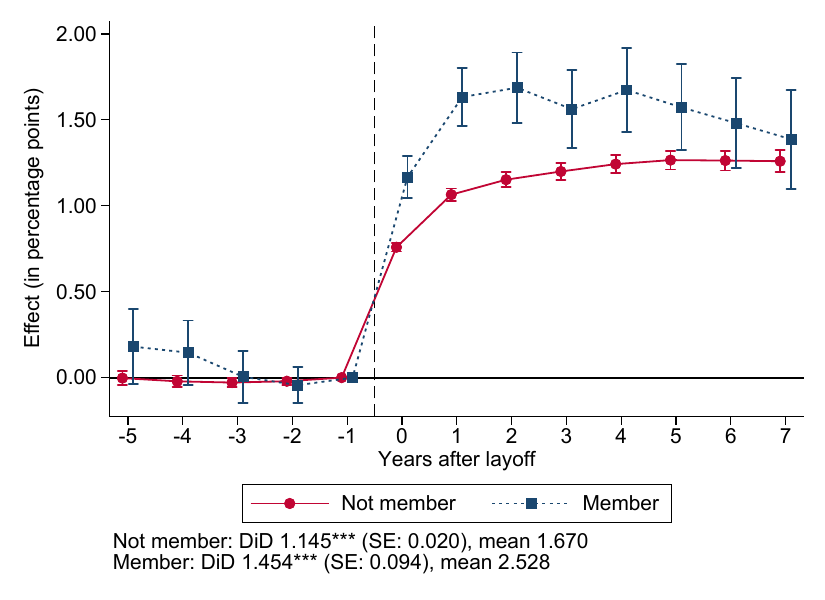} &\includegraphics[width=0.5\linewidth]{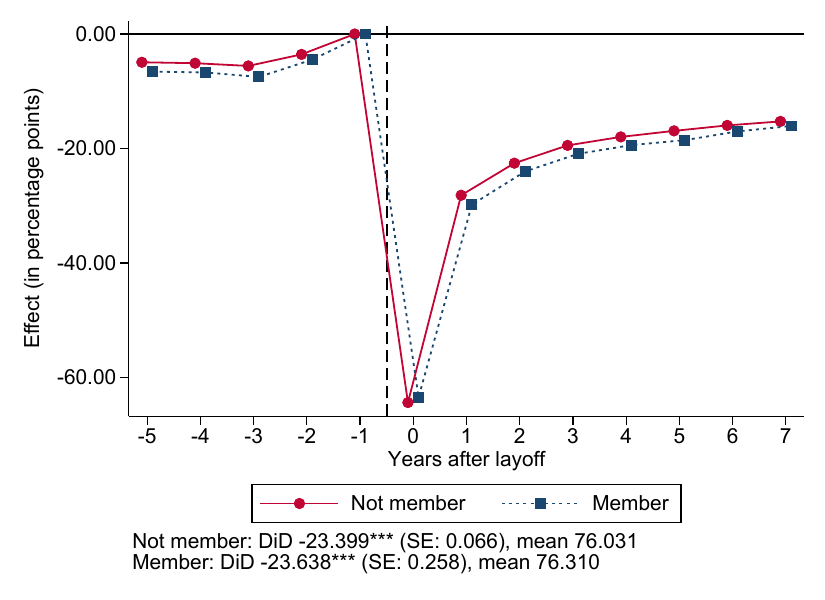} \\ 
				\end{tabular}
			\end{footnotesize}
			\begin{tablenotes} \scriptsize
			\item \textit{Note:}
                The figure reports event-study coefficients $\delta_k$, estimated in model (\ref{model:es_yearly}), for the effect of job loss on the likelihood to be formally employed at the end of the year in the public sector (Panel A) and in the private sector (Panel B), separately for individuals who were or were not member of political party before the layoff. The vertical bars depict 95\% confidence intervals based on standard errors clustered at the individual level. Below each graph, the DiD coefficient from a static version of model (\ref{model:es_yearly}), its standard error, and the mean of the control group across all post-treatment periods are reported.
                \end{tablenotes}
	\end{threeparttable}}
\end{figure}

\vspace{2em}

\begin{figure}[h!]
	\caption{Distribution of new party memberships over time}  \label{fig:hist_affil}
	\centerline{
		\begin{threeparttable}
                \centering
			\includegraphics[width=0.65\linewidth]{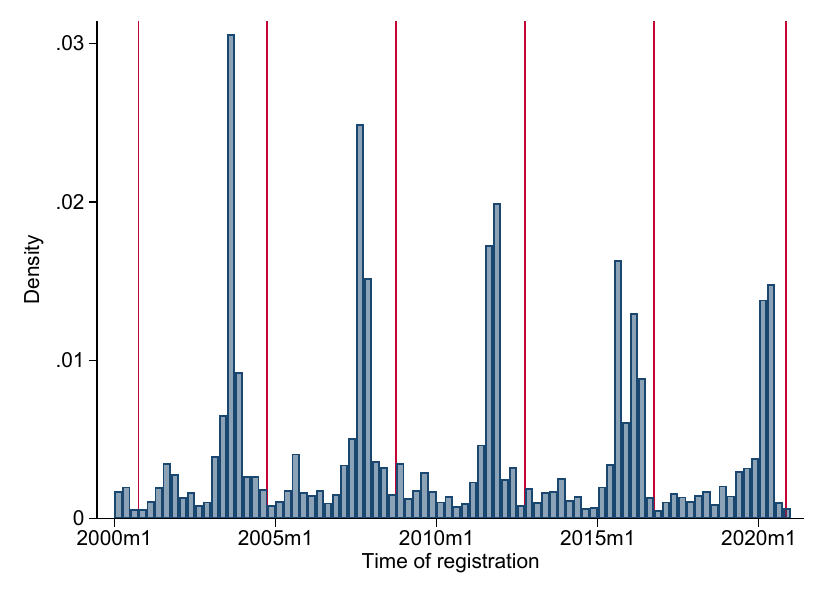}
			\begin{tablenotes} \scriptsize
			\item \textit{Note:}
			The figure depicts the monthly density of new party memberships over time. We include all affiliations in Brazil aggregated at the monthly level between January 2000 and December 2020. Red lines indicate local election months.
			\end{tablenotes}
	\end{threeparttable}}
\end{figure}

\clearpage

\begin{table}[h!]
	\centerline{
		\begin{threeparttable}
			\caption{Campaign donations of councilor candidates}
			\label{tab:sumstat_donations}
\begin{tabular}{lrrrrrr}
\toprule[1.5pt]
                    &        2004&        2008&        2012&        2016&        2020&         All\\
\midrule
\multicolumn{7}{l}{\textbf{Total donations}} \\
Mean                &    3,715.55&    4,861.25&    5,649.81&    3,427.54&    3,467.26&    4,200.86\\
P10                 &      202.69&      211.58&      212.38&      230.70&      214.83&      213.61\\
P25                 &      520.92&      617.11&      613.78&      534.02&      519.33&      552.17\\
P50                 &    1,403.30&    1,833.70&    2,123.80&    1,468.55&    1,327.48&    1,586.85\\
P75                 &    3,506.12&    4,752.17&    5,932.49&    3,741.29&    3,276.06&    4,169.63\\
P90                 &    7,994.55&   10,655.85&   13,199.48&    7,435.77&    7,450.96&    9,300.73\\
[1em]
\multicolumn{7}{l}{\textbf{Donations from own resources}} \\
Share$>$0             &        0.61&        0.68&        0.55&        0.56&        0.46&        0.56\\
Mean                &    1,632.70&    1,718.96&    1,704.98&    1,225.80&      624.07&    1,313.14\\
P10                 &        0.00&        0.00&        0.00&        0.00&        0.00&        0.00\\
P25                 &        0.00&        0.00&        0.00&        0.00&        0.00&        0.00\\
P50                 &      318.93&      439.03&      196.81&      213.61&        0.00&      168.13\\
P75                 &    1,811.53&    1,819.59&    1,792.28&    1,281.64&      681.86&    1,275.73\\
P90                 &    4,297.07&    4,557.80&    4,955.54&    3,226.50&    1,494.49&    3,539.67\\
[1em]
Observations                   &  202,534&  284,031&  359,753&  395,179&  415,691&1,657,188\\
\hline\hline
\end{tabular}
			\begin{tablenotes} \scriptsize 
				\item \textit{Note:} 
                    The table shows summary statistics for campaign donations received by candidates for local council in the elections 2000 to 2020. We report total donation values and the value of donations that come from candidates' own resources. All values are in BRL (CPI 2018) and are winsorized at the 1\% level. The data is obtained from the \textit{Tribunal Superior Eleitoral}. 
			\end{tablenotes}
		\end{threeparttable}
	}
\end{table}

\clearpage

\begin{table}[h!]\centering \caption{Summary statistics: workers with and without unique names} \label{tab:sum_stat_nameuniqueness}
\begin{threeparttable}
\begin{tabular}{l*{3}{c}} 
\toprule[1.5pt]
					& (1) & (2) & (3) \\
                    &     Unique&     Non-unique&    Std Diff \\
\midrule
\multicolumn{4}{l}{\textbf{Socio-demographic characteristics}} \\
Years of education  &       10.72&        9.97&       0.25\\
Age                 &       33.84&       34.52&        -0.10\\
Male (\%)           &       65.78&       73.86&        -0.18\\
[1em]
\multicolumn{4}{l}{\textbf{Job and firm characteristics}} \\ 
Earnings (per month, BRL, CPI 2018)&     2158.15&     1977.51&       0.09\\
Tenure (months)     &       32.62&       32.95&        -0.01\\
Manager (\%)        &        2.70&        1.77&       0.06\\
Firm size           &      610.80&      619.27&        -0.01\\
[1em]
\multicolumn{4}{l}{\textbf{Political outcomes before layoff (2005-2008, \%)}} \\ 
Candidate            &        0.11&        0.12&        -0.00\\
\quad Left party               &        0.04&        0.04&        -0.00\\
\quad Center party             &        0.03&        0.03&       0.00\\
\quad Right party              &        0.04&        0.05&        -0.00\\
\quad Incumbent party     &        0.01&        0.01&       0.00\\
\quad Not incumbent party &        0.10&        0.11&        -0.00\\
\quad Union-affiliated party&        0.03&        0.03&        -0.00\\
\quad Not union-affiliated party&        0.04&        0.04&       0.00\\
[1em]
Observations		& 944,214 & 937,353 &  \\
\hline\hline
\end{tabular}
\begin{tablenotes} \scriptsize 
    \item \textit{Note:} 
    The table reports the average characteristics of displaced workers in our estimation sample who have and who do not have a unique name within Brazil, and the standardized difference between the two groups.
\end{tablenotes}
\end{threeparttable}
\end{table}

\clearpage

\renewcommand{\thesection}{B}
\renewcommand{\thetable}{B.\arabic{table}}
\setcounter{table}{0}
\renewcommand{\thefigure}{B.\arabic{figure}}
\setcounter{figure}{0}

\section{Appendix to Section \ref{sec:results}}

\begin{table}[h!]\centering \caption{Summary statistics: party characteristics } \label{tab:sum_stat_party}
\begin{threeparttable}
\begin{tabular}{l*{3}{c}} 
\toprule[1.5pt]
					& (1) & (2) & (3)  \\
                    &     Treated&     Control&    Std Diff\\
\midrule                    
\textbf{Party membership (2008, \%)}  &        7.47&        7.30&       0.01\\ 
\quad Left party     &        2.66&        2.52&       0.01\\
\quad Center party        &        2.40&        2.40&       0.00\\
\quad Right party    &        2.41&        2.38&       0.00\\
\quad Incumbent party     &        0.78&        0.82&        -0.00\\
\quad Not incumbent party &        6.69&        6.49&       0.01\\
\quad Union-affiliated party&        2.81&        2.74&       0.00\\
\quad Not union-affiliated party&        2.25&        2.18&       0.00\\
[0.5em]
\textbf{New party membership (2005-2008, \%)} &        1.99&        1.89 & 0.01\\
\quad Left party          &        0.73&        0.67&       0.01\\
\quad Center party        &        0.57&        0.58&        -0.00\\
\quad Right party         &        0.68&        0.65&       0.00\\
\quad Incumbent party     &        0.23&        0.24&        -0.00\\
\quad Not incumbent party &        1.76&        1.65&       0.01\\
\quad Union-affiliated party&        0.71&        0.67&       0.00\\
\quad Not union-affiliated party&        0.59&        0.57&       0.00\\
[0.5em]
\textbf{Candidate (2005-2008, \%)} &        0.11&        0.10&       0.00\\
\quad Left party          &        0.04&        0.04&       0.00\\
\quad Center party        &        0.03&        0.03&       0.00\\
\quad Right party         &        0.04&        0.03&       0.01\\
\quad Incumbent party     &        0.01&        0.01&       0.00\\
\quad Not incumbent party &        0.10&        0.09&       0.01\\
\quad Union-affiliated party&        0.03&        0.03&        -0.00\\
\quad Not union-affiliated party&        0.04&        0.03&       0.00\\
[1em]
Observations		& 944,214 & 944,214 & \\
\hline\hline
\end{tabular}
\begin{tablenotes} \scriptsize 
    \item \textit{Note:} 
    The table reports the average pre-layoff political outcomes of treated workers who are displaced from a mass layoff firm and matched control workers who are not displaced in the same calendar year, and the standardized difference between the two groups. The sample only includes workers with a unique name within the country.
    The ideological classification into `left', `center' and `right' parties follows \cite{Colonnelli2022} and is shown in Appendix Table \ref{tab:class_parties_ideo}. The `incumbent' party is the party of the elected mayor in the individual's municipality in 2008, and `not incumbent' refers to all other parties. The `union-affiliated' parties are PT, PDT, PSB, PCB, PSD, and MDB, and `not union-affiliated' refer to all other left or center parties.
\end{tablenotes}
\end{threeparttable}
\end{table}

\clearpage

\begin{figure}[h!]
	\caption{Effects of job loss on future separations}  \label{fig:es_separations}
	\centerline{
		\begin{threeparttable}
			\begin{footnotesize}
				\begin{tabular}{cc}
					\textbf{A. Mass layoff} & \textbf{B. Any job separation}\\
					\includegraphics[width=0.5\linewidth]{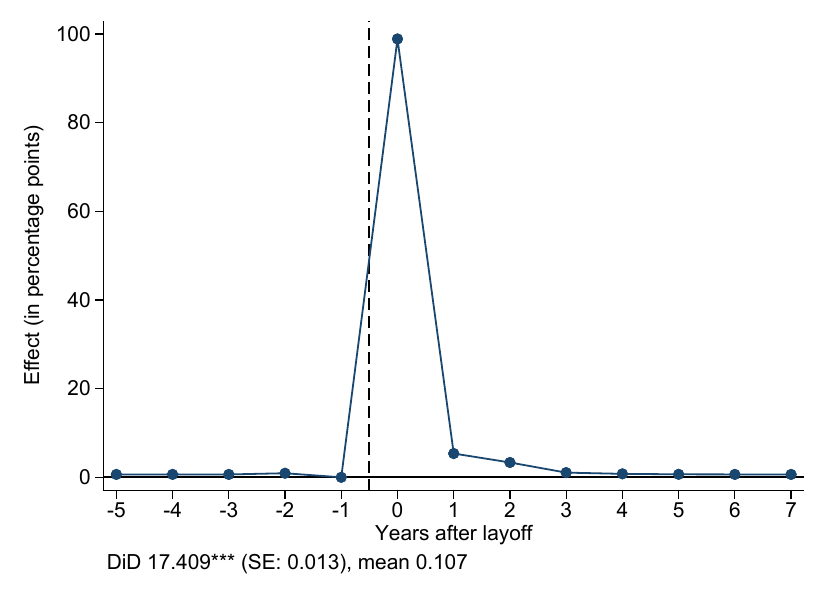} &\includegraphics[width=0.5\linewidth]{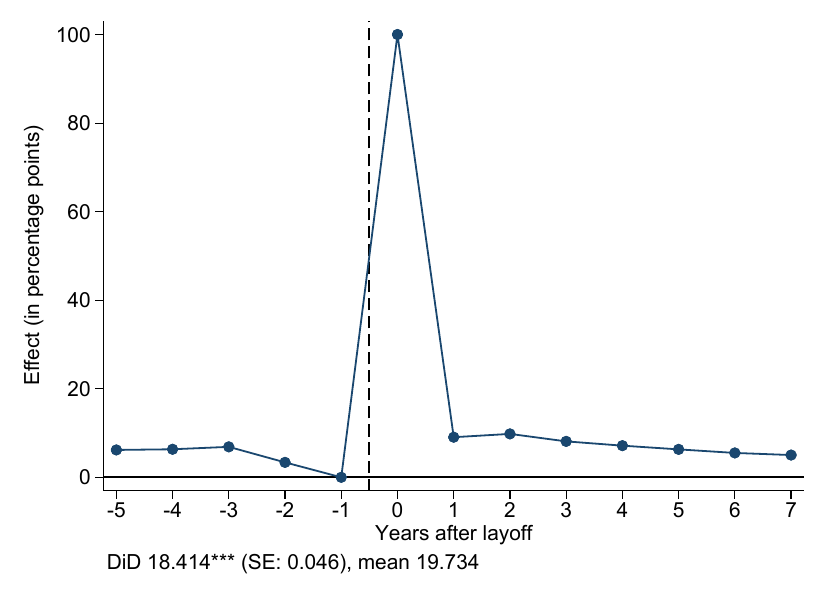} \\ 
				\end{tabular}
			\end{footnotesize}
			\begin{tablenotes} \scriptsize
			\item \textit{Note:}
                The figure reports event-study coefficients $\delta_k$, estimated in model (\ref{model:es_yearly}), for the effect of job loss on the likelihood of being displaced in a mass layoff (Panel A) and of having any job separation (Panel B). $N = 24,549,564$ individual-year observations. The graphs include vertical bars that depict 95\% confidence intervals based on standard errors clustered at the individual level (too small to be visible). Below each graph, the DiD coefficient from a static version of model (\ref{model:es_yearly}), its standard error, and the mean of the control group in the post-treatment period are reported.
                \end{tablenotes}
	\end{threeparttable}}
\end{figure}

\clearpage

\begin{table}[h!]
	\centerline{
		\begin{threeparttable}
			\caption{Effect of job loss: alternative mass layoff definitions}
			\label{tab:robust_mldefinition}
			\begin{footnotesize}
				\begin{tabular}{lcccccccc}
					\toprule[1.5pt] 
					& (1) & (2) & (3)& (4)& (5)& (6)& (7) & (8) \\
					\midrule \\[-2.0ex]
					Layoff share: 	& $\geq$ 30\% & $\geq$ 50\% & $\geq$ 75\% & closure & $\geq$ 30\% & $\geq$ 30\% & $\geq$ 30\% & Excl.  \\
					[0.5em]
					Firm size: 		& $\geq$ 30   & $\geq$ 30   & $\geq$ 30   & $\geq$ 30& $\geq$ 50   & $\geq$ 100  & $\geq$ 250 & volatile \\	
					\midrule
					\multicolumn{9}{l}{\textbf{[A] Outcome: new party membership (\%)}} \\ 
					[0.5em]
$\delta_{DiD}$ 		&       0.175\sym{***}&       0.143\sym{***}&       0.135\sym{***}&       0.201\sym{***}&       0.171\sym{***}&       0.141\sym{***}&       0.147\sym{***}&       0.129\sym{***}\\
               &     (0.018)         &     (0.029)         &     (0.046)         &     (0.072)         &     (0.022)         &     (0.027)         &     (0.039)         &     (0.026)               \\
[0.5em]
Control mean   &       1.407         &       1.444         &       1.454         &       1.452         &       1.413         &       1.417         &       1.423         &       1.382              \\
Relative effect&        12.4\%       &         9.9\%       &         9.3\%       &        13.9\%       &        12.1\%       &        10.0\%       &        10.3\%       &         9.3\%           \\
Observations   &  11,317,632         &   5,681,304         &   2,850,012         &     1,506,516         &   9,085,176         &   6,652,320         &   4,298,994         &   5,446,026               \\
					[0.5em]
					\midrule
					\multicolumn{9}{l}{\textbf{[B] Outcome: candidacy (\%)}} \\ 
					[0.5em]
$\delta_{DiD}$   &       0.041\sym{***}&       0.050\sym{***}&       0.055\sym{***}&       0.053\sym{**}  &       0.039\sym{***}&       0.041\sym{***}&       0.044\sym{***}&       0.034\sym{***}\\
                 &     (0.006)         &     (0.009)         &     (0.015)         &     (0.023)         &     (0.007)         &     (0.009)         &     (0.013)         &     (0.008)                \\
[0.5em]
Control mean   &       0.229         &       0.237         &       0.233         &       0.233         &       0.227         &       0.225         &       0.226         &       0.223                \\
Relative effect&        18.1\%       &        21.2\%       &        23.5\%       &        22.8\%       &        17.4\%       &        18.2\%       &       19.3\%        &        15.5\%               \\
Observations   &  11,317,632         &   5,681,304         &   2,850,012         &     1,506,516         &   9,085,176         &   6,652,320         &   4,298,994         &   5,446,026              \\
[0.5em]
\hline\hline
				\end{tabular}
			\end{footnotesize}
			\begin{tablenotes} \scriptsize 
				\item \textit{Note:} 
				The table presents robustness checks for the effect of job loss on the likelihood of newly registering as a party member (Panel A) and running for local councilor (Panel B), as estimated from model (\ref{model:did}) using different definitions of mass layoffs. Columns (1) to (3) progressively increase the minimum layoff share, and column (4) restricts the treated group to individuals laid off in a plant closure (defined as a minimum layoff share of 95\%). Columns (5) to (7) increase the minimum size of firms used to define mass layoffs. Column (8) excludes establishments with an employment growth of more than 30\% in at least one of the two last years before the mass layoff (using the minimum layoff share and firm size as in our baseline definition in column (1)). Standard errors clustered at the individual level are in parentheses. All coefficients, standard errors, and control means have been scaled by 100, such that effects are interpreted in terms of percentage points. \sym{*} \(p<0.10\), \sym{**} \(p<0.05\), \sym{***} \(p<0.01\)
			\end{tablenotes}
		\end{threeparttable}
	}
\end{table}

\clearpage

\begin{table}[h!]
	\centerline{
		\begin{threeparttable}
			\caption{Effect of job loss: alternative specifications}
			\label{tab:robust_spec}
				\begin{tabular}{lcccc}
					\toprule[1.5pt] 
					& (1) & (2) & (3)& (4) \\
					\midrule \\[-2.0ex]
					Fixed effects: 	& Cycle & Municipality    & 2dgt industry   & Municipality \\
					    		&       & $\times$ cycle  & $\times$ cycle  & $\times$ 2dgt industry \\	
					    		&       &                 &                 & $\times$ cycle \\	
					\midrule
					\multicolumn{5}{l}{\textbf{[A] Outcome: new party membership (\%)}} \\ 
					[0.5em]
$\delta_{DiD}$ 	&       0.151\sym{***}&       0.150\sym{***}&       0.178\sym{***}&       0.175\sym{***}\\
               &     (0.015)         &     (0.016)         &     (0.016)         &     (0.018)         \\
[0.5em]
Control mean   &       1.409         &       1.408         &       1.409         &       1.407         \\
Relative effect&        10.7\%       &        10.7\%       &        12.6\%       &        12.4\%       \\
Observations   &  11,330,568         &  11,329,716         &  11,330,568         &  11,317,632         \\
					[0.5em]
					\midrule
					\multicolumn{5}{l}{\textbf{[B] Outcome: candidacy (\%)}} \\ 
					[0.5em]
$\delta_{DiD}$ &       0.033\sym{***}&       0.041\sym{***}&       0.041\sym{***}&       0.041\sym{***}\\
               &     (0.005)         &     (0.005)         &     (0.006)         &     (0.006)         \\
[0.5em]
Control mean   &       0.229         &       0.229         &       0.229         &       0.229         \\
Relative effect&        14.5\%       &        18.1\%       &        18.1\%       &        18.1\%         \\
Observations   &  11,330,568         &  11,329,716         &  11,330,568         &  11,317,632         \\
[0.5em]
\hline\hline
				\end{tabular}
			\begin{tablenotes} \scriptsize 
				\item \textit{Note:} 
				The table presents robustness checks for the effect of job loss on the likelihood of newly registering as a party member (Panel A) and running for local councilor (Panel B), as estimated from model (\ref{model:did}) using different definitions of mass layoffs. Columns (1) to (3) progressively increase the minimum layoff share, and column (4) restricts the treated group to individuals laid off in a plant closure. Columns (5) to (7) increase the minimum size of firms used to define mass layoffs. Column (8) excludes establishments with an employment growth of more than 30\% in at least one of the two last years before the mass layoff (using the minimum layoff share and firm size as in our baseline definition in column (1)). Standard errors clustered at the individual level are in parentheses. All coefficients, standard errors, and control means have been scaled by 100, such that effects are interpreted in terms of percentage points. \sym{*} \(p<0.10\), \sym{**} \(p<0.05\), \sym{***} \(p<0.01\)
			\end{tablenotes}
		\end{threeparttable}
	}
\end{table}

\clearpage

\begin{figure}[h!]
	\caption{Effect of job loss on running for local councilor: unique vs. non-unique names}  \label{fig:es_cand_unique}
	\centerline{
		\begin{threeparttable}
                \centering
			\includegraphics[width=0.65\linewidth]{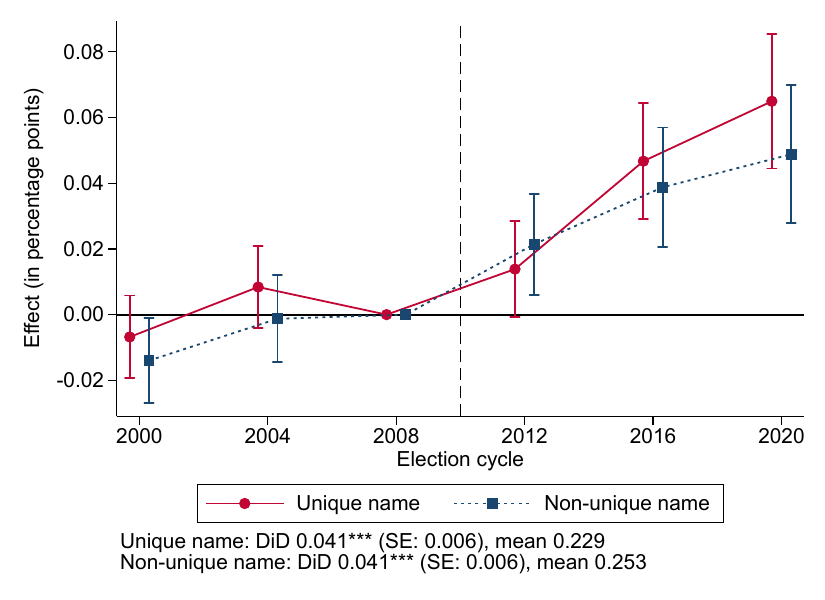}
			\begin{tablenotes} \scriptsize
			\item \textit{Note:}
			The figure reports event-study coefficients $\delta_c$, estimated in model (\ref{model:es}), for the effect of job loss on the likelihood of running for local councilor, separately for individuals with unique and non-unique names in Brazil. The vertical bars depict 95\% confidence intervals based on standard errors clustered at the individual level. Below the graph, the DiD coefficient from model (\ref{model:did}), its standard error, and the mean of the control group in the post-treatment period (average across 2012, 2016, and 2020) are reported.
			\end{tablenotes}
	\end{threeparttable}}
\end{figure}

\vspace{2em}

\begin{figure}[h!]
	\caption{Heterogeneity by layoff year}  \label{fig:es_layoffyear}
	\centerline{
		\begin{threeparttable}
			\begin{footnotesize}
				\begin{tabular}{cc}
					\textbf{A. New party membership} & \textbf{B. Candidacy}\\
					\includegraphics[width=0.55\linewidth]{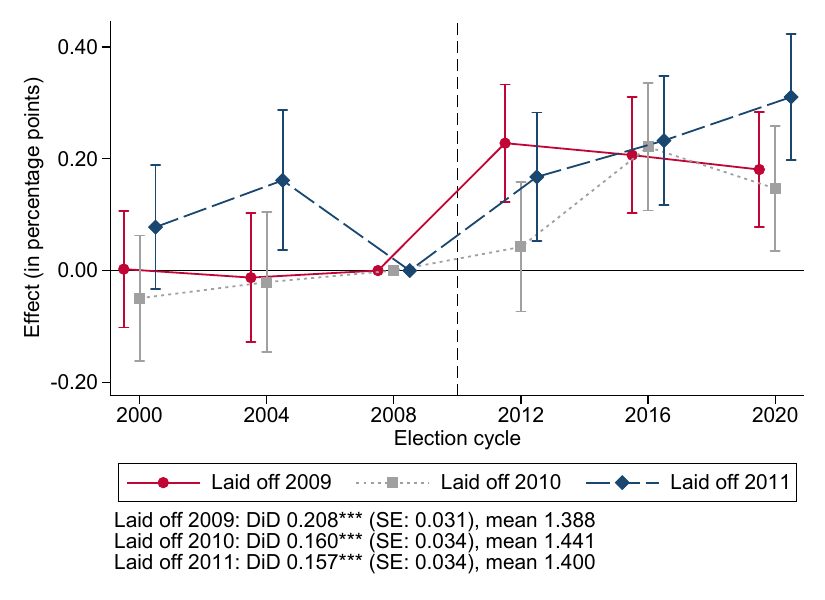} &\includegraphics[width=0.55\linewidth]{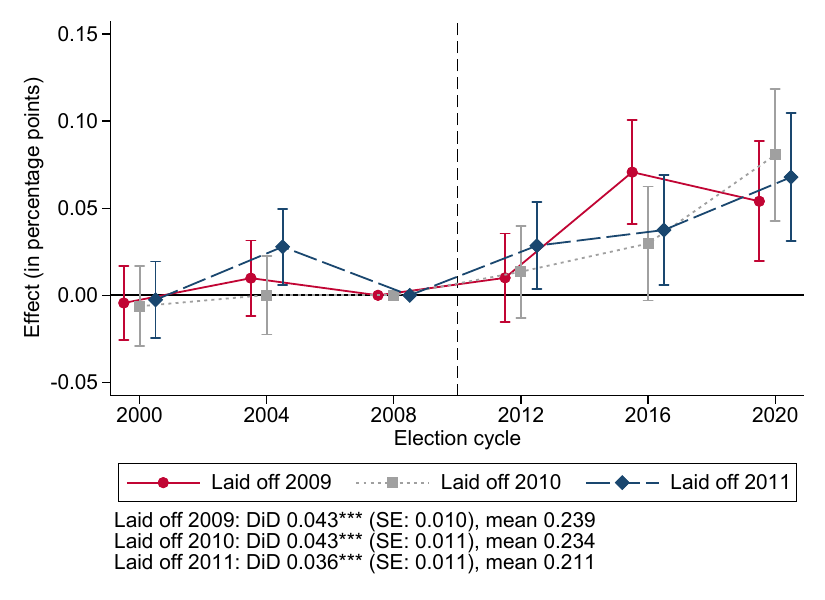} \\ 
				\end{tabular}
			\end{footnotesize}
			\begin{tablenotes} \scriptsize
			\item \textit{Note:}
				The figure reports event-study coefficients $\delta_c$, estimated in model (\ref{model:es}), for the effect of job loss on the likelihood of newly registering as a party member (Panel A) and running for local councilor (Panel B), separately for individuals laid off in 2009, 2010, and 2011 (and their matched control individuals). The vertical bars depict 95\% confidence intervals based on standard errors clustered at the individual level. Below each graph, the DiD coefficient from model (\ref{model:did}), its standard error, and the mean of the control group in the post-treatment period (average across 2012, 2016, and 2020) are reported.
			\end{tablenotes}
	\end{threeparttable}}
\end{figure}

\clearpage

\begin{figure}[h!]
	\caption{Heterogeneity by municipality characteristics}  \label{fig:did_hetero_municipal}
	\centerline{
		\begin{threeparttable}
			\begin{footnotesize}
				\begin{tabular}{cc}
					\textbf{A. New party membership} & \textbf{B. Candidacy}\\
					\includegraphics[width=0.55\linewidth]{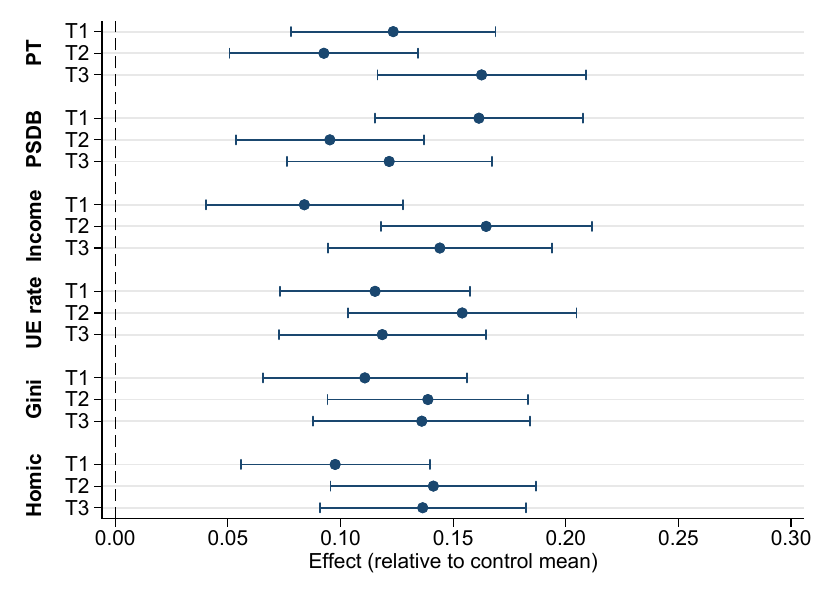} &\includegraphics[width=0.55\linewidth]{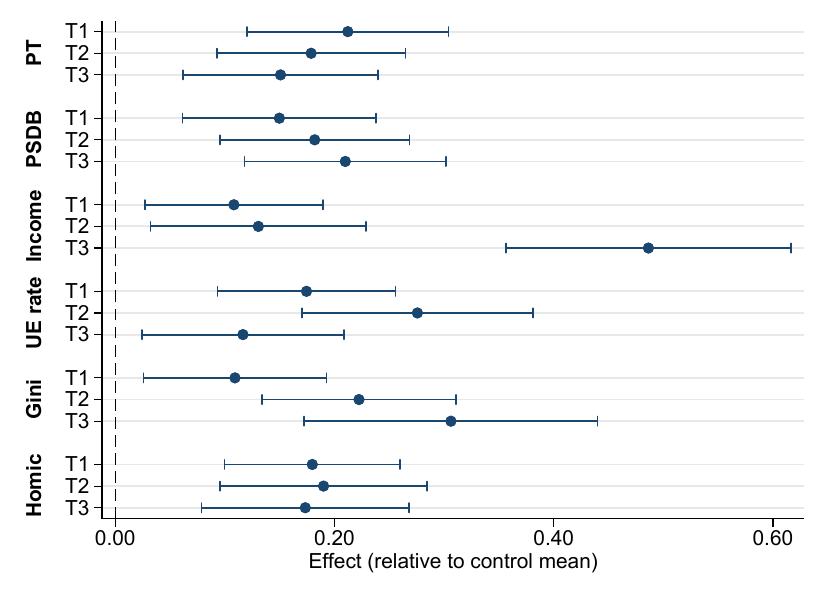} \\ 
				\end{tabular}
			\end{footnotesize}
			\begin{tablenotes} \scriptsize
			\item \textit{Note:}
                The figure reports DiD coefficients, estimated in model (\ref{model:did}), for the effect of job loss on the likelihood of newly registering as a party member (Panel A) and running for local councilor (Panel B) in municipalities with different characteristics. We split the sample by terciles of each variable. `PT' and `PSDB' represent the share of votes for Luiz In\'acio Lula da Silva and Geraldo Alckmin in each municipality in the runoff of the 2006 Brazilian presidential election. `Income' denotes the average household income in each municipality, `UE rate' is the municipal unemployment rate, and `Gini' refers to the municipality income Gini index. All these variables are obtained from the 2010 Brazilian census. `Homic' refers to the homicide rate in each municipality per 100,000 inhabitants as of 2010, sourced from DATASUS. The horizontal bars depict 95\% confidence intervals based on standard errors clustered at the individual level. All coefficients and standard errors are scaled by the group-specific mean of the control group in the post-treatment period (average across 2012, 2016, and 2020). 			\end{tablenotes}
	\end{threeparttable}}
\end{figure}

\clearpage

\begin{figure}[h!]
	\caption{Labor market effects by education level}  \label{fig:es_laboroutcomes_education}
	\centerline{
		\begin{threeparttable}
			\begin{footnotesize}
				\begin{tabular}{cc}
					\textbf{A. Labor earnings} & \textbf{B. Any employment}\\
					\includegraphics[width=0.5\linewidth]{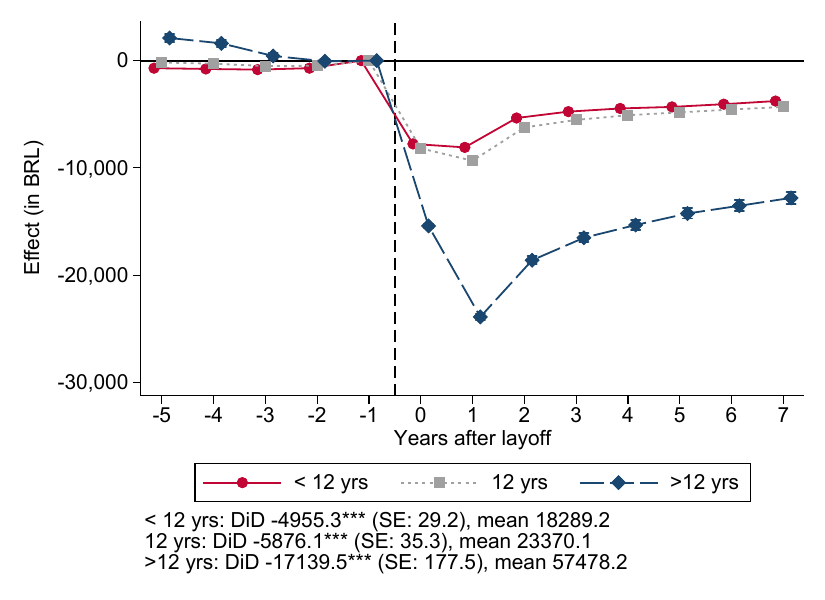} &\includegraphics[width=0.5\linewidth]{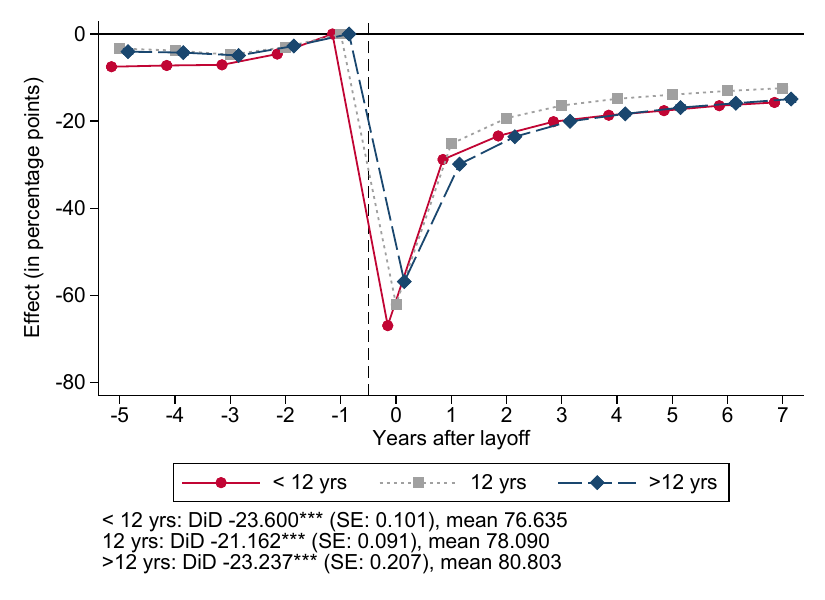} \\ [1em] 
 					\textbf{C. Private-sector employment} & \textbf{D. Public-sector employment}\\
					\includegraphics[width=0.5\linewidth]{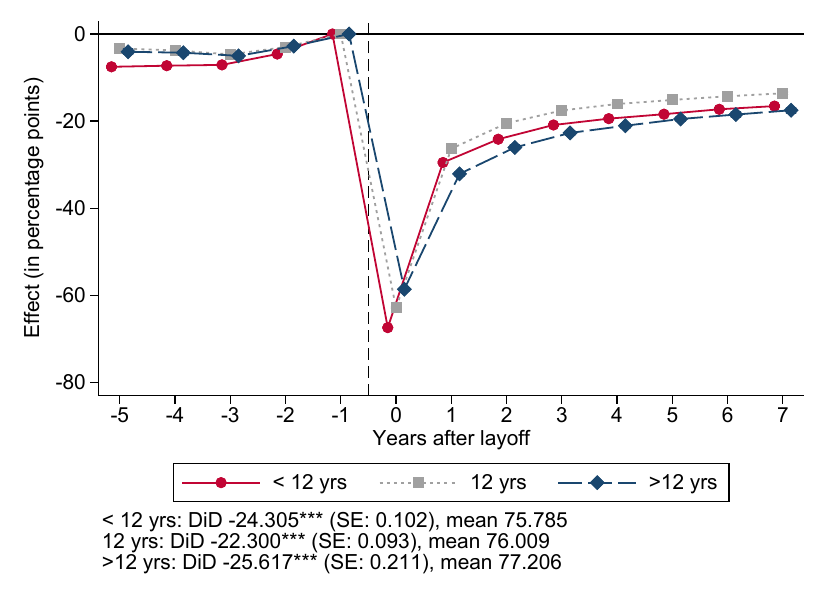} &\includegraphics[width=0.5\linewidth]{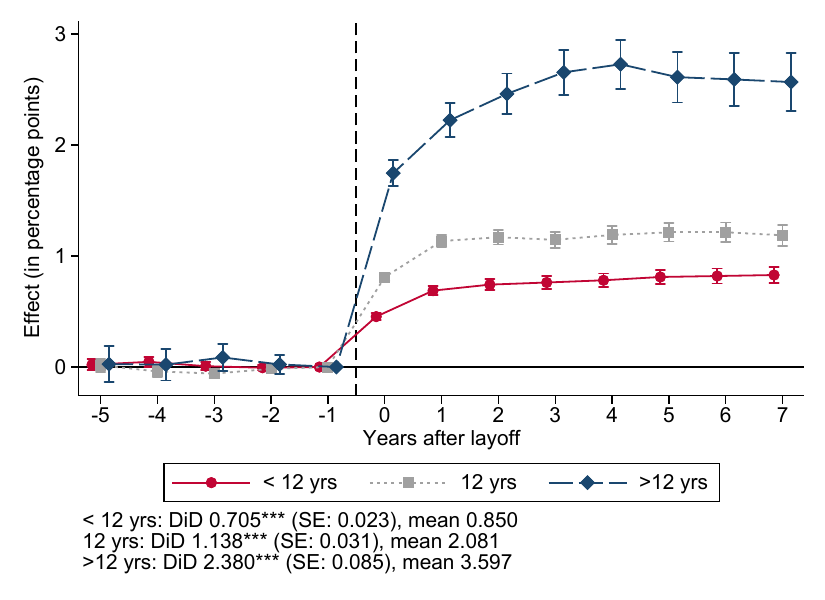} \\ 
				\end{tabular}
			\end{footnotesize}
			\begin{tablenotes} \scriptsize
			\item \textit{Note:}
                The figure reports event-study coefficients $\delta_k$, estimated in model (\ref{model:es_yearly}), for the effect of job loss on yearly formal labor earnings (Panel A), the likelihood of being employed in any formal job at the end of the year (Panel B), and the likelihoods of being employed in the private sector (Panel C) and public sector (Panel D), separately for individuals with no degree, elementary school, or middle school ($<$ 12 years), high school (12 years), and university education ($>$ 12 years). The vertical bars depict 95\% confidence intervals based on standard errors clustered at the individual level. Below each graph, the DiD coefficient from a static version of model (\ref{model:es_yearly}), its standard error, and the mean of the control group across all post-treatment periods are reported.
                \end{tablenotes}
	\end{threeparttable}}
\end{figure}

\clearpage

\begin{table}[h!]
	\centerline{
		\begin{threeparttable}
			\caption{Councilor wage caps}
			\label{tab:counc_wage_caps}
\begin{tabular}{lcc}
\toprule[1.5pt]
Population      & Maximum wage       & Maximum wage \\
                & in \% of state     & in BRL, 2004 \\
                & deputy salary      & \\
\midrule
1-10,000        & 20\%  & 1927.1   \\
10,001-50,000   & 30\%  & 2890.6   \\
50,001-100,000  & 40\%  & 3854.2   \\
100,001-300,000 & 50\%  & 4817.7   \\
300,001-500,000 & 60\%  & 5781.2   \\
above 500,000   & 75\%  & 7226.6  \\ 
\hline\hline
\end{tabular}
			\begin{tablenotes} \scriptsize 
				\item \textit{Note:} 
				The table shows the maximum monthly wage that councilors can earn by municipality population size. Wage caps are shown in \% of state deputy salaries, as defined by the Constitutional Amendment No 25. in 2000, and in absolute values, as estimated by \cite{Ferraz2011} based on the 2004 federal deputy salary.
			\end{tablenotes}
		\end{threeparttable}
	}
\end{table}

\vspace{2em}

\begin{table}[h!]
	\centerline{
		\begin{threeparttable}
			\caption{Ideological classification of political parties}
			\label{tab:class_parties_ideo}
\begin{tabular}{ccc}
\toprule[1.5pt] 
\textbf{Left} & \textbf{Center} & \textbf{Right} \\
\midrule \\[-2.0ex]
PT            & MDB             & PP             \\
PDT           & PSDB            & DEM            \\
PSB           & PTB             & PL             \\
CIDADANIA     & AVANTE          & PSC            \\
PCDOB         & PSD             & REPUBLICANOS   \\
PV            &                 & PSL            \\
PMN           &                 & PTC            \\
PSOL          &                 & DC             \\
PCB           &                 & PODE           \\
PSTU          &                 & PRTB           \\
PCO           &                 & PRP            \\ 
SD            &                 & PHS            \\ 
PROS          &                 & PATRIOTA       \\ 
PPL           &                 & NOVO           \\ 
PMB           &                 &               \\ 
REDE          &                 &               \\ 
UP            &                 &               \\ 
\hline\hline
\end{tabular}
			\begin{tablenotes} \scriptsize 
				\item \textit{Note:} 
				The table shows the ideological classification of political parties, based on \cite{Colonnelli2022}.
			\end{tablenotes}
		\end{threeparttable}
	}
\end{table}

\clearpage

\begin{figure}[h!]
	\caption{Heterogeneity by party ideology}  \label{fig:es_ideo}
	\centerline{
		\begin{threeparttable}
			\begin{footnotesize}
				\begin{tabular}{cc}
					\textbf{A. New party membership} & \textbf{B. Candidacy}\\
					\includegraphics[width=0.55\linewidth]{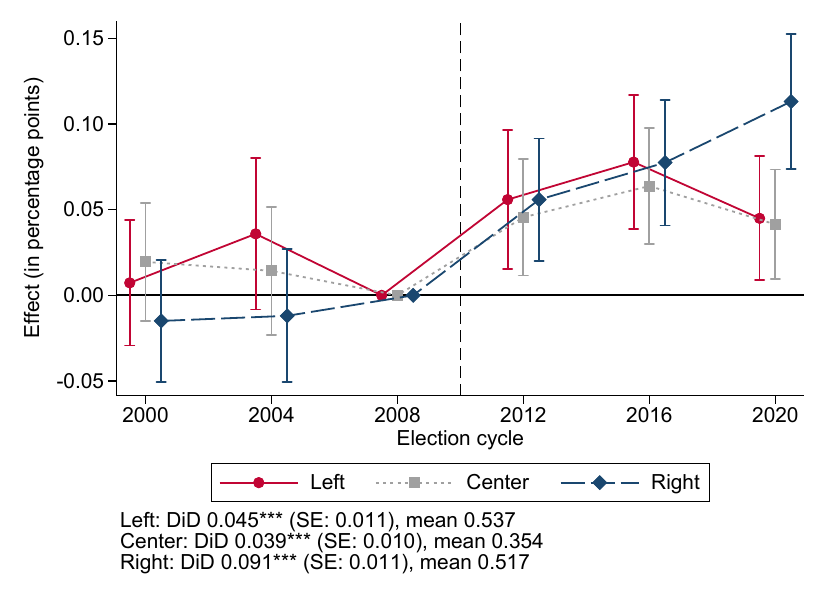} &\includegraphics[width=0.55\linewidth]{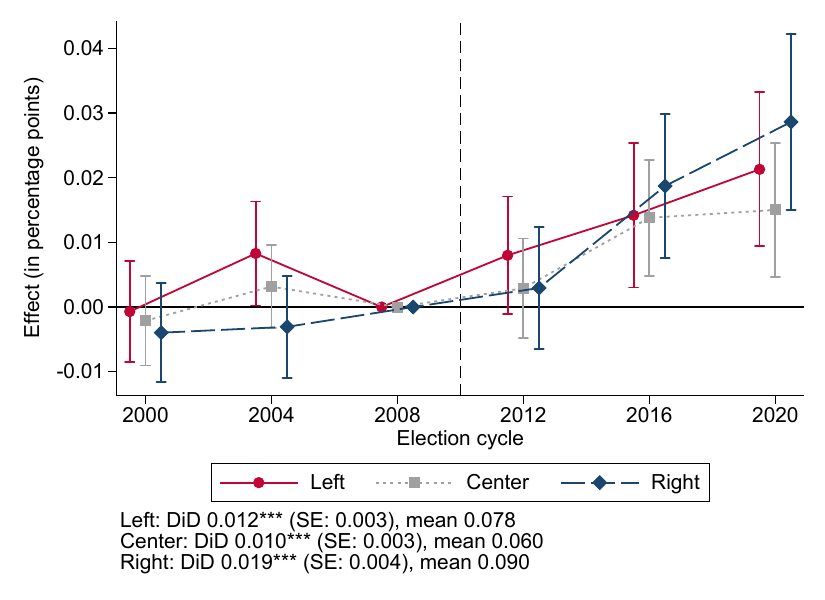} \\ 
				\end{tabular}
			\end{footnotesize}
			\begin{tablenotes} \scriptsize
			\item \textit{Note:}
				The figure reports event-study coefficients $\delta_c$, estimated in model (\ref{model:es}), for the effect of job loss on the likelihood of newly registering as a party member (Panel A) and running for local councilor (Panel B) in left, center, and right parties. The ideological classification of parties follows \cite{Colonnelli2022} and is shown Table \ref{tab:class_parties_ideo}. The vertical bars depict 95\% confidence intervals based on standard errors clustered at the individual level. Below each graph, the DiD coefficient from model (\ref{model:did}), its standard error, and the mean of the control group in the post-treatment period (average across 2012, 2016, and 2020) are reported.
			\end{tablenotes}
	\end{threeparttable}}
\end{figure}

\vspace{2em}

\begin{figure}[h!]
	\caption{Heterogeneity by party local incumbency}  \label{fig:es_incumbent}
	\centerline{
		\begin{threeparttable}
			\begin{footnotesize}
				\begin{tabular}{cc}
					\textbf{A. New party membership} & \textbf{B. Candidacy}\\
					\includegraphics[width=0.55\linewidth]{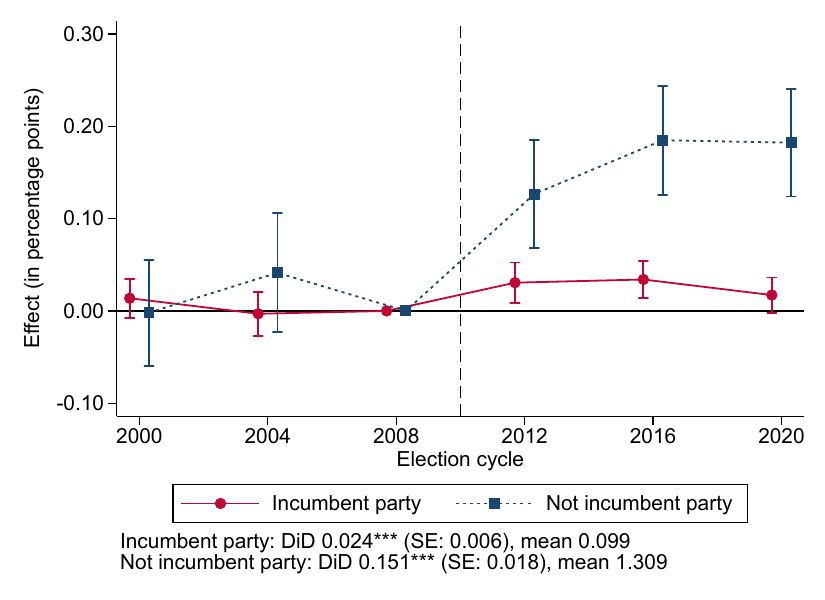} &\includegraphics[width=0.55\linewidth]{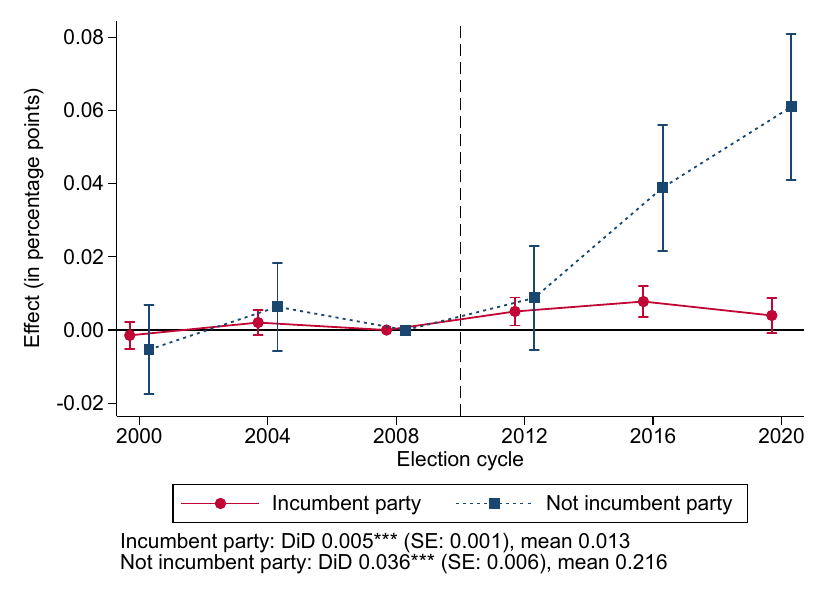} \\ 
				\end{tabular}
			\end{footnotesize}
			\begin{tablenotes} \scriptsize
			\item \textit{Note:}
				The figure reports event-study coefficients $\delta_c$, estimated in model (\ref{model:es}), for the effect of job loss on the likelihood of newly registering as a party member (Panel A) and running for local councilor (Panel B). Results are shown separately for memberships and candidacies in the party of the elected mayor in 2008 and in other non-incumbent parties. The vertical bars depict 95\% confidence intervals based on standard errors clustered at the individual level. Below each graph, the DiD coefficient from model (\ref{model:did}), its standard error, and the mean of the control group in the post-treatment period (average across 2012, 2016, and 2020) are reported. 
			\end{tablenotes}
	\end{threeparttable}}
\end{figure}

\clearpage

\begin{figure}[h!]
	\caption{Heterogeneity by party union affiliation}  \label{fig:es_union}
	\centerline{
		\begin{threeparttable}
			\begin{footnotesize}
				\begin{tabular}{cc}
					\textbf{A. New party membership} & \textbf{B. Candidacy}\\
					\includegraphics[width=0.55\linewidth]{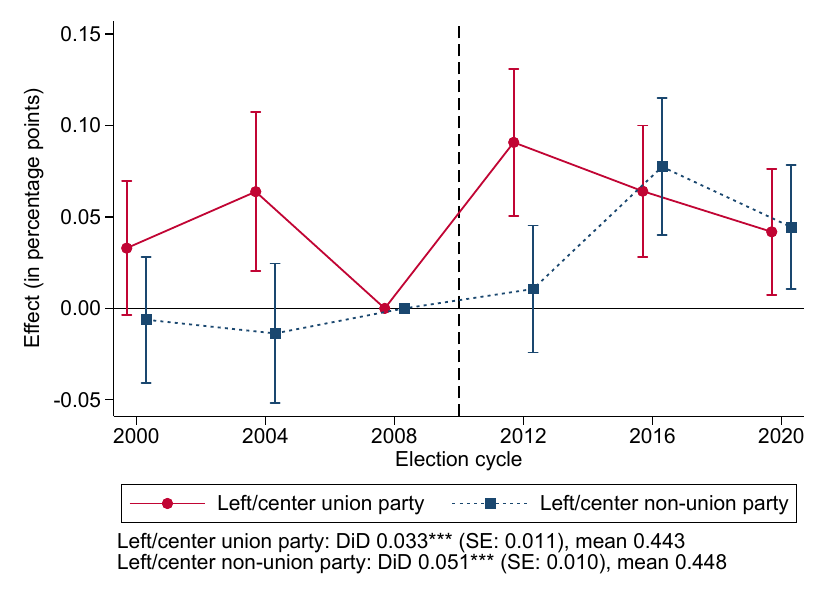} &\includegraphics[width=0.55\linewidth]{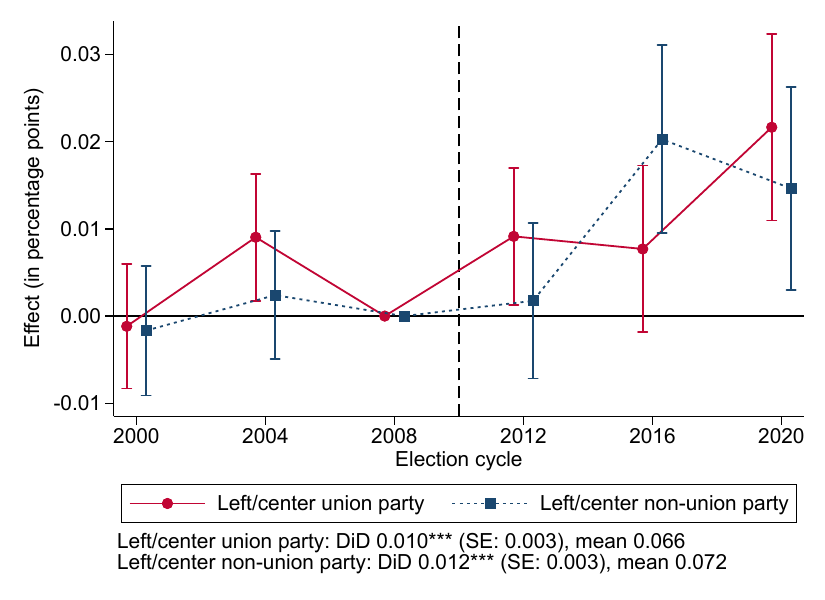} \\ 
				\end{tabular}
			\end{footnotesize}
			\begin{tablenotes} \scriptsize
			\item \textit{Note:}
				The figure reports event-study coefficients $\delta_c$, estimated in model (\ref{model:es}), for the effect of job loss on the likelihood of newly registering as a party member (Panel A) and running for local councilor (Panel B). Results are shown separately for memberships and candidacies in parties historically affiliated with labor unions (PT, PDT, PSB, PCB, PSD, and MDB) and all other left or center parties (defined in Table \ref{tab:class_parties_ideo}). The vertical bars depict 95\% confidence intervals based on standard errors clustered at the individual level. Below each graph, the DiD coefficient from model (\ref{model:did}), its standard error, and the mean of the control group in the post-treatment period (average across 2012, 2016, and 2020) are reported. 
			\end{tablenotes}
	\end{threeparttable}}
\end{figure}

\clearpage

\begin{figure}[h!]
	\caption{Heterogeneity by party}  \label{fig:did_party}
	\centerline{
		\begin{threeparttable}
			\begin{footnotesize}\begin{center}
					\textbf{A. New party membership} \\
     				\includegraphics[width=\linewidth]{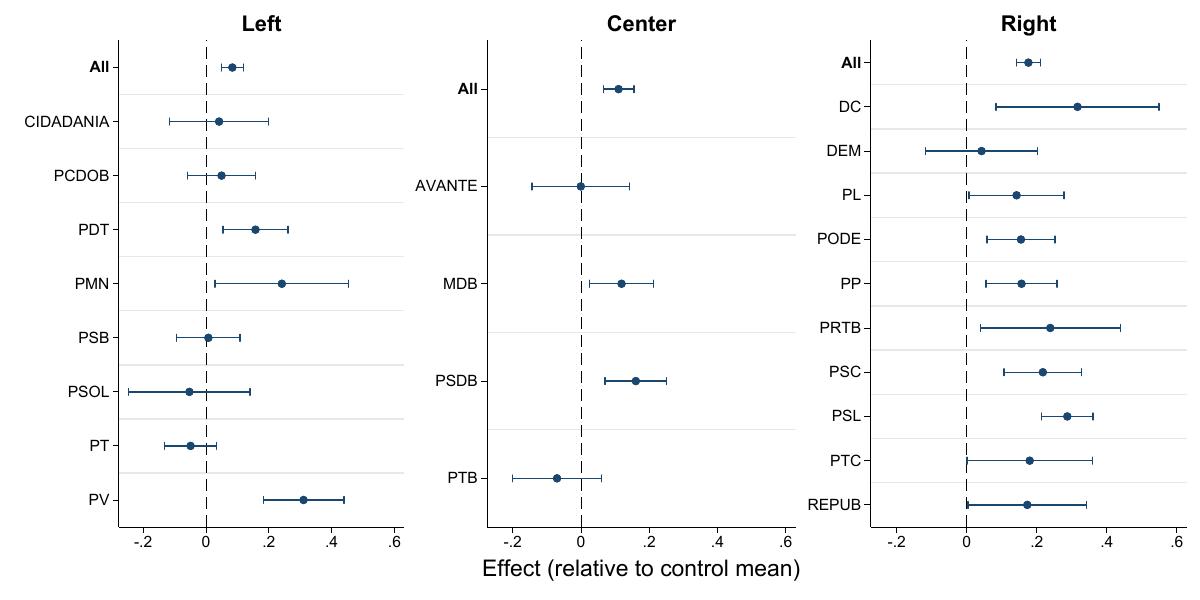} \\
                        \vspace{1em}
                         \textbf{B. Candidacy}\\
					\includegraphics[width=\linewidth]{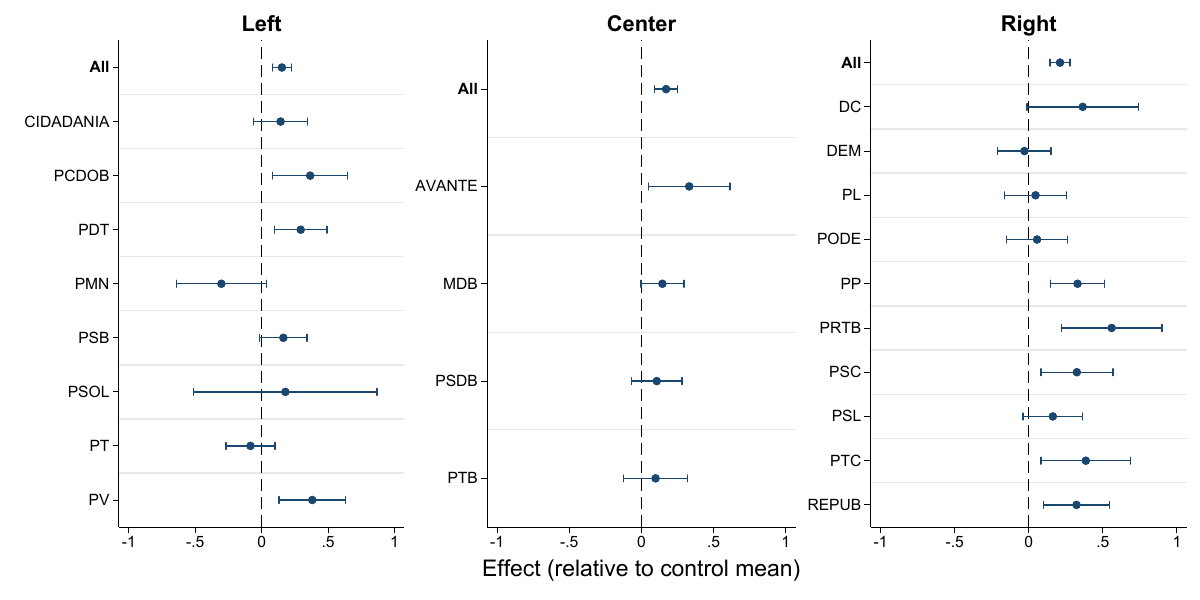}  
			\end{center}\end{footnotesize}
			\begin{tablenotes} \scriptsize
			\item \textit{Note:}
                The figure reports DiD coefficients, estimated in model (\ref{model:did}), for the effect of job loss on the likelihood of newly registering as a party member (Panel A) and running for local councilor (Panel B) in each of Brazil's major parties. The ideological classification of parties follows \cite{Colonnelli2022}. The horizontal bars depict 90\% confidence intervals based on standard errors clustered at the individual level. All coefficients and standard errors are scaled by the mean of the control group in the post-treatment period (average across 2012, 2016, and 2020). 
			\end{tablenotes}
	\end{threeparttable}
        }
\end{figure}

\clearpage

\renewcommand{\thesection}{C}
\renewcommand{\thetable}{C.\arabic{table}}
\setcounter{table}{0}
\renewcommand{\thefigure}{C.\arabic{figure}}
\setcounter{figure}{0}

\section{Appendix to Section \ref{sec:ui}}

\begin{figure}[h!]
	\caption{Layoff date cyclicality}  \label{fig:ui_cycles}
	\centerline{
		\begin{threeparttable}
			\begin{footnotesize}
				\begin{tabular}{cc}
					\textbf{A. Layoff day in month} & \textbf{B. Density around cutoff}\\
					\includegraphics[width=0.55\linewidth]{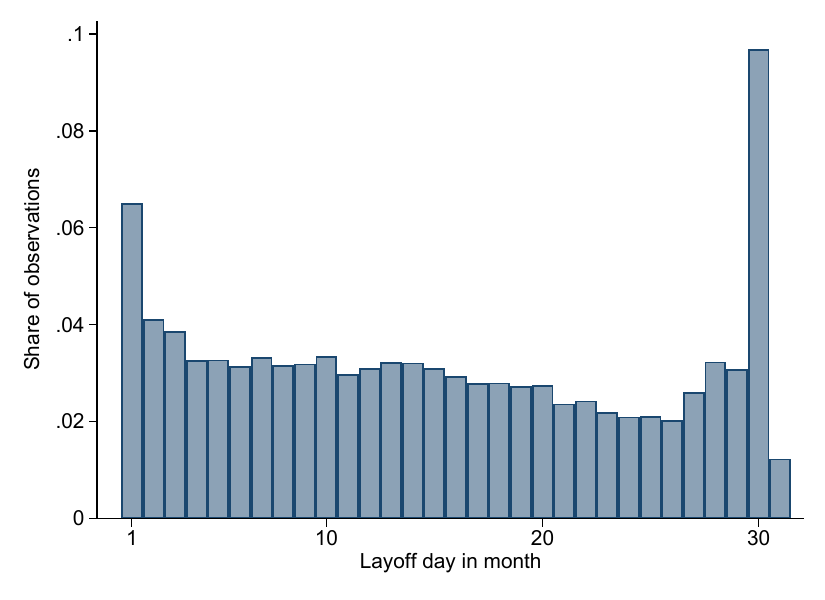} &\includegraphics[width=0.55\linewidth]{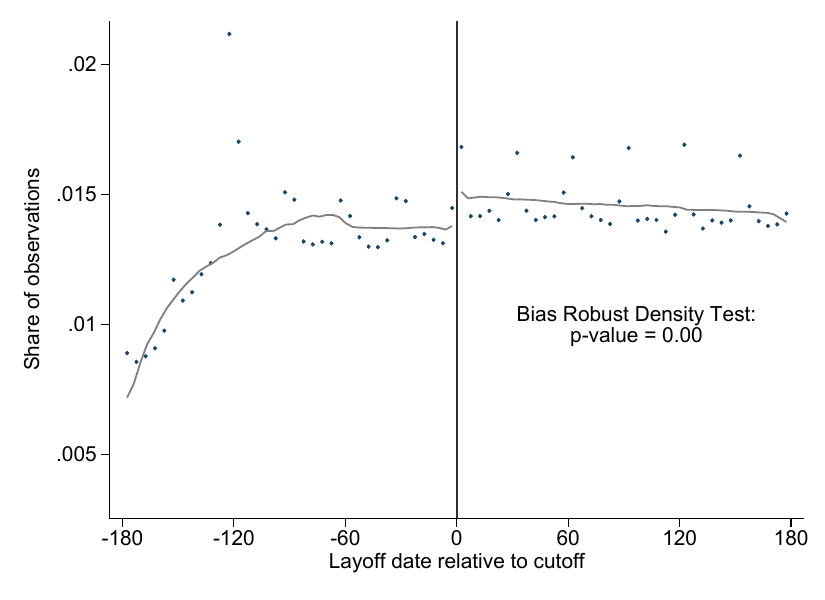} \\ 
				\end{tabular}
			\end{footnotesize}
			\begin{tablenotes} \scriptsize
			\item \textit{Note:}
                    Panel A reports a histogram for the distribution of layoff dates by calendar day in each month. Panel B reports the density of time between the current and the last layoff date around the 16-month cutoff date for eligibility for unemployment benefits, based on the initial sample that includes all layoff dates. The line represents a local linear polynomial smoothing using a 60-day bandwidth. Results for the bias-robust density test proposed by \cite{cattaneo2018, cattaneo2020} are also shown.
			\end{tablenotes}
	\end{threeparttable}}
\end{figure}

\vspace{2em}

\begin{figure}[h!]
	\caption{Layoff date density}  \label{fig:ui_density}
	\centerline{
		\begin{threeparttable}
                \centering
			\includegraphics[width=0.65\linewidth]{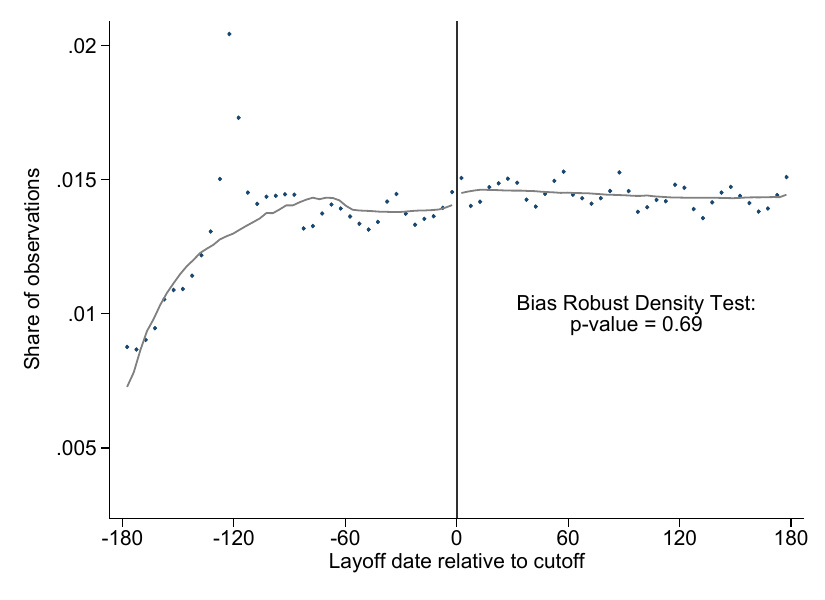}
			\begin{tablenotes} \scriptsize
			\item \textit{Note:}
			The figure reports the density of time between the current and the last layoff date around the 16-month cutoff date for eligibility for unemployment benefits, based on our estimation sample that excludes all workers who were displaced on the first or the last day of each month in their last layoff. The line represents a local linear polynomial smoothing using a 60-day bandwidth. Results for the bias-robust density test proposed by \cite{cattaneo2018, cattaneo2020} are also shown.
			\end{tablenotes}
	\end{threeparttable}}
\end{figure}

\clearpage

\begin{figure}[h!]
	\caption{Balance of covariates near the UI eligibility cutoff}  \label{fig:ui_covariates}
	\centerline{
		\begin{threeparttable}
                \centering
                \hspace*{-1.8cm}
			\includegraphics[width=1.2\linewidth]{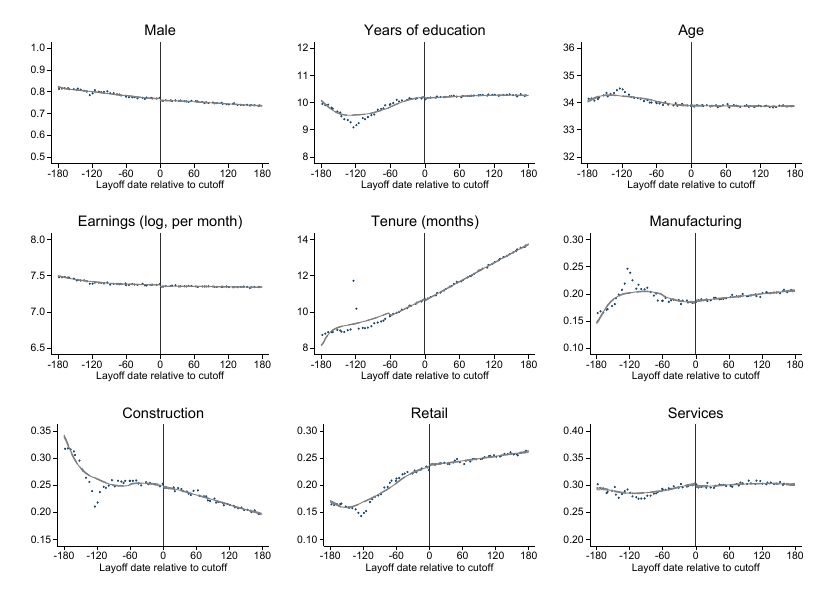}
			\begin{tablenotes} \scriptsize
			\item \textit{Note:}
			The figure plots the average characteristics of workers around the 16-month cutoff date for eligibility for unemployment benefits. All characteristics are measured just before the layoff. Dots show averages for 5-day bins. The lines represent a local linear polynomial smoothing using a 60-day bandwidth, together with 95\% confidence intervals.
			\end{tablenotes}
	\end{threeparttable}}
\end{figure}

\clearpage

\begin{table}[h!]
	\centerline{
		\begin{threeparttable}
			\caption{Effect of UI eligibility on party membership: alternative specifications}
			\label{tab:ui_robust_member}
			\begin{footnotesize}
				\begin{tabular}{lcccccccccc}
					\toprule[1.5pt] 
					& (1) & (2) & (3)& (4)& (5)& (6)& (7) & (8) & (9) & (10) \\
					\midrule \\[-2.0ex]
					Bandwidth (days): & 30 & opt & 30 & 60 & 90 & opt & 90 & 120 & 150 & opt \\
					[0.5em]
					Polynomial order: & 0 & 0 & 1 & 1 & 1 & 1 & 2 & 2 & 2 & 2 \\	
					\midrule
					\multicolumn{11}{l}{\textbf{[A] Full sample}} \\ 
					[0.5em]
$\beta$ &       0.092\sym{*}  &       0.097\sym{**} &       0.200\sym{**} &       0.150\sym{**} &       0.122\sym{**} &       0.162\sym{**} &       0.174\sym{*}  &       0.153\sym{**} &       0.150\sym{**} &       0.173\sym{*}  \\
               &     (0.051)         &     (0.049)         &     (0.102)         &     (0.072)         &     (0.059)         &     (0.075)         &     (0.089)         &     (0.077)         &     (0.069)         &     (0.090)         \\
[0.5em]
Control mean   &       3.500         &       3.493         &       3.463         &       3.467         &       3.479         &       3.454         &       3.465         &       3.461         &       3.458         &       3.450         \\
Relative effect&         2.6\%         &         2.8\%         &         5.8\%         &         4.3\%         &         3.5\%         &         4.7\%         &         5.0\%         &         4.4\%         &         4.3\%         &         5.0\%         \\
Observations   &     529,098         &     565,362         &     529,098         &   1,047,683         &   1,560,276         &     976,135         &   1,560,276         &   2,083,546         &   2,590,023         &   1,507,096         \\
					[0.5em]
					\midrule
					\multicolumn{9}{l}{\textbf{[B] Low educated ($<$ 12 years of education)}} \\ 
					[0.5em]
$\beta$   &       0.033         &       0.017         &       0.128         &       0.066         &      -0.031         &       0.091         &       0.100         &       0.008         &       0.016         &       0.112         \\
               &     (0.077)         &     (0.074)         &     (0.153)         &     (0.109)         &     (0.089)         &     (0.112)         &     (0.133)         &     (0.115)         &     (0.104)         &     (0.136)         \\
[0.5em]
Control mean   &       3.479         &       3.473         &       3.417         &       3.454         &       3.487         &       3.449         &       3.471         &       3.483         &       3.469         &       3.432         \\
Relative effect&         0.9\%         &         0.5\%         &         3.8\%         &         1.9\%         &        -0.9\%         &         2.6\%         &         2.9\%         &         0.2\%         &         0.5\%         &         3.3\%         \\
Observations   &     229,191         &     244,890         &     229,191         &     454,665         &     682,272         &     430,963         &     682,272         &     922,954         &   1,157,173         &     650,326         \\
					[0.5em]
					\midrule
					\multicolumn{9}{l}{\textbf{[C] High educated ($\geq$ 12 years of education)}} \\ 
					[0.5em]
$\beta$  &       0.137\sym{**} &       0.137\sym{**} &       0.255\sym{*}  &       0.216\sym{**} &       0.244\sym{***}&       0.231\sym{**} &       0.230\sym{*}  &       0.262\sym{**} &       0.259\sym{***}&       0.214\sym{*}  \\
               &     (0.068)         &     (0.068)         &     (0.136)         &     (0.097)         &     (0.080)         &     (0.094)         &     (0.119)         &     (0.104)         &     (0.093)         &     (0.121)         \\
[0.5em]
Control mean   &       3.517         &       3.517         &       3.498         &       3.476         &       3.467         &       3.459         &       3.462         &       3.446         &       3.443         &       3.460         \\
Relative effect&         3.9\%         &         3.9\%         &         7.3\%         &         6.2\%         &         7.0\%         &         6.7\%         &         6.6\%         &         7.6\%         &         7.5\%         &         6.2\%         \\
Observations   &     299,907         &     299,907         &     299,907         &     593,018         &     878,004         &     641,120         &     878,004         &   1,160,592         &   1,432,850         &     848,505         \\
[0.5em]
\hline\hline
				\end{tabular}
			\end{footnotesize}
			\begin{tablenotes} \scriptsize 
				\item \textit{Note:} 
				The table presents robustness checks for the effect of unemployment benefit eligibility on the likelihood of newly registering as a party member when using different bandwidths and polynomial orders of the running variable. `opt' refers to the optimal bandwidth proposed by \cite{calonico2014}. Standard errors clustered at the individual level are in parentheses. The table also reports the control mean of the outcome at the cutoff and the effect sizes scaled by the control mean. All coefficients, standard errors, and control means have been scaled by 100, such that effects are interpreted in terms of percentage points. \sym{*} \(p<0.10\), \sym{**} \(p<0.05\), \sym{***} \(p<0.01\)
			\end{tablenotes}
		\end{threeparttable}
	}
\end{table}

\clearpage

\begin{table}[h!]
	\centerline{
		\begin{threeparttable}
			\caption{Effect of UI eligibility on running for local councilor: alternative specifications}
			\label{tab:ui_robust_candidate}
			\begin{footnotesize}
				\begin{tabular}{lcccccccccc}
					\toprule[1.5pt] 
					& (1) & (2) & (3)& (4)& (5)& (6)& (7) & (8) & (9) & (10) \\
					\midrule \\[-2.0ex]
					Bandwidth (days): & 30 & opt & 30 & 60 & 90 & opt & 90 & 120 & 150 & opt \\
					[0.5em]
					Polynomial order: & 0 & 0 & 1 & 1 & 1 & 1 & 2 & 2 & 2 & 2 \\	
					\midrule
					\multicolumn{11}{l}{\textbf{[A] Full sample}} \\ 
					[0.5em]
$\beta$   &       0.025\sym{*}  &       0.020         &       0.055\sym{**} &       0.028         &       0.030\sym{*}  &       0.030         &       0.031         &       0.024         &       0.025         &       0.030         \\
               &     (0.013)         &     (0.013)         &     (0.027)         &     (0.019)         &     (0.016)         &     (0.020)         &     (0.023)         &     (0.020)         &     (0.018)         &     (0.024)         \\
[0.5em]
Control mean   &       0.447         &       0.453         &       0.431         &       0.449         &       0.444         &       0.447         &       0.450         &       0.446         &       0.451         &       0.447         \\
Relative effect&         5.6\%         &         4.3\%         &        12.9\%         &         6.2\%         &         6.7\%         &         6.7\%         &         7.0\%         &         5.3\%         &         5.6\%         &         6.6\%         \\
Observations   &   1,035,237         &   1,105,868         &   1,035,237         &   2,048,674         &   3,058,455         &   1,908,388         &   3,058,455         &   4,101,059         &   5,107,090         &   2,953,238         \\
					[0.5em]
					\midrule
					\multicolumn{9}{l}{\textbf{[B] Low educated ($<$ 12 years of education)}} \\ 
					[0.5em]
$\beta$   &       0.001         &      -0.007         &       0.035         &      -0.006         &      -0.004         &      -0.003         &      -0.002         &      -0.008         &      -0.001         &      -0.004         \\
               &     (0.018)         &     (0.018)         &     (0.037)         &     (0.026)         &     (0.021)         &     (0.027)         &     (0.032)         &     (0.028)         &     (0.025)         &     (0.033)         \\
[0.5em]
Control mean   &       0.419         &       0.424         &       0.396         &       0.428         &       0.417         &       0.427         &       0.435         &       0.427         &       0.423         &       0.426         \\
Relative effect&         0.3\%         &        -1.6\%         &         8.9\%         &        -1.4\%         &        -0.9\%         &        -0.6\%         &        -0.5\%         &        -1.9\%         &        -0.2\%         &        -1.0\%         \\
Observations   &     497,269         &     531,075         &     497,269         &     986,755         &   1,485,874         &     934,928         &   1,485,874         &   2,021,245         &   2,540,083         &   1,414,964         \\
					[0.5em]
					\midrule
					\multicolumn{9}{l}{\textbf{[C] High educated ($\geq$ 12 years of education)}} \\ 
					[0.5em]
$\beta$  &       0.047\sym{**} &       0.047\sym{**} &       0.075\sym{**} &       0.061\sym{**} &       0.063\sym{***}&       0.072\sym{***}&       0.063\sym{*}  &       0.054\sym{*}  &       0.053\sym{**} &       0.063\sym{*}  \\
               &     (0.019)         &     (0.019)         &     (0.038)         &     (0.027)         &     (0.023)         &     (0.026)         &     (0.034)         &     (0.029)         &     (0.026)         &     (0.034)         \\
[0.5em]
Control mean   &       0.473         &       0.473         &       0.463         &       0.468         &       0.465         &       0.461         &       0.463         &       0.464         &       0.473         &       0.462         \\
Relative effect&        10.0\%         &        10.0\%         &        16.2\%         &        13.0\%         &        13.6\%         &        15.6\%         &        13.5\%         &        11.6\%         &        11.2\%         &        13.6\%         \\
Observations   &     537,968         &     537,968         &     537,968         &   1,061,919         &   1,572,581         &   1,148,350         &   1,572,581         &   2,079,814         &   2,567,007         &   1,519,603         \\
[0.5em]
\hline\hline
				\end{tabular}
			\end{footnotesize}
			\begin{tablenotes} \scriptsize 
				\item \textit{Note:} 
				The table presents robustness checks for the effect of unemployment benefit eligibility on the likelihood of running for local councilor when using different bandwidths and polynomial orders of the running variable. `opt' refers to the optimal bandwidth proposed by \cite{calonico2014}. Standard errors clustered at the individual level are in parentheses. The table also reports the control mean of the outcome at the cutoff and the effect sizes scaled by the control mean. All coefficients, standard errors, and control means have been scaled by 100, such that effects are interpreted in terms of percentage points. \sym{*} \(p<0.10\), \sym{**} \(p<0.05\), \sym{***} \(p<0.01\)
			\end{tablenotes}
		\end{threeparttable}
	}
\end{table}

\clearpage

\begin{figure}[h!]
	\caption{Effect of UI eligibility: permutation tests}  \label{fig:ui_ptests}
	\centerline{
		\begin{threeparttable}
  			\begin{footnotesize}
				\begin{tabular}{cc}
                        \multicolumn{2}{c}{\textbf{A. Full sample}}\\ [0.5em]
					New party membership & Candidacy \\
					\includegraphics[width=0.35\linewidth]{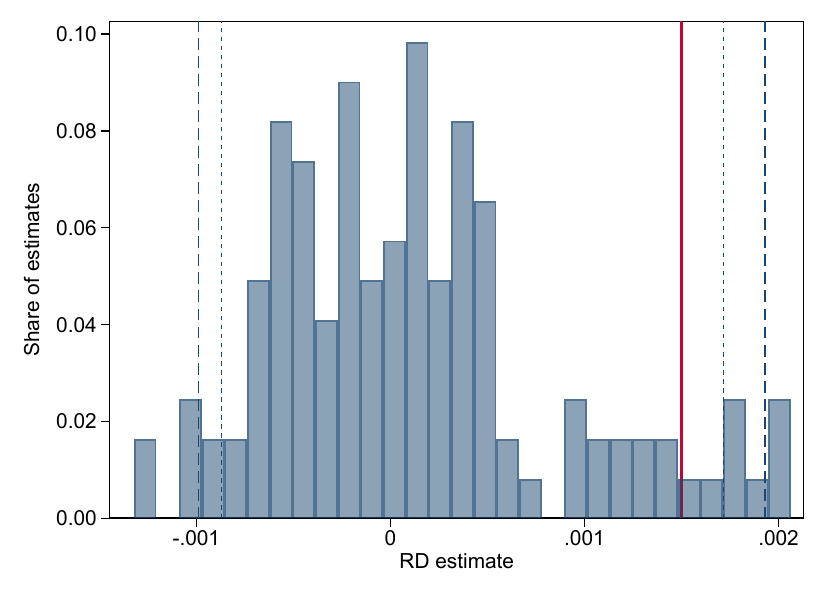}
                        & \includegraphics[width=0.35\linewidth]{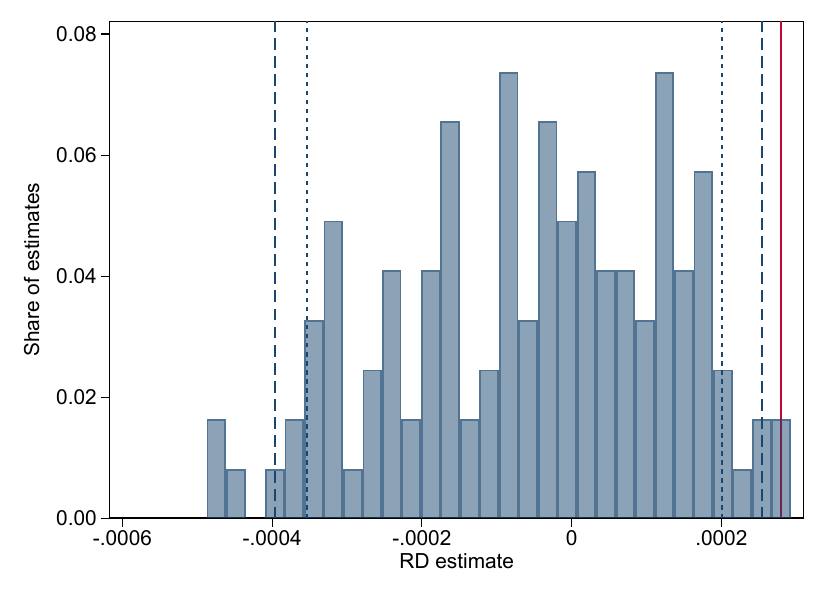}   \\ [1em]
                        \multicolumn{2}{c}{\textbf{B. Low educated ($<$ 12 years of education)}}\\ [0.5em]
					New party membership & Candidacy \\
					\includegraphics[width=0.35\linewidth]{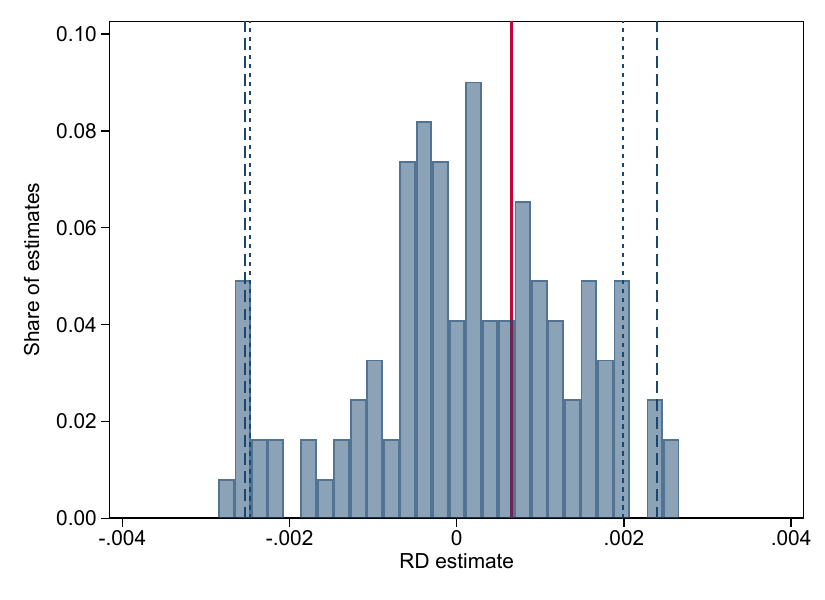}
                        & \includegraphics[width=0.35\linewidth]{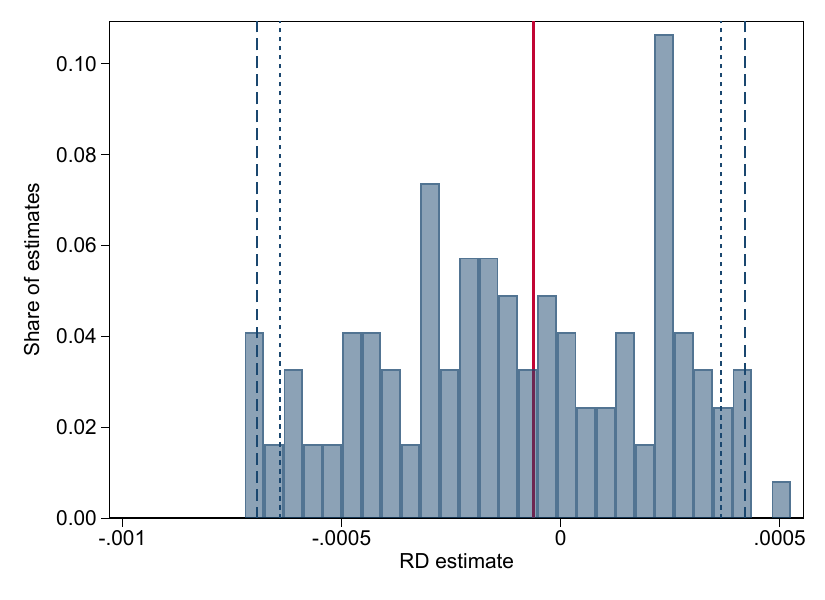}   \\ [1em]        
                        \multicolumn{2}{c}{\textbf{C. High educated ($\geq$ 12 years of education)}}\\ [0.5em]
					New party membership & Candidacy \\
					\includegraphics[width=0.35\linewidth]{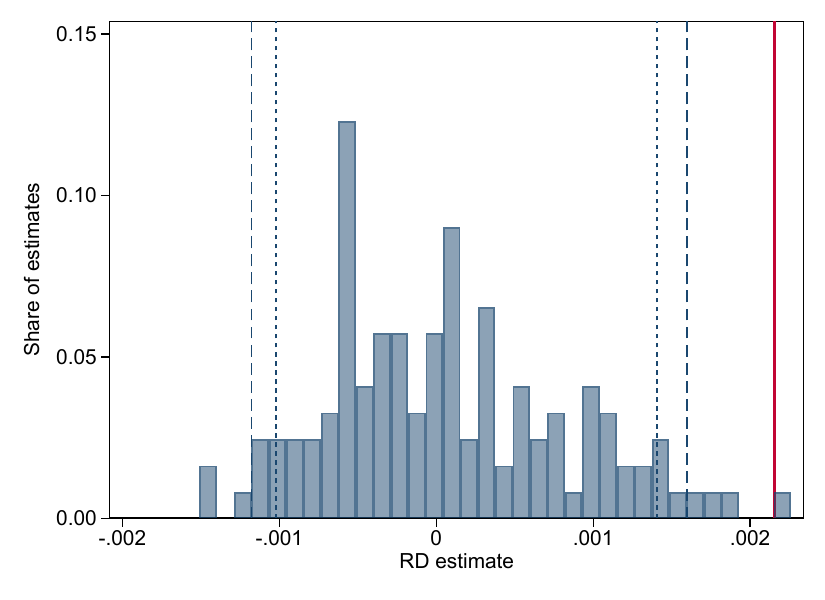}
                        & \includegraphics[width=0.35\linewidth]{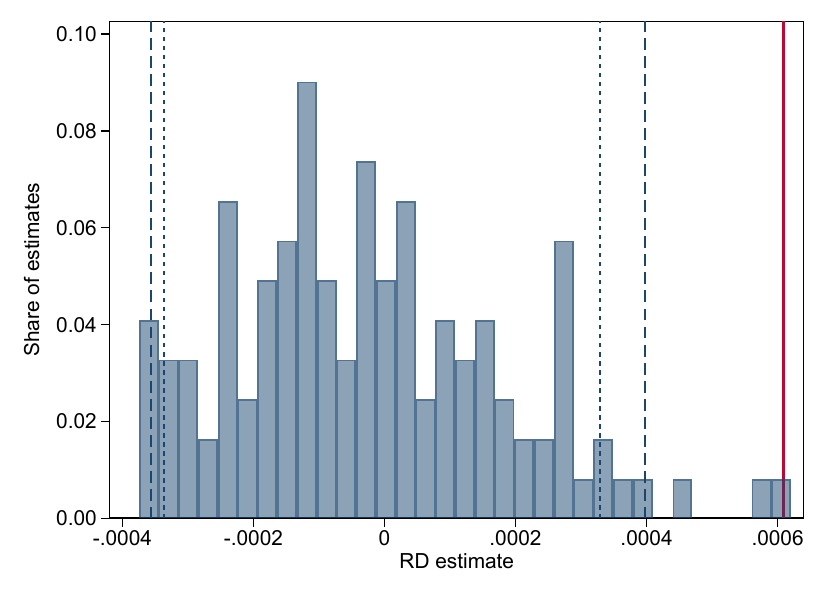}   \\ [1em]     
				\end{tabular}
			\end{footnotesize}
			\begin{tablenotes} \scriptsize
			\item \textit{Note:}
			The figures compare RD estimates for the effect of UI eligibility on the likelihood of newly registering as a party member and running for local councilor at the true 16-month cutoff (vertical red line) with the distribution of estimates at all possible placebo cutoffs within 180 days away from the true cutoff. The dashed (dotted) lines indicate the 2.5 (5) and 97.5 (95) percentiles of the distribution of placebo cutoffs. Estimates are based on the local linear regression model (\ref{model:rdd}) with a bandwidth of 60 days. Panel A shows results among all workers in the sample, and Panels B and C distinguish workers with less than vs. at least 12 years of education. 
			\end{tablenotes}
	\end{threeparttable}}
\end{figure}

\clearpage

\renewcommand{\thesection}{D}
\renewcommand{\thetable}{D.\arabic{table}}
\setcounter{table}{0}
\renewcommand{\thefigure}{D.\arabic{figure}}
\setcounter{figure}{0}

\section{Appendix to Section \ref{sec:implications}}

\begin{table}[h!]
    \centering
    \begin{threeparttable}
    \caption{Summary of AKM estimation}
    \label{tab:akm_summary}
    \begin{tabular}{l*{1}{c}}
    \toprule[1.5pt] 
    \multicolumn{2}{l}{\textbf{Largest connected set}} \\
Share of sample  &     0.983 \\
Number of worker-year observations &  189,819,461 \\
Number of establishments &    3,245,750 \\
Number of workers &   48,146,471 \\
Share of workers with any job change &     0.504 \\
SD of log wages &     0.804 \\ 
\\
\multicolumn{2}{l}{\textbf{Summary of estimation results}} \\
SD of worker effects  &     0.638 \\
SD of establishment effects &     0.331 \\
SD of covariates  &     0.159 \\
SD of residuals &     0.223 \\
Correlation of worker and establishment effects & 0.311  \\
Adjusted $R^2$ of the model &     0.895 \\
Root Mean Squared Error of the model &     0.261  \\
\\
\multicolumn{2}{l}{\textbf{Share of variance of log wages due to}}\\
Worker effects &     0.629  \\
Establishment effect &     0.169   \\
Covariance of worker and establishment effects &    0.203      \\ 
\hline\hline
\end{tabular}
        \begin{tablenotes} \scriptsize 
            \item \textit{Note:} The table summarizes the results from estimating the model $y_{it} = \alpha_i + \psi_{\boldsymbol{j}(i,t)}+X_{it}'\beta+\varepsilon_{it}$, where the dependent variable is the log hourly wage of worker $i$ at year $t$. The model includes worker fixed effects $\alpha_i$ and establishment fixed effects $\psi_{\boldsymbol{j}(i,t)}$, as well as a set of covariates $X_{it}$ which contains year dummies and quadratic and cubic age, all interacted with dummies for five educational groups and with gender. The model is estimated on the largest connected set of establishments and workers aged 18-60 in RAIS between 2004 and 2008.
        \end{tablenotes}
    \end{threeparttable}
\end{table}

\clearpage
\begin{table}[h!]
	\centerline{
		\begin{threeparttable}
			\caption{Correlation of candidate characteristics with misconduct during candidacy and electoral performance}
			\label{tab:validation}
				\begin{tabular}{lcccc}
					\toprule[1.5pt] 
					& (1) & (2) & (3) & (4) \\
					\midrule \\[-2.0ex]
				 & \multicolumn{2}{c}{All candidates} & \multicolumn{2}{c}{Candidates matched to RAIS} \\ \cmidrule(lr){2-3} \cmidrule(lr){4-5}
					& Misconduct & Elected  & Misconduct & Elected \\	
					\midrule
Years of education & -0.0017\sym{***}&   0.0045\sym{***} &  -0.0014\sym{***}&    0.0051\sym{***} \\
                   &    (0.0000)     &    (0.0001)       &     (0.0001)     &    (0.0001)    \\
[0.5em]
Male               & -0.0057\sym{***}&   0.1009\sym{***} &  -0.0012\sym{***}&    0.0715\sym{***}\\
                   &    (0.0004)     &    (0.0006)       &     (0.0006)     &    (0.0009)    \\
[0.5em]
Age                & 0.0001\sym{***} &  -0.0004\sym{***} &  0.0002\sym{***} &    -0.0013\sym{***}\\
                   &    (0.0000)     &    (0.0000)       &     (0.0000)     &    (0.0001)    \\
[0.5em]
Log wage 	  &             &    &  -0.0025\sym{***}&      0.0377\sym{***}  \\
              &             &    &  (0.0004)         &    (0.0007)                \\
[0.5em]
AKM person effect   & & &   -0.0029\sym{***}&      0.0364\sym{***} \\
                  &             &         &      (0.0003)         &    (0.0005)                \\
[0.5em]
AKM person effect residual  &   & &   -0.0016\sym{***}&      0.0230\sym{***} \\
              &             &              &    (0.0002)         &    (0.0005)                \\
[1em]
Mean dep. variable    &  0.0358         &      0.1181   &    0.0313         &      0.1242    \\
Observations      & 980,130 & 980,130    & 454,816  & 454,816  \\
[0.5em]
\hline\hline
				\end{tabular}
			\begin{tablenotes} \scriptsize 
				\item \textit{Note:} The table shows correlations between various characteristics of councilor candidates and indicators for engaging in misconduct behavior during candidacy (columns (1) and (3)) and for being elected (columns (2) and (4)). The misbehavior variable captures irregularities detected by TSE during the elections in 2016 and 2020 that resulted in candidacy annulment, including abuse of power, failure to meet registration requirements, vote buying, prohibited conduct, ineligibility under the ``Clean Record Act'' (\textit{Lei da Ficha Limpa}), illicit use of campaign resources, objections to the candidacy, rejection by the party or coalition, and invalidation of the party. See Section \ref{sec:implications_individual} for details on the measurements of the various candidate characteristics. In columns (1) and (2), the sample consists of all individuals who run for local councilor in the election years 2016 and 2020. Since TSE reports the socio-demographic information of candidates, we can show their correlation with misconduct and election outcomes for the full population of councilor candidates. For the wage measures shown in columns (3) and (4), we are limited to the subsample of candidates who appear in RAIS before the layoff period, i.e., those with at least one formal work contract between 2004 and 2008 (about 46\% of all candidates).. Each cell shows the result of a bivariate OLS regression. Robust standard errors are in parentheses. \sym{*} \(p<0.10\), \sym{**} \(p<0.05\), \sym{***} \(p<0.01\)
			\end{tablenotes}
		\end{threeparttable}
	}
\end{table}

\clearpage
\begin{table}[h!]
	\centerline{
		\begin{threeparttable}
			\caption{Municipality-level labor demand and gender of councilor candidates}
			\label{tab:bartik_gender}
				\begin{tabular}{lcccc}
					\toprule[1.5pt] 
					& (1) & (2) & (3)& (4) \\
					\midrule \\[-2.0ex]
                        Outcome: & \multicolumn{4}{c}{$\Delta$ Share male (ppt) of ...} \\\cmidrule(lr){2-5}
					& \multicolumn{2}{c}{Councilor candidates} & \multicolumn{2}{c}{Elected councilors}\\	
					\midrule
					\multicolumn{5}{l}{\textbf{[A] OLS}} \\ 
					[0.5em]
$\Delta$ Employment rate (ppt)  &       0.078\sym{***}&       0.061\sym{***}&       0.100\sym{**} &       0.063         \\
                                &     (0.020)         &     (0.020)         &     (0.050)         &     (0.051)         \\
[0.5em]
Observations                    &  16,650             &  16,650             &  16,650             & 16,650         \\
					[0.5em]
					\midrule
					\multicolumn{5}{l}{\textbf{[B] 2SLS (Bartik IV)}} \\ 
					[0.5em]
$\Delta$ Employment rate (ppt)	&       0.631\sym{***}&       0.631\sym{***}&       0.370\sym{**} &       0.354\sym{*}  \\
                                &     (0.071)         &     (0.082)         &     (0.159)         &     (0.189)         \\
[0.5em]
KP F-stat                       &       301.7         &       250.3         &       301.7         &     250.3         \\
Observations                    &  16,650             &  16,650             &  16,650             &  16,650        \\
[0.5em]
\midrule
Fixed effects  & Year & State $\times$ year & Year & State $\times$ year \\
[0.5em]
\hline\hline
				\end{tabular}
			\begin{tablenotes} \scriptsize 
				\item \textit{Note:} 
				The table presents OLS and 2SLS results from model (\ref{model:local}) for the effect of formal employment rates on the male share of local councilors. The sample contains all municipal elections in 2004, 2008, 2012, and 2016. The dependent variable measures the percentage point change across subsequent elections in the male share of individuals running (columns (1) and (2)) and being elected (columns (3) and (4)) for local council. The explanatory variable is the percentage point change in the municipal formal employment-to-population rate. For the 2SLS results, changes in employment rates are instrumented with Bartik instruments that are constructed by interacting baseline differences in municipalities' industry composition with national industry employment growth rates (see equation (\ref{bartik_iv})). All regressions are weighted by municipal population. Standard errors clustered at the municipality level are in parentheses. \sym{*} \(p<0.10\), \sym{**} \(p<0.05\), \sym{***} \(p<0.01\)
			\end{tablenotes}
		\end{threeparttable}
	}
\end{table}

\clearpage
\begin{table}[h!]
	\centerline{
		\begin{threeparttable}
			\caption{Municipality-level labor demand and age of councilor candidates}
			\label{tab:bartik_age}
				\begin{tabular}{lcccc}
					\toprule[1.5pt] 
					& (1) & (2) & (3)& (4) \\
					\midrule \\[-2.0ex]
                        Outcome: & \multicolumn{4}{c}{$\Delta$ Average age of ...} \\\cmidrule(lr){2-5}
					& \multicolumn{2}{c}{Councilor candidates} & \multicolumn{2}{c}{Elected councilors}\\	
					\midrule
					\multicolumn{5}{l}{\textbf{[A] OLS}} \\ 
					[0.5em]
$\Delta$ Employment rate (ppt)  &       0.022\sym{***}&       0.001         &       0.037\sym{**} &       0.027         \\
                                &     (0.008)         &     (0.009)         &     (0.017)         &     (0.018)         \\
[0.5em]
Observations                    &  16,650             &  16,650             &  16,650             & 16,650         \\
					[0.5em]
					\midrule
					\multicolumn{5}{l}{\textbf{[B] 2SLS (Bartik IV)}} \\ 
					[0.5em]
$\Delta$ Employment rate (ppt)	&       0.081\sym{***}&       0.020         &       0.115\sym{**} &       0.111\sym{*}  \\
                                &     (0.027)         &     (0.032)         &     (0.053)         &     (0.062)         \\
[0.5em]
KP F-stat                       &       301.7         &       250.3         &       301.7         &     250.3         \\
Observations                    &  16,650             &  16,650             &  16,650             &  16,650        \\
[0.5em]
\midrule
Fixed effects  & Year & State $\times$ year & Year & State $\times$ year \\
[0.5em]
\hline\hline
				\end{tabular}
			\begin{tablenotes} \scriptsize 
				\item \textit{Note:} 
				The table presents OLS and 2SLS results from model (\ref{model:local}) for the effect of formal employment rates on the average age of local councilors. The sample contains all municipal elections in 2004, 2008, 2012, and 2016. The dependent variable measures the change across subsequent elections in the average age of all individuals running (columns (1) and (2)) and being elected (columns (3) and (4)) for local council. The explanatory variable is the percentage point change in the municipal formal employment-to-population rate. For the 2SLS results, changes in employment rates are instrumented with Bartik instruments that are constructed by interacting baseline differences in municipalities' industry composition with national industry employment growth rates (see equation (\ref{bartik_iv})). All regressions are weighted by municipal population. Standard errors clustered at the municipality level are in parentheses. \sym{*} \(p<0.10\), \sym{**} \(p<0.05\), \sym{***} \(p<0.01\)
			\end{tablenotes}
		\end{threeparttable}
	}
\end{table}

\end{document}